\documentclass[twoside,report,seriffont, 11pt,numbers=noenddot]{tugrazbooklet}
\colorlet{main}{black}

\usepackage[utf8]{inputenc}
\usepackage[T1]{fontenc}
\usepackage[nottoc,numbib]{tocbibind}
\usepackage[backend=biber,defernumbers=true,maxbibnames=199]{biblatex} % ONE Bibliography in total
\usepackage[title,toc,titletoc,page]{appendix}
\usepackage{nameref}
\usepackage{overpic}

\usepackage{pdfpages}

\usepackage{amsmath,amsfonts,amssymb,bm,cancel,trfsigns}

\newcommand{\ra}{{\mathrm{a}}}

\usepackage{float, placeins, subfig}
\usepackage[framed, numbered]{matlab-prettifier}

\usepackage{tikz}

\makeatletter % DEFINITION OF MATHCOLOR ENVIRONMENT
\def\mathcolor#1#{\@mathcolor{#1}}
\def\@mathcolor#1#2#3{%
  \protect\leavevmode
  \begingroup
    \color#1{#2}#3%
  \endgroup
}
\makeatother

\usepackage[framemethod=TikZ]{mdframed}
\definecolor{IGTEyellow}{RGB}{245,213,10}
\definecolor{Tool}{RGB}{66, 165, 245}

\newcommand{\exmplLine}{black!15} % color and transparency of the example box

\newcounter{exmpl}[chapter]
\renewcommand{\theexmpl}{\thechapter.\arabic{exmpl}}

\newcommand{\learningLine}{IGTEyellow} % make lines in IGTE yellow
\newenvironment{learningobj}[1][]{%
\mdfsetup{%
frametitle={%
\tikz[baseline=(current bounding box.east),outer sep=0pt]
\node[anchor=east,rectangle,fill=\learningLine]
{\strut \sffamily #1};}}
\mdfsetup{innertopmargin=10pt,linecolor=\learningLine,%
linewidth=2pt,topline=true,%
frametitleaboveskip=\dimexpr-\ht\strutbox\relax
}
\begin{mdframed}[]\relax%
}{\end{mdframed}}

\newcommand{\toolLine}{Tool} % make lines in IGTE yellow
\newcounter{tool}[chapter]
\renewcommand{\thetool}{\thechapter.\arabic{tool}}

\DeclareUnicodeCharacter{0301}{O}

\usepackage[per-mode=fraction]{siunitx} % per-mode=fraction OR per-mode=symbol
\usepackage{multirow,booktabs}

\usepackage[LabelsAligned]{currvita}

\DeclareSourcemap{
  \maps[datatype=bibtex]{
    \map[overwrite]{
      \perdatasource{Biblatex_Schoder.bib}
      \step[fieldset=keywords, fieldvalue={,inbib}, append]
    }
  }
}

\usepackage[en-US]{datetime2}
\makeatletter
\newcommand{\monthyeardate}{%
  \@dtm@month/\@dtm@year
}
\makeatother

\newcommand{\myversion}{Graz} % Document Version Number

\begin{document}

% Metadata displayed on titlepage and (optional parameter) headline/footline:
% \title[FEM, BEM, and PINN: Overview on Numerical Methods]{FEM, BEM, and PINN: Overview on Numerical Methods}
\title[Aeroacoustics]{Aeroacoustics}

\subtitle{Theory and methods for analyzing flow-induced sound generation of technical and biological applications}
\author[S. Schoder]{Dr. Stefan Schoder}
\date{\myversion,~\monthyeardate}
\institute{Group of Aeroacoustics and Vibroacoustics (AVG)\\Institute of Fundamentals and Theory in Electrical Engineering (IGTE)\\Faculty of Electrical and Information Engineering\\Graz University of Technology}
\logobar{\includegraphics{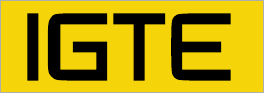}}

% Select filename for figures on titlepage (optional):
% Title page style depends on which parameters are set
\titlegraphic{Figures/schema_aiaa22}
\instituteacronym{institute/igte}

\pagenumbering{roman}

%%%%%%%%%%%%%%%%%%%%%%%%%%%%%%%%%%%%%%%%%%%%%%%%%%%%%%%%%%%%%%%%%%%%%%%%%%%%%%%%%5
\maketitle

%%%%%%%%%%%%%%%%%%%%%%%%%%%%%%%%%%%%%%%%%%%%%%%%%%%%%%%%%%%%%%%%%%%%%%%%%%%%%%%%%
\clearpage

\vspace*{\fill}
\textit{\textbf{Aeroacoustics} studies the origin (generation), the propagation (including inside flowing and heterogeneous media), and mitigation of flow-induced sound.}

\bigskip

\paragraph{Habilitationsschrift}
Diese Arbeit wurde von Stefan Schoder zur Erlangung der \textit{Lehrbefugnis} im Feld \textbf{"Aeroakustik"} geschrieben und an der \textbf{Technischen Universität Graz}, Fakultät für Elektrotechnik und Informationstechnik eingereicht.

\paragraph{Habilitation thesis}
This manuscript was written by Stefan Schoder to attain the \textit{venia docendi} in the scientific field \textbf{"Aeroacoustics"}, was submitted to the \textbf{Graz University of Technology}, Faculty of Electrical and Information Engineering.

\vspace*{\fill}

%%%%%%%%%%%%%%%%%%%%%%%%%%%%%%%%%%%%%%%%%%%%%%%%%%%%%%%%%%%%%%%%%%%%%%%%%%%%%%%%%
\cleardoublepage
% !TeX spellcheck=en_US
% !TeX root=main.tex
%%%%%%%%%%%%%%%%%%%%%%%%%%%%%%%%%%%%%%%%%%%%%%%%%%%%%%%%%%%%%%%%%%%%%%%%%%%%%%%%%
\chapter*{Summary}
%\todo{auf meine erungenschaften im bereich eingehen nach einer kurzen einleitung}

Flow instabilities, wave propagation phenomena, and structural interaction are current topics of the field "Flow acoustics" also named "Aeroacoustics"\footnote{The theory of flow-induced sound (as a general theory for fluids) and the specific topic of aeroacoustics (as the theory of sound generation and propagation in the terrestrial atmosphere) are both used in this habilitation thesis. Most manuscripts in literature use these two terms in research and teaching synonymously. Most of the applications and theories presented herein deal with air as propagation medium and, as a consequence, in parts of the cited literature this distinction is not presented consistently.}. 
Assuming the theory of classical mechanics, aeroacoustic applications are modeled by the conservation equations and suitable material models. In particular, the continuity equation, the Navier-Stokes equation, energy conservation, and the Navier equation are coupled. Depending on the field of application (e.g., slow flow speeds in relation to the speed of sound), further assumptions can simplify the calculation considerably. A systematic derivation of the models according to physical accuracy and calculation efficiency allows us to categorize a computational aeroacoustic model into a hierarchy of models in terms of accuracy, applicability and computational effort. In the simplest case, flow acoustics is described by analytical models in the form of scale models (class~1), like the eighth power law of Lighthill or the methods of VDI 2081 and VDI 3731 for technical sound emissions. Class~2 models (e.g. Sharland, Költzsch, stochastic noise generation and radiation, random particle mesh method) allow empirical factors to be incorporated, which are based on experience (such as fan noise) and allow prediction of the sound. Class~3 models use a numerical decoupling of flow, acoustics, and structure. Thus, this class of models describe a pure forward coupling from the higher energy containing flow field to the sound field (e.g. see \cite{schoder2019hybrid,schoder2020computational}). Finally, to solve the full fluid-structure-acoustic interaction numerically, the field equations are solved in a coupled manner (class~4). The class~4 models are characterized by high computational effort and are physically most general but struggle with considerable numerical challenges (e.g. see \cite{schoder2022aeroacoustic}).

First, we look at beautiful aspects of aeroacoustics, such as music, and phonation of humans and animals. In classical musical instruments, energy is usually transferred in concentrated form to acoustic waves via mechanical or fluid dynamic excitation. The produced sound waves are then filtered by a cavity (resonating body) and perceived by the ear as distinctive and clear sound. %The human voice also functions according to this principle. 
The lungs cause the vocal folds to vibrate due to an excess pressure. This periodically constricting movement and the fluid flow forms the source signal for the human voice. Via the resonating upper airways, the sound waves are filtered and become a pleasant-sounding voice. Besides, vowels, consonants, and fricatives are formed by shaping of the mouth and throat area. As in music, we perceive the sound with our ears. The ear (acoustic perception instrument of the human being) converts the acoustic wave with the help of vibroacoustic coupling at the eardrum and successive translation in the cochlea located in the inner ear into a nerve signal that can be processed by the human brain.
Spoken language is an important form of human communication. Therefore, people suffering from voice disorders are severely restricted in their everyday life and quality of life. Within the project \textit{Numerical computation of the human voice source}, which is part of this habilitation, a numerical model for supporting the treatment of voice disorders was developed. 

Fluid-structure-acoustic interaction is the basis for a multitude of parasitic acoustic effects of technical and medical applications, which influence our daily life through unwanted acoustic emissions.
Among the disturbing emittents are transportation (e.g., automobile, airplane, helicopter, railway), moving machines of all kinds, production, and manufacturing (e.g., furnaces, separation plants, valves). Each of these applications emits noise which is essentially based on the same physical situation as previously presented for the musical instrument and human phonation. Fluid motion and mechanical vibration energy is transferred to acoustic waves. The sound waves are then filtered by a cavity (resonating body) and perceived by the ear as distinctive sound. 

The present work shows with selected publications how numerical methods can be used for the analysis of flow-induced sound for specific applications. In particular, the theory, models, and applications are discussed based on recent publications. %class~ 3 models are presented in detail, and the corresponding numerical methods are explained. This approach is particularly advantageous for low Mach number flows (due to the scale discrepancy present). Applications in the field of the human voice and acoustic signature of fans are demonstrated by publications. Based on the research activity, a holistic value concept is presented with which the value of research, teaching, and collaborations can be evaluated. This shows how students benefit from the use of the open-source software "openCFS" and better understand and learn the subject of flow-induced sound. This balancing act of theory, model, and simulation with application to a concrete problem shows: how the methodological toolbox of an engineer can be applied to flow-induced sound. Furthermore,  sufficient understanding of the theory of "flow-induced sound" is essential for the interpretation and evaluation of simulation results to arrive at an appropriate innovation.

\chapter*{Kurzfassung}
%\todo{auf meine erungenschaften im bereich eingehen nach einer kurzen einleitung}

Strömungsinstabilitäten, Wellenausbreitungsphenomäne und Strukturinteraktion sind aktuelle Themengebiete des Fachgebietes „Strömungsakustik“. Mit Annahme der klassischen Mechanik werden alle diese Anwendung mit der Theorie der Erhaltungsgleichungen und der geeigneten Materialmodelle gelöst (im Speziellen der Kopplung der Kontinuitätsgleichung, der Navier-Stokes Gleichung, der Energieerhaltung und der Navier-Gleichung). Je nach Anwendungsgebiet (zum Beispiel niedrige Strömungsgeschwindigkeit in Relation zur Schallgeschwindigkeit) können weitere Annahmen die strömungsakustische Berechnung vereinfachen. Die systematische Ableitung der Modelle nach physikalischer Genauigkeit und der Berechnungseffizienz lässt die Kategorisierung in eine Hierarchie von Modellen zu. Im einfachsten Fall wird Strömungsakustik über analytische Modelle in Form von Skalenmodellen beschrieben (Klasse~1), wie das Potenzgesetz von Lighthill oder die Methoden der VDI 2081 und VDI 3731 für technische Schallemissionen. Klasse~2 Modelle (z.B. Sharland, Költzsch, stochastic noise generation and radiation (SNGR) Methode, random particle mesh (RPM) Methode) lassen empirische Faktoren einfließen, die durch Erfahrung in speziellen Situation (wie zum Beispiel bei Lüfterlärm) eine gute Prädiktion des Schalles erlauben. Die Klasse~3 Modelle bedienen sich einer numerischen Entkopplung von Strömung, Akustik und Struktur. Somit beschreibt diese Art von Modell eine reine Vorwärtskopplung vom (meist) energiereicherem Feld auf das Schallfeld (wie z.B. in \cite{schoder2019hybrid,schoder2020computational} präsentiert). Abschließend und um die volle Fluid-Struktur-Akustik-Interaktion numerisch aufzulösen, werden die Feldgleichungen gekoppelt gelöst (Klasse~4). Die Klasse~4 Modelle zeichnen sich durch sehr hohen Rechenaufwand aus und sind physikalisch sauber definiert, aber kämpfen mit erheblichen numerischen Herausforderungen (wie z.B. in  \cite{schoder2022aeroacoustic} diskutiert). 

Betrachten wir zuerst die schönen Seiten der Strömungsakustik, die Musik und die Lautbildung des Menschen und der Tierwelt. Im klassischen Instrumentenbau wird meist Energie konzentriert über mechanische oder fluiddynamische Anregung auf das Schallfeld übertragen. Die so produzierte Schallwelle wird anschließend durch einen Hohlraum (Resonanzkörper) moduliert und als markanter und klarer Ton  vom Ohr wahrgenommen. Genau nach diesem Prinzip funktioniert auch die menschliche Stimme. Die Lunge bringt durch den produzierten Überdruck die angestellten Stimmlippen mithilfe der entstandenen Strömung zum Schwingen. Diese periodisch abschnürende Bewegung und Strömung bilden das Grundsignal für die Akustik der Stimme. Über den Resonanzkörper Lunge und den Rachenraum wird die Akustik verstärkt und zum wohlig klingenden Gesang. Darüber hinaus werden durch die Formung des Mundraumes und Rachenraumes die Vokale gebildet. Wie in der Musik nehmen wir mit dem Ohr den Schall wahr. Das Ohr ("Akustisches Wahrnehmungsorgan" des Menschen) setzt die Schallwelle mithilfe der Vibroakustik-Kopplung am Trommelfell und das Gehör in ein für das menschliche Hirn verarbeitbares Nervensignal um.
Diese gesprochene Sprache ist eine grundlegende Art der menschlichen Kommunikation. Deshalb sind Menschen, die an einer Stimmerkrankung leiden, in ihrem Alltag und ihrer Lebensqualität stark eingeschränkt. Im Rahmen des Projektes \textit{Numerical computation of the human voice source}, das Teil dieser Habilitation ist, wurde ein numerisches Modell zur Unterstützung bei der Behandlung von Stimmerkrankungen entwickelt. 

Insbesondere
auf die Fluid-Struktur-Akustik Interaktion beruhen eine Vielzahl an parasitären Effekten technischer und medizinischer Anwendungen, die durch Akustik-Emissionen unser tägliches Leben beeinflussen.
Zu den technischen störenden Anwendungen zählen Anwendungen im Transportwesen (z.B. Automobil, Flugzeug, Hubschrauber, Eisenbahn), bewegte Maschinen aller Art, die Produktion und Fertigung (z.B. Öfen, Abscheideanlagen, Ventile). Jeder dieser Anwendungen beruht im Wesentlichen auf demselben Schallemissionsszenario wie zuvor beim Musikinstrument und der Sprache vorgestellt. Energie wird konzentriert über mechanische oder fluiddynamische Anregung auf das Schallfeld übertragen. Die produzierte Schallwelle wird anschließend durch einen Hohlraum (Resonanzkörper) verstärkt und als markanter Ton vom Ohr wahrgenommen. 

Die vorliegende Arbeit zeigt mit ausgewählten Publikationen, wie numerische Verfahren zur Analyse von strömungsakustischen Aufgabenstellungen eingesetzt werden können und wo die aktuellen Herausforderungen liegen.

\chapter*{Enclosed Original Works}
The following scientific contributions published as peer-reviewed articles in journals are included as part of this cumulative Habilitation thesis\footnote{Separated into three topics discussed later: Theory and Computational Methods, Technical Applications and Benchmarking, Fluid-Structure-Acoustic Interaction with application to Human Phonation.}:

\begin{center}
\begin{tabular}{ c m{10cm} m{3cm} }
\toprule
 \cite{schoder2020helmholtz} & \textbf{Stefan Schoder}, Klaus Roppert, and Manfred Kaltenbacher. “Helmholtz’s decomposition for compressible flows and its application to computational aeroacoustics”. In: SN Partial Differential Equations and Applications 1.6 (2020), pp. 1–20. doi: 10.1007/s42985-020- 00044-w. & \includegraphics[scale=0.13]{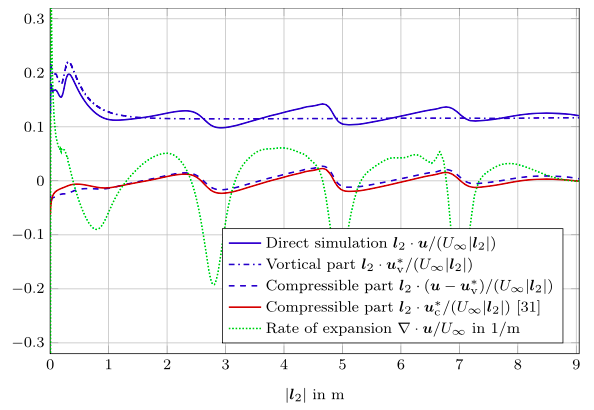}
 \\  
  & \textcolor{white}{\tiny{.}}
 \\ 
 \cite{schoder2019revisiting} & \textbf{Stefan Schoder}, Florian Toth, C Freidhager, and M Kaltenbacher. “Revisiting infinite mapping layer for open domain problems”. In: Journal of Computational Physics 392
(2019), pp. 354–367. doi: 10.1016/j.jcp.2019.04.067. & \includegraphics[scale=0.18]{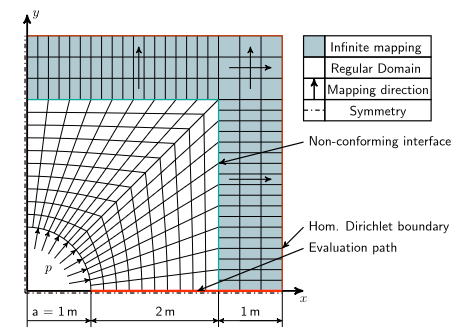} \\  
  & \textcolor{white}{\tiny{.}}
 \\ 
  \cite{maurerlehner2022aeroacoustic} & Paul Maurerlehner, \textbf{Stefan Schoder}, Johannes Tieber, Clemens Freidhager, Helfried Steiner, Günter Brenn, Karl-Heinz Schäfer, Andreas Ennemoser, and Manfred Kaltenbacher. “Aeroacoustic formulations for confined flows based on incompressible flow data”. In: Acta Acustica 6 (2022), p. 45. doi: 10.1051/aacus/2022041. & \includegraphics[scale=0.2]{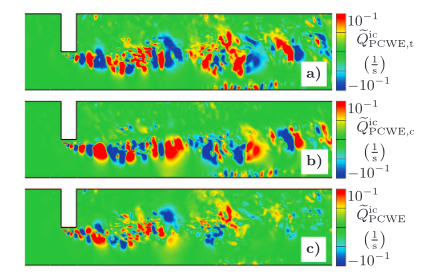}\\  
  & \textcolor{white}{\tiny{.}}
 \\ 
   \cite{Schoder2022b} &  \textbf{Stefan Schoder}, Étienne Spieser, Hugo Vincent, Christophe Bogey, and Christophe Bailly. “Acoustic modeling using the aeroacoustic wave equation based on Pierce's operator”. AIAA Journal (2023) accessed June 8, 2023. doi: 10.2514/1.J062558. & \includegraphics[scale=0.15]{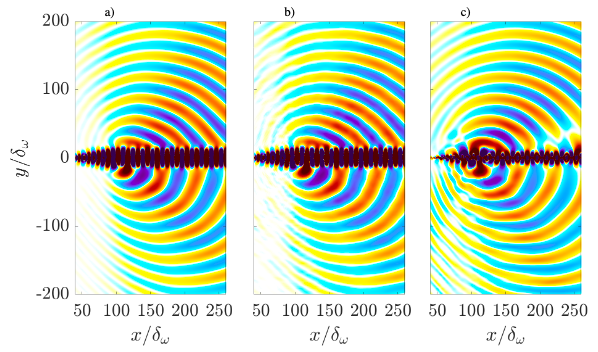}\\  

\end{tabular}
\end{center}

\begin{center}
\begin{tabular}{ c m{10cm} m{3cm} }
%\toprule
\toprule
   \cite{schoder2020computational} & \textbf{Stefan Schoder}, Clemens Junger, and Manfred Kaltenbacher. “Computational aeroacoustics of the EAA benchmark case of an axial fan”. In: Acta Acustica 4.5 (2020), p. 22. doi: 10.1051/aacus/2020021. \textbf{Under the 5 most cited papers in the journal 2021.} & \includegraphics[scale=0.25]{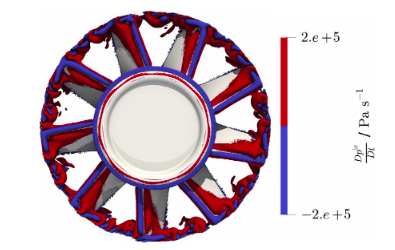}\\ 
     & \textcolor{white}{\tiny{.}}
 \\ 
   \cite{schoder2021application} & \textbf{Stefan Schoder}, Andreas Wurzinger, Clemens Junger, Michael Weitz, Clemens Freidhager,
Klaus Roppert, and Manfred Kaltenbacher. “Application limits of conservative source
interpolation methods using a low Mach number hybrid aeroacoustic workflow”. In:
Journal of Theoretical and Computational Acoustics 29.1, 2050032 (2021). doi: 10.1142/
S2591728520500322. & \includegraphics[scale=0.2]{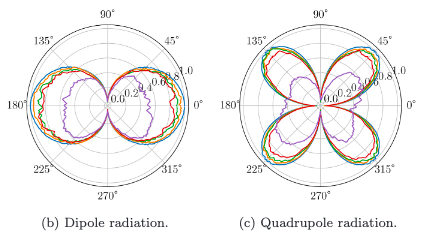} \\ 
  & \textcolor{white}{\tiny{.}}
 \\ 
      \cite{tieghi2023machine} & Lorenzo Tieghi, Stefan Becker, Alessandro Corsini, Giovanni Delibra, \textbf{Stefan Schoder}, and Felix Czwielong. “Machine-learning clustering methods applied to detection of noise sources in low-speed axial fan”. In: Journal of Engineering for Gas Turbines and Power 145.3 (2023), p. 031020. doi: 10.1115/1.4055417. & \includegraphics[scale=0.2]{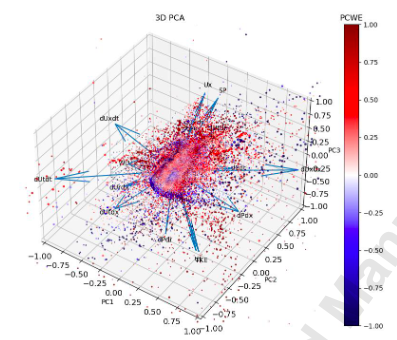}\\
  & \textcolor{white}{\tiny{.}}
 \\ 
      \cite{fluids8040116} & \textbf{Stefan Schoder}, Jakob Schmidt, Andreas Fürlinger, Roppert Klaus, and Maurerlehner Paul. “An Affordable Acoustic Measurement Campaign for Early Prototyping Applied to Electric Ducted Fan Units”. Fluids 8, no. 4: 116 (2023), doi: 10.3390/fluids8040116. & \includegraphics[scale=0.2]{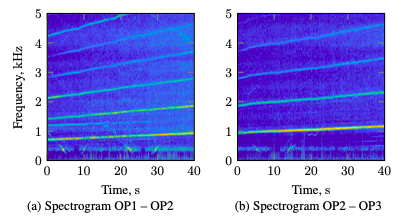}\\ 
\toprule
   \cite{schoder2021aeroacoustic} & \textbf{Stefan Schoder}, Paul Maurerlehner, Andreas Wurzinger, Alexander Hauser, Sebastian Falk, Stefan Kniesburges, Michael Döllinger, and Manfred Kaltenbacher. “Aeroacoustic sound source characterization of the human voice production-perturbed convective wave equation”. In: Applied Sciences 11.6 (2021), p. 2614. doi: 10.3390/app11062614. \textbf{Under the 5 most cited papers in the journals field 2022.} & \includegraphics[scale=0.1]{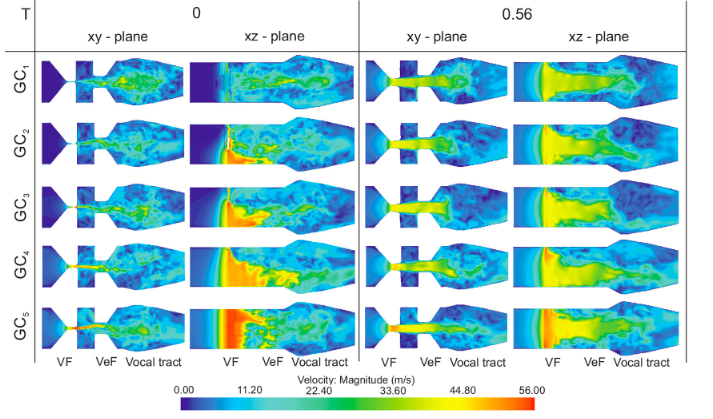}\\ 
  & \textcolor{white}{\tiny{.}}
 \\ 
   \cite{falk20213d} &  Sebastian Falk, Stefan Kniesburges, \textbf{Stefan Schoder}, Bernhard Jakubaß, Paul Maurerlehner, Matthias Echternach, Manfred Kaltenbacher, and Michael Döllinger. “3D-FV- FE aeroacoustic larynx model for investigation of functional based voice disorders”. In: Frontiers in physiology 12 (2021), p. 226. doi: 10.3389/fphys.2021.616985. & \includegraphics[scale=0.12]{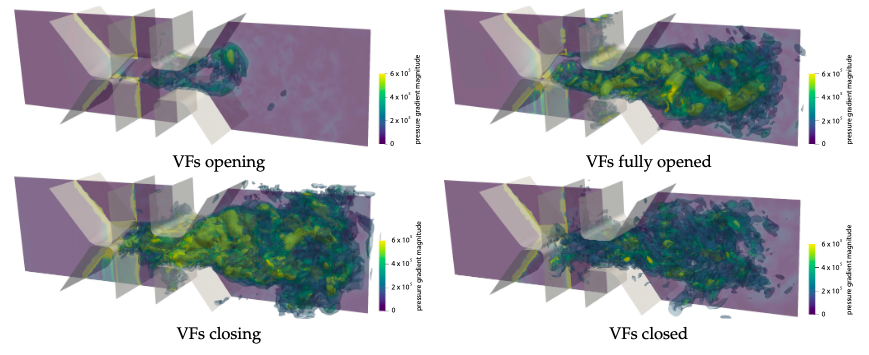}\\ 
\end{tabular}
\end{center}

\begin{center}
\begin{tabular}{ c m{10cm} m{3cm} }
%\toprule
   \cite{lasota2023anisotropic} &  Martin Lasota, Petr Šidlof, Paul Maurerlehner, Manfred Kaltenbacher, and \textbf{Stefan Schoder}. "Anisotropic minimum dissipation subgrid-scale model in hybrid aeroacoustic simulations of human phonation." The Journal of the Acoustical Society of America 153.2 (2023): 1052-1063. doi: 10.1121/10.0017202. & \includegraphics[scale=0.25]{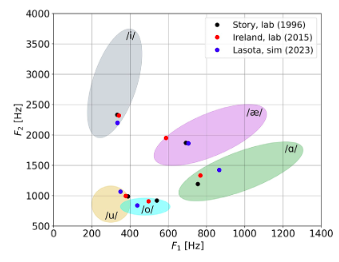}\\ 
   \toprule
\end{tabular}
\end{center}

\cleardoublepage
\tableofcontents

%%%%%%%%%%%%%%%%%%%%%%%%%%%%%%%%%%%%%%%%%%%%%%%%%%%%%%%%%%%%%%%%%%%%%%%%%%%%%%%%%
\cleardoublepage
\pagenumbering{arabic}

%%%%%%%%%%%%%%%%%%%%%%%%%%%%%%%%%%%%%%%%%%%%%%%%%%%%%%%%%%%%%%%%%%%%%%%%%%%%%%%%%
\cleardoublepage
% !TeX spellcheck=en_US
% !TeX root=main.tex
%%%%%%%%%%%%%%%%%%%%%%%%%%%%%%%%%%%%%%%%%%%%%%%%%%%%%%%%%%%%%%%%%%%%%%%%%%%%%%%%%
\chapter{Introduction} %1p

Working in the field of flow-induced sound, I am repeatedly faced with two statements or questions: (1) "What is aeroacoustics?", often heard from the scientific community, and (2) "This sounds complicated. What should I imagine, if you talk about aeroacoustics?", when talking to people outside of the scientific community.

From my perspective, the answer to the first question is rather simple - per definition. Since, people asking are mostly interested on the analytical definition of what the physical research domain is about. Before I give an answer, I provide some of the definitions that have tried to explain aeroacoustics precisely.
\begin{itemize}
\item Dictionary, Merriam-Webster \cite{Webster}: \textit{The study of the generation and propagation of sound specifically in an atmosphere.}
\item Collins English Dictionary  \cite{Collins}: \textit{The study of the generation and transmittance of sound by fluid flow.}
\item Hierschberg and Rienstra \cite{rienstra2004introduction}: \textit{Aero-acoustics provides such approximations (non-linearity, production, disparity of scales) and at the same time a definition of the acoustical field as an extrapolation of an ideal reference flow.}
\item Howe \cite{howe2003theory}: \textit{it is now widely recognized that any
mechanism that produces sound can actually be formulated as a problem of
aerodynamic sound.}
\item  Wim De Roeck \cite{de2007hybrid}: \textit{Aero-acoustic problems are concerned with the analysis of both the generation and dissipation of acoustic waves by an unsteady turbulent aerodynamic field as well as the propagation of acoustic waves in a moving medium. There is an interaction between the unsteady aerodynamic variables and the acoustic variables resulting in an energy transport between the turbulent flow field and the acoustic field.}
\item Oriol Guasch \cite{guasch2007computational}: \textit{Aeroacoustics has emerged from the fields of acoustics and fluid mechanics and is concerned with sound generated by unsteady and/or turbulent flows and also by their interaction with solid boundaries.}
\end{itemize}
To conclude and to answer the first question, I would reply that \textit{Aeroacoustics studies the origin (generation), the propagation (including inside flowing and heterogeneous media), and mitigation of flow-induced sound.} 

Regarding the second statement, most people know the sound of howling wind. This type of sound is airborne and is a type of parasitic sound source that the scientific field of aeroacoustics studies in technical acoustic applications. It is also the fundamental mechanism of wind instruments and the human voice. Regarding this definition and the scope of the field, the habilitation thesis at hand reviews validated computational procedures to predict aeroacoustic sound and is structured in three main topics. %Each topic is motivated in section 1.1.

\begin{learningobj}[Selected Topics]
\textbf{Topic 1: Theory and Computational Methods} 
\\

\textbf{Topic 2: Technical Applications and Benchmarking}
\\

\textbf{Topic 3: Fluid-Structure-Acoustic Interaction} 

\end{learningobj}
%The selection of these three particular 

\section{Selected Challenges in Aeroacoustics} %3.5p
\begin{figure}[ht!]
    \centering
    \includegraphics[width=0.86\textwidth]{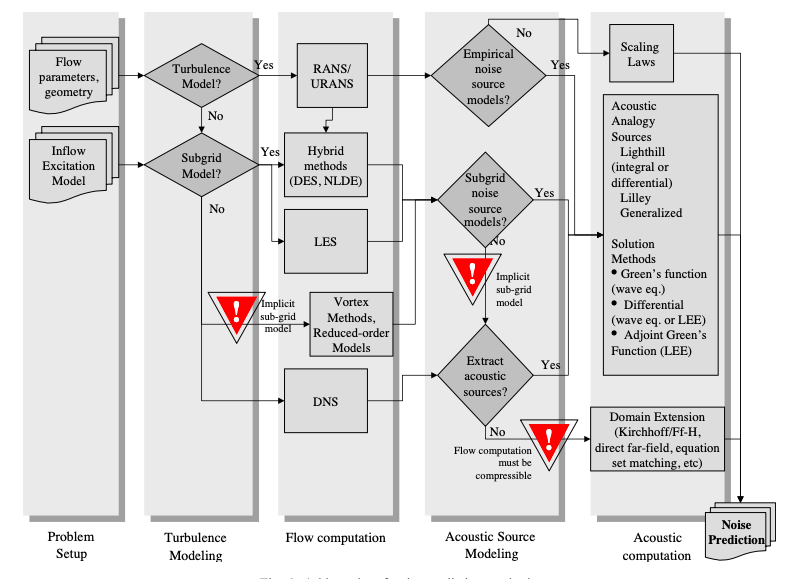}
    \caption{A hierarchy of noise prediction methods. \cite{colonius2004computational}}
    \label{fig:methods}
\end{figure}

The prediction of flow-induced sound encompasses a set of challenges that go beyond current modeling techniques of classical coupled-field physics \cite{schoder2019hybrid}. A hierarchy of sound prediction methods indicating some pitfalls is depicted in figure \ref{fig:methods}.
Some challenges stem directly from the physical fields involved, and some are primarily of computational nature. For instance, an unresolved question is the exact definition of the acoustic field in arbitrary flows, which is discussed in the following.
% TODO ... Definition of acoustics ... Leading to question in the nearfield of the superior method direct noise computation
\paragraph{Definition of acoustics}
A hardship of defining acoustics comes from seeking a distinction between the acoustic field and the non-acoustic part of the physical quantities that describe the flow field \cite{Fedorchenko2000,Morfey1973}. 
    Noise generation mechanisms \cite{Chu1958} and sound propagation \cite{Doak1954,Golubev1998} are intrinsically coupled. This fact is considered by the concept of acoustic-vortical-entropy waves in a moving fluid.
    At a low Mach number, the incompressible flow part is a valid basis for constructing a decisive separation of the fields \cite{kaltenbacher2017aiaa,Ribner1962}. In general, the restriction to just a solenoidal flow may be oversimplifying \cite{Fedorchenko1997}. The clear distinction between propagation effects and sound generation mechanism is tedious. 

Inside the source region, the precise definition of acoustic perturbation quantities is not general and until now ambiguous; it is a fact that the same physical quantities describe acoustics and fluid dynamics. There is a specific scenario where acoustics is defined properly \cite{spieser2020modelisation}: 
\begin{itemize}
    \item Acoustics is properly defined \cite{michalke1975turbulence} when the ratio of the fluid vorticity magnitude and the acoustic frequency tends towards zero \cite{legendre2015interactions}.
\end{itemize}
Ideally, no vortical flow is present. This explains why most aeroacoustic theories introduce a potential background flow. Or at very high frequencies of the sound waves (high-frequency limit), the criterion \cite{legendre2015interactions} is also fulfilled. As a critical consequence, seeking a general definition of acoustics in arbitrary moving media seems extraordinary challenging \cite{mcintyre1981wave}. However, with the use of hybrid methods \cite{schoder2019hybrid} and solving a system of equations, this was already achieved for incompressible modeled low Mach number flows \cite{kaltenbacher2017aiaa} and arbitrary flows for very small Mach numbers \cite{ewert2021hydrodynamic}. For any other flow situation, aeroacoustics raises unresolved practical and fundamental questions about the \textbf{definition of acoustics} and the \textbf{correct energy transfer mechanism}. \textbf{These two Gordian knots may be resolved by fundamental research in fluid dynamics and acoustics.} 
One method to circumvent both modeling issues (the definition of acoustics and the energy transformation mechanism) is to use a direct simulation of flow and sound with a compressible medium \cite{Bailly2010,Brionnaud,Fares2016,Frank2016,Sanders2016}. Compared to aeroacoustic wave equations, this method is more general since fewer model assumptions are made when computing flow and acoustics with a single simulation. This direct numerical simulation (DNS) resolves both fields and models both mechanisms simultaneously represented by a united field, but at the end of such a DNS, the questions for a human exposed to noise remains: 
\begin{itemize}
    \item What proportion of the fluctuations propagate and can be perceived as sound? %(Definition of sound)
    \item What proportion of the flow is effectively converted into acoustic energy? %(Energy transformation)
\end{itemize}
Therefore, DNS is an ideal candidate if acoustics should be predicted in steady flow regions and the energy conversation process is not a subject of the study. Until now, the huge demand on computational resources limits the applicability in aeroacoustics \cite{kempf2022zonal}.

\paragraph{Direct simulation of flow and sound}
Several challenges arise from the need to resolve several physical scales. %Special numerical methods have been proposed to overcome particular problem challenges for an effective and accurate computation of the radiated sound \cite{colonius2004computational,schoder2019hybrid,crighton1993computational,hardin1993regarding}.
% Spatial, amplidtude... making it difficult for a direct noise simulation to predict it correctly ... Low dispersion schemes FD, DG, Lattice boltzmann
For instance, the direct simulation of flow and sound (DNS) of aeroacoustic applications faces some additional practical problems based on the different physical characteristics of flow and sound fields as well as in numerical issues \cite{colonius2004computational,crighton1993computational,hardin1993regarding}. These difficulties include 
\begin{itemize}
    \item a large disparity between the energy in the flow and the radiated acoustic field (with a maximum proportion of radiated sound of 1\% for very high flow velocities \cite{FfowcsWilliams1963,Lighthill1963}) and
    \item a large disparity between the smallest turbulent length scales and the wavelength of the generated acoustic waves (ratio between turbulent length scale and smallest acoustic wavelength $~10^{-4}$) \cite{Bogey2003,Freund2000}.
    %\item and avoidance of the reflection of outgoing waves on the truncated free-field boundary of the computational domain (which is important to have a valid aero-acoustic model).
\end{itemize} 
These disparities between flow and the sound field are challenging to simulate in one aeroacoustic simulation model for practical applications at low Mach number (Ma) and high Reynolds number (Re) flows. Thus, hybrid aeroacoustic models (class~3) have been found to be useful over the last decade \cite{schoder2019hybrid}. Heavy loads of computational power and storage are needed still for a hybrid aeroacoustic model, but considerably less than for a DNS (by a factor of Re$/$Ma).
% resolve turbulent scales ..., DNS, LES, ... which leads to a large computational demand for the Dnoise simulation
% and high storage requeirements and rigorous defintiiton of the acoustic equation and the source term for an hybrid approach

%Leading to the idea of lighthill to search for the mechanism of energy transfer from the flow to the acosutic ...

% Definition of acosutics variables with an underlying background flow ..., acosutic potential, pressure, acoustic particle velocity, acoustic intensity, acoustic energy, ...

% Go beyound stationary situations to transient effects, since they are frequently occuring ...

%Take the JTCA paper and add a majority of the stuff provided there ...

\paragraph{Towards practical application}
% FANS
In recent years, my research in aeroacoustics has been focused on developing validated high-fidelity hybrid aeroacoustic computational workflows based on computationally efficient physical models. Primarily, the hybrid aeroacoustic workflow, in conjunction with the perturbed convective wave equation (PCWE), is employed. This method was chosen due to the following criteria; The PCWE computes a purely acoustic field \cite{maurerlehner2022aeroacoustic}, has an easy-to-compute source term \cite{schoder2019hybrid}, requires relatively low storage amount and a reduced number of unknowns compared to multiple equation acoustic models \cite{kaltenbacher2017aiaa}, coupling to vibro-acoustics is straightforward \cite{maurerlehner2022aeroacoustic,1ccbd3dfdc534dc080870d7f2c452f51}, the scalar source term is beneficial for machine learning algorithms \cite{tieghi2023machine} and easy to visualize \cite{schoder2020computational,schoder2021aeroacoustic} and the sources can be computed from any incompressible unsteady flow simulation, as it relies on the incompressible pressure and the mean velocity \cite{lasota2023anisotropic,maurerlehner2022aeroacoustic,schoder2020computational}.
Due to these specific properties of the PCWE, the main challenge was to establish a valid computational workflow and prove its usefulness over several examples \cite{lasota2023anisotropic,schoder2020computational,maurerlehner2022aeroacoustic}. Providing these reliable results with the PCWE for several applications led to a wide acceptance of this model and we further shed light into particular problems of flow-induced sound \cite{schoder2022aeroacoustic,Schoder2022b}. Overall, the validation involved the generation of many experimental datasets by collaborations. In conclusion, our findings were only possible because of the distinct composition of our research environment, having the computational models and the experimental data in place (which involves the respective measurement facilities). The last is usually a major limitation.

\paragraph{Benchmarking datasets}
%Benchmarking
As summarized in \cite{hornikx2015platform}, no easy-accessible datasets and standardized validation procedures exist for an out-of-the-box validation in aeroacoustics. In the aeroacoustic field, NASA initiated four benchmark workshops on relevant problems between 1995 and 2003 \cite{nasaBM}. Researchers contributed to the workshops with results from their methods applied to the benchmark problems. Initiatives to set up benchmarks existed but are outdated in terms of the availability and resolution of the datasets. This has become a significant hurdle for aeroacoustic development since the computation of such a dataset often involves the development of a flow solver that takes at least a year \cite{CFS}. Verifying this implementation to generate a trusted simulation is a matter of years. To this end, the computations often take several weeks to be performed on high-performance computing (HPC) clusters. Ergo, the non-availability of datasets critically constrains the development of flow-induced sound research. 

Additionally, the efficiency and variety of numerical methods have been increasing drastically. As a result, newly developed solution techniques are compared to state-of-the-art methods without state-of-the-art implementations of these methods. Often these state-of-the-art implementations for the reference method are out of their expertise scope. However, these comparisons are essential for the development of new methods and must be done in an objective manner (e.g., when comparing the finite element method (FEM) and the boundary element method, one should use a fair cost comparison~\cite{harari1992cost}).

Only research networks having both, the datasets and the numerical methods, in place will be able to push the limits of physical understanding. This substantially limits the validation capabilities and 
%Both issues raise practical questions during the validation of aeroacoustic models: % in  while the development of computers rising the processor power and graphic processing units.
% \begin{learningobj}[Fundamental Questions]
% \textbf{Q1.} Where do we get reliable simulation and experimental data from to put evidence on our research?\\ %(Definition of sound)
% \textbf{Q2.} What is the  state-of-the-art method? %(Energy transformation)
% \end{learningobj}
% Partly due to the lack of a common way of comparing techniques, results and performance are differently compared between different methods, and not always fairly. For example, a fair approach to a cost comparison between . Finally, when developers of a new method try to compare their development with another method, they often cannot use state-of-the-art implementations for the reference method since they are out of their expertise scope. 
\textbf{two main barriers for benchmarking aeroacoustic are identified:}
\begin{learningobj}[Current Barriers in Benchmarking]
\begin{enumerate}
    \item No clearly defined and widely accepted benchmark cases for aeroacoustics exist \cite{hornikx2015platform}.
    \item No consistent validation method exists to detect missing information or invalid model assumptions (e.g., neglecting non-linearity, missing flow-acoustic interactions, wrong boundary conditions).
\end{enumerate}
\end{learningobj}
A framework for open and easy-accessible benchmark datasets, as well as well-defined configurations for comparison of results from various techniques, is of high value and part of my research roadmap. During the course of closing this gap, we conducted detailed studies on a developed fan noise benchmark \cite{schoder2020computational,schoder2021application}. The impact of this benchmark motivates us to issue the used validation cases of the project as high-quality and easy-accessible benchmark datasets for aeroacoustics. Recently, the collaboration partner C. Bogey started to build a jet noise database \cite{Bogey2022,Bogey2022a}.

\paragraph{One huge step forward: Fluid-Structure-Acoustic-Interaction Modeling}
% As recognized by Phillips \cite{PhillipsEQ} and Lilley \cite{Lilley1974}, the source terms responsible for mean flow-acoustics interactions should be part
% of a convective wave operator.
% Therefore, an aeroacoustic approach based on a
% systematic decomposition of the field properties emerged. This circumvents that sources depend on the acoustic solution and provides a rigorous definition of acoustics. Ribner \cite{Ribner1962} decomposed the fluctuating pressure in a pseudo pressure and an acoustic pressure part $p' =
% p^0 + p^\mathrm{a} $. Hardin and Pope \cite{HP1994} formulated their
% viscous/acoustic splitting technique expansion about the
% incompressible flow (EIF). The EIF
% formulation was modified and applied to several examples \cite{Shen1999,Shen1999a,Slimon1999}. 
% Ewert and Schröder \cite{Ewert2003} proposed a different technique leading to the acoustic perturbation equations (APE). Instead of filtering the flow field, the source terms of the
% wave equations are filtered according to the characteristic properties of the acoustic modes. The acoustic modes are obtained from LEE. 
Recent advances of the aeroacoustic theory led to \cite{ewert2021hydrodynamic}, which resolves a clear definition of the forward and backward coupling process from the flow to the acoustic field. In \cite{Hueppe} a computationally efficient reformulation of the APE-2 system was derived, the perturbed convective wave equation (PCWE). Subsequently, this model was applied to the human voice production (phonation), which is a complex process consisting of the interaction between the tracheal airflow and the vocal folds (fluid-structure-acoustic interaction process). However, there is a main limitation of the PCWE, when applying it to non-predictable time dependent geometrical changes as featured by the phonation process. The limitation is that the PCWE relies on the definition of a mean flow field for the computation of the material derivatives, which is infeasible for these types of flows \cite{1ccbd3dfdc534dc080870d7f2c452f51}. As a consequence, my latest efforts of model derivation led to an improved version of the PCWE including second-order physical effects to account for changing domains and as a consequence for moving vocal folds.

\newpage
\begin{learningobj}[Executive Summary]
\textbf{Topic 1 (Chapter 1.2): Theory and Computational Methods} 

The theory of scalar acoustic wave equation concerning low Mach number flows (in-compressible) and a first attempt to a subsonic aeroacoustic model has been successfully verified. Furthermore, an outlook on a subsonic model called \textit{compressible flow data based variant of the PCWE} is outlined as going beyond state of the art in the future. The proofs and ideas collected during the last years resulted in an ERC Starting Grant proposal to advance aeroacoustic theories and computational methods. For the developed methods, respective models have been implemented in \textit{openCFS}, validated and published. % accordingly as shown in the extensive list of contributions inside the appendix. 
 My work is presented by selected articles in Chapter 2.\\

\textbf{Topic 2 (Chapter 1.3): Technical Applications and Benchmarking}

We provided a solid foundation for hybrid aeroacoustic model with special attention to the PCWE model. Regarding benchmarking datasets, which was issued previously for fan noise we showed how systematic validation of an aeroacoustic simulation works. Based on this work, we get frequent requests to contribute to others people and companies research. With increasing demand on electric propulsion units, we are focusing on these type of applications in the future. Furthermore, we plan to issue further benchmarking datasets in aeroacoustics, for instance duct aeroacoustic which is a bridging of topic 2 to topic 3.
My work is presented by selected articles in Chapter 3.\\

\textbf{Topic 3 (Chapter 1.4): Fluid-Structure-Acoustic interaction} 

Based on the application of human phonation, we investigate the possibility of using the PCWE for FSAI processes. In doing so, it is planned to issue a benchmark dataset in the next months (submitted to Acta Acustica). Furthermore, in the outlook of this chapter a formulation of the PCWE is presented that can account for variable domain size, as the vocal folds are oscillating when the human is speaking. Based on the promising investigation on the structure of the PCWE, the investigations on functional voice disorders and the studies on the influence of turbulence and subgrid-scale contributions to the noise emissions, a FWF/DFG grant application was submitted to further investigate the human phonation process. 
My work is presented by selected articles in Chapter 4.
\end{learningobj}

\newpage
\section{Theory and Computational Methods}

Murphy and King \cite{Murphy2014} defined acoustic noise as the unwanted sound %(including to one from aviation and transportation) 
that can negatively disrupt human or animal life. Environmental noise is a form of pollution that has to be avoided, controlled, regulated, and reduced as much as possible because of its negative impact on animals and human beings \cite{Murphy2014}. A healthy human hears in an aged-dependent audibility range of roughly 20 Hz to 20~kHz. Exposure of a human being to acute noise within the audibility range may lead to increased stress hormone levels and changes in body functions, like sleeplessness \cite{Jhanwar2016} and hearing loss \cite{Organization2021}. The World Health Organisation (WHO \cite{Organization2018}) defines 40 dB(A) at night as a critical value above which human health suffers \cite{Murphy2014}. Globally, the WHO estimates that about 1.5 billion people are affected by severe noise exposure today and have a partial or permanent loss of hearing \cite{Organization2021}. %Continuous exposure to at least  85-90 dB(A) leads to a progressive loss of hearing, with an increasing threshold of hearing sensitivity which makes understanding speech harder \cite{Jhanwar2016}. %https://www.who.int/publications/i/item/world-report-on-hearing

Consequently, in 2014 the European Commission proposed the regulation (EU) 540/2014 (Amending Directive 2007/46/EC and repealing Directive 70/157/EEC) to reduce noise in industry and traffic. In particular, a noise reduction of 25 \% for the automotive industry is imposed. In 2016, the International Civil Aviation Organization (ICAO) council announced that aircraft noise will be reduced significantly between 2020 and 2036. The ICAO council assumes that the number of people exposed to over “day night average sound level of 55 dB” will decrease by more than a million people \cite{ICAO2016}. %It is expected that under Committee on Aviation Environmental Protection (CAEP) 13 new heavy noise reduction measures are developed to sustain community acceptance of the aviation industry \cite{ICAO2022}.
This noise reduction aim contradicts recent ideas of the aviation industry flying faster than the speed of sound (e.g., \href{https://www.nasa.gov/specials/Quesst/}{NASA's~X-59}). If an airplane exceeds than the speed of sound, a sonic boom is generated. The sonic boom is an intense burst of noise with a pressure level of about 110 - 120 dB (overflight), therefore strongly annoying humans and disturbing animals. 
\begin{learningobj}[The Challenge]
Today, there is a significant gap between the industry's noise reduction necessities and the accurate modeling of acoustic emissions using computationally efficient descriptions of physical models even for flows slower than the speed of sound. A detailed understanding of sound, the energy transformation to acoustic emissions, and the generation of computationally lean high-fidelity models are essential to investigate noise at its origin and propose effective mitigation strategies. 
\end{learningobj}

\noindent Out of curiosity and with today's focus on reducing the impact of noise, many great discoveries have been made in the fields of acoustics and fluid dynamics \cite{Crighton1981}. %Sound in a compressible fluid has been closely related to aerodynamics  ever since. 
In 1746, d'Alembert discovered the one-dimensional wave equation and proposed its fundamental solution. Within ten years, Euler proposed the three-dimensional wave equation by connecting first-order principles \cite{Spieser2020}. This milestone leveled the ground for almost all future developments connected to sound propagation in media.
Lagrange \cite{Lagrange1781} mentioned source terms for the wave equation (based on the acoustic potential) being the convective gradient $\bm u \cdot \nabla Q$ of a source potential~$Q$ and a flow velocity $\bm u$. %Later Helmholtz remarked that  
%a volume force $\nabla Q$ acts on the fluid, derived from a potential $Q$ . 
According to Helmholtz \cite{Helmholtz1860}, a source can be written as the partial time derivative of a potential $\frac{\partial Q}{\partial t}$. Using $Q=-p^{\mathrm{ic}}$, one recovers Ribner's wave equation \cite{Ribner1962}. 
After Helmholtz' work, it took nearly 100 years for the next big step in the field of aeroacoustics, conducted by Lighthill \cite{Lighthill1951}. 

Since then, major progress has been made, but until now, acoustics has not been defined properly in arbitrary moving media and is  %(vortices, turbulent structures, inhomogeneities, and flow variations). 
%Inside the source region, the precise definition of acoustic perturbation quantities is not general and 
until now ambiguous. It is a fact that the same physical quantities describe acoustics and fluid dynamics. There are only some cases where acoustics is defined properly \cite{Spieser2020}. Currently, acoustics is properly defined when the ratio of the fluid vorticity magnitude and the acoustic frequency tends towards zero \cite{legendre2015interactions}. %As a critical consequence, seeking a general definition of acoustics in arbitrary moving media seems exceptionally challenging and may be misleading \cite{mcintyre1981wave,Tam2001}. %Legender Chapter 6

\begin{figure}
    \centering
    \includegraphics[width=0.99\textwidth, trim=0cm 0cm 0cm 0cm,clip]{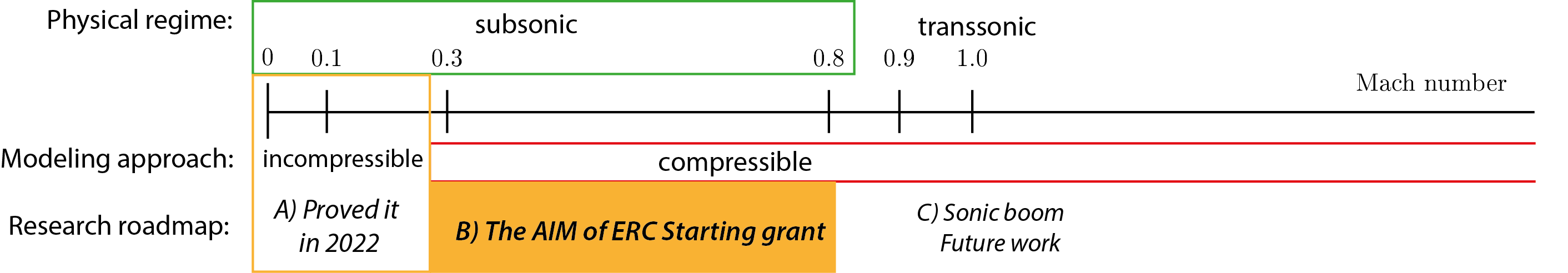}
    \caption{Definition of the terms used and the flow regime with respect to the Mach number. My research roadmap is advancing over time with the Mach number. In previous years, we proved the aims for low Mach numbers. With the current grant application, we tackle the subsonic range and, in the future, investigate transonic flows and the sonic boom.}
    \label{fig:Zahlenstreifen}
\end{figure}

Figure \ref{fig:Zahlenstreifen} shows the physical terminology of flows with respect to the flow speed and the Mach number. \textit{Subsonic} describes all flows slower than the speed of sound. \textit{Incompressible} flows are subsonic flows at very low flow speeds (typically lower than 30\% of the speed of sound).
Recently, acoustics was defined by using hybrid aeroacoustic methods \cite{schoder2019} and solving a system of equations \cite{Kaltenbacher2017,ewert2021hydrodynamic} for incompressible flows. We showed that the definition of acoustics in \cite{Kaltenbacher2017} is valid for incompressible confined flows \cite{maurerlehner2022aeroacoustic}. However, for larger Mach number flows, aeroacoustics raises unresolved practical and fundamental questions about the definition of acoustics and the correct energy transfer mechanism. \textbf{This open question may be resolved by fundamental research in fluid dynamics and acoustics using the hybrid aeroacoustic method and extend it to subsonic Mach number flows (see the orange box "AIM of current grant applications"\footnote{This current aim is submitted as an ERC Starting grant application.} which is a part of the future work on my research roadmap in Fig. \ref{fig:Zahlenstreifen}).} 

% \begin{figure}
% \floatbox[{\capbeside\thisfloatsetup{capbesideposition={left,top},capbesidewidth=6cm}}]{figure}[\FBwidth]
% {\caption{Definition of modeling terms and the flow regime with respect to the Mach number. The embedding of our research is highlighted consistently gaining insights and advance to more challenging areas of flow-induced sound production.}
%     \label{fig:Zahlenstreifen}}
% {\includegraphics[width=0.55\textwidth, trim=0cm 0cm 0cm 0cm,clip]{Zahlenstreifen.png}}
% \end{figure}

\begin{figure}[ht!]
\centering
\begin{overpic}[scale=0.6,trim=0 4 0 30,clip]{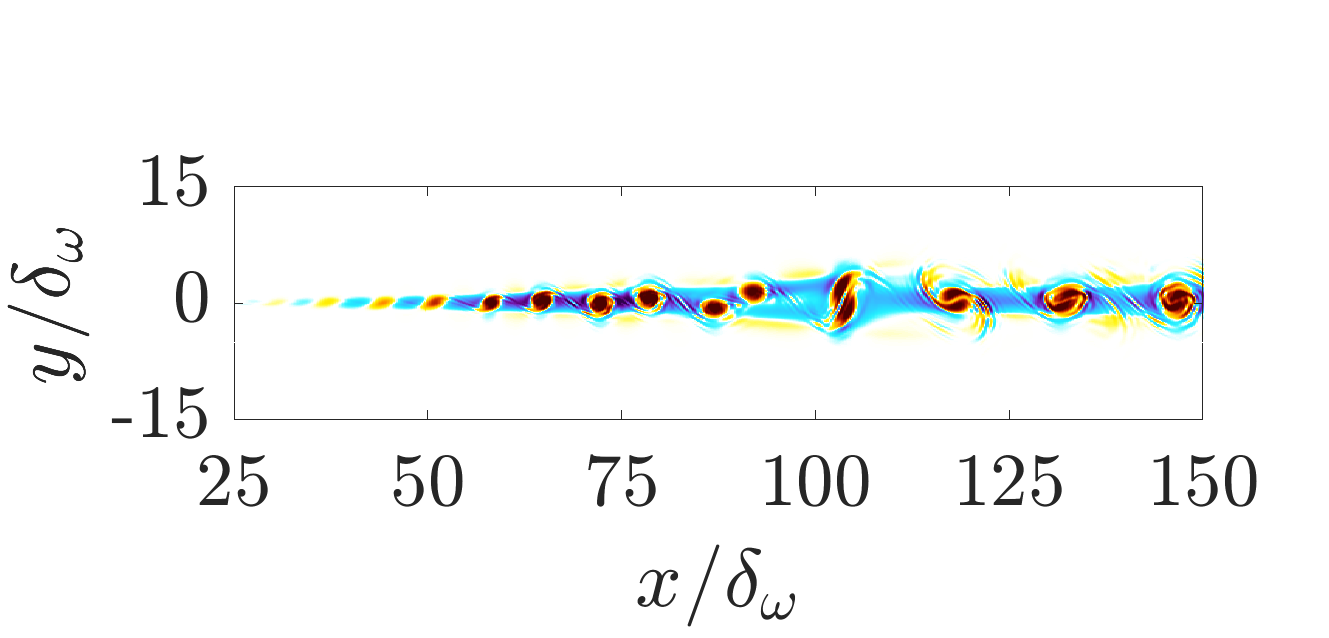}% 57
\put(21,17) {a)}%
\end{overpic}
\begin{overpic}[scale=0.6,trim=55 4 0 30,clip]{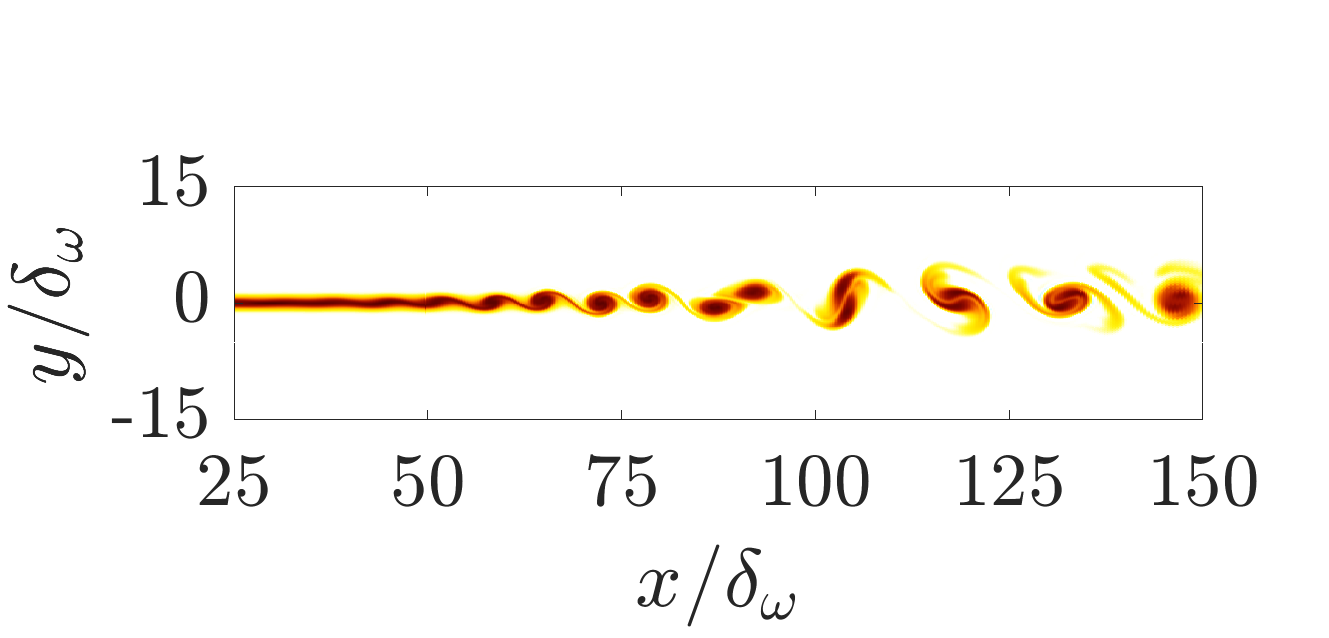}%  92
\put(6,19) {b)}%
\end{overpic}
\caption{Snapshot of a) the source term of Lighthill's equation and b) the vorticity \cite{Schoder2022b}.}\label{fig:Lighthill}
\end{figure}

\begin{figure}[ht!]
\centering
\begin{overpic}[scale=0.9,trim=0 30 0 51,clip]{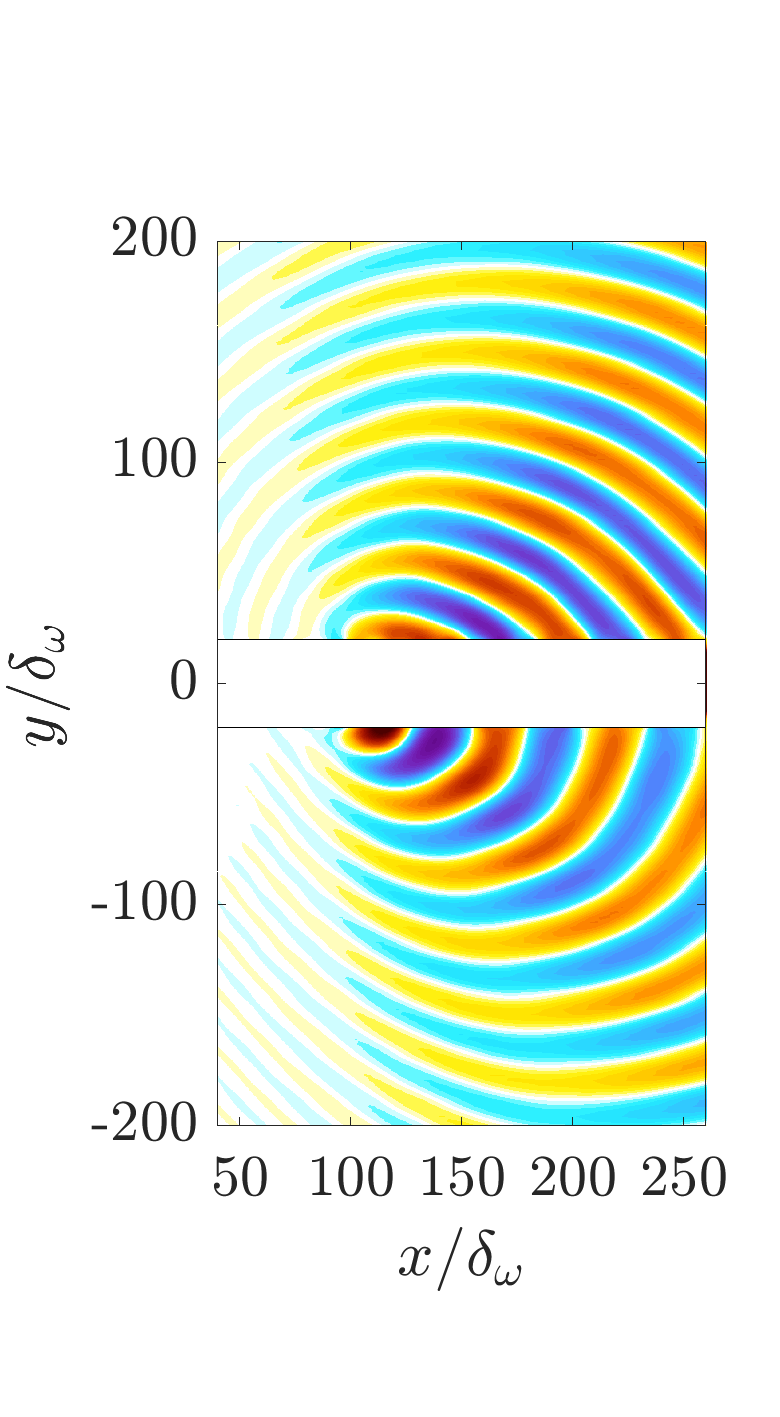}%
\put(29,99) {a)}%
\end{overpic}
\begin{overpic}[scale=0.9,trim=50 30 0 51,clip]{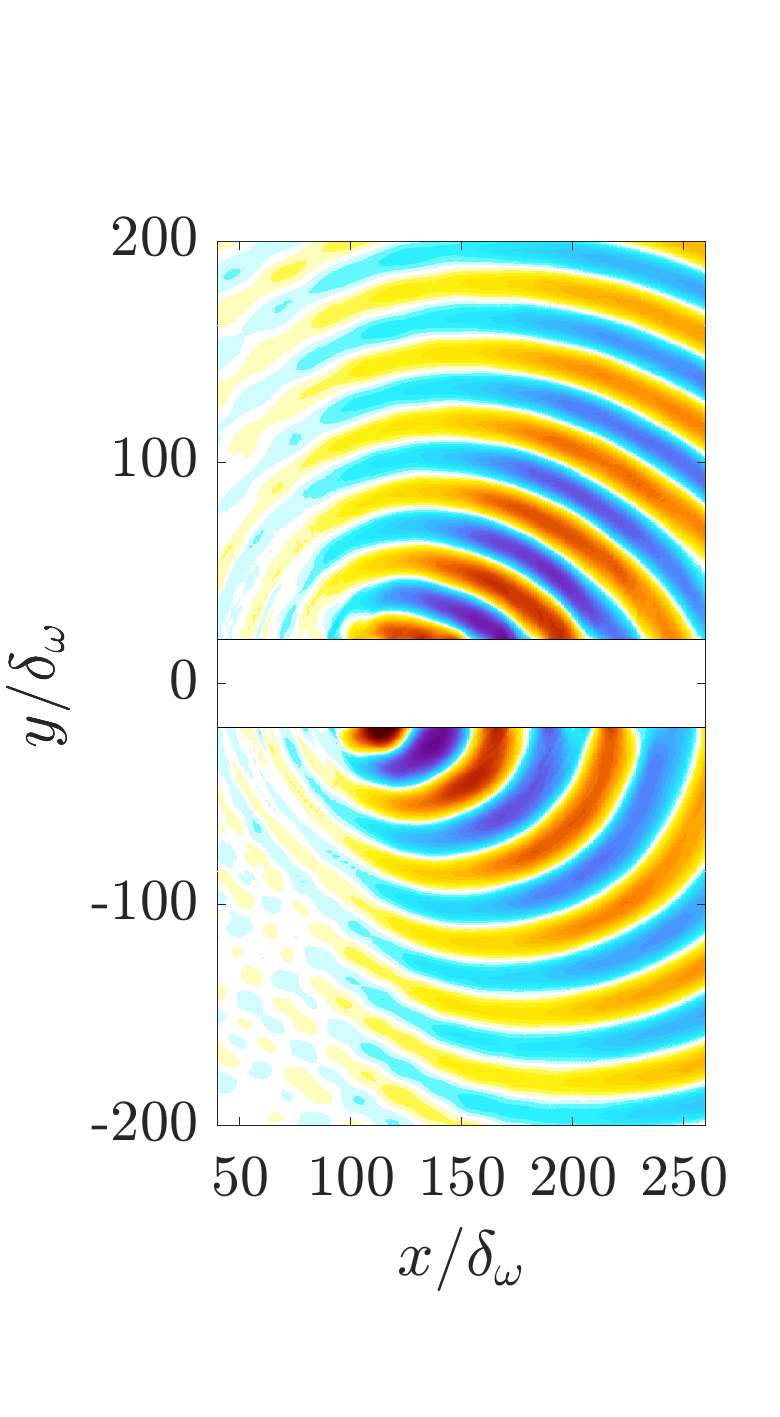}%
\put(10,99) {b)}%33_Lighthill_wave_p_T_x2_pml4_spatialSourceBlending
%Better? 233_Lighthill_wave_p_T_x2_pml4_spatialSourceBlending-Cutcell
\end{overpic}
\begin{overpic}[scale=0.9,trim=50 30 0 51,clip]{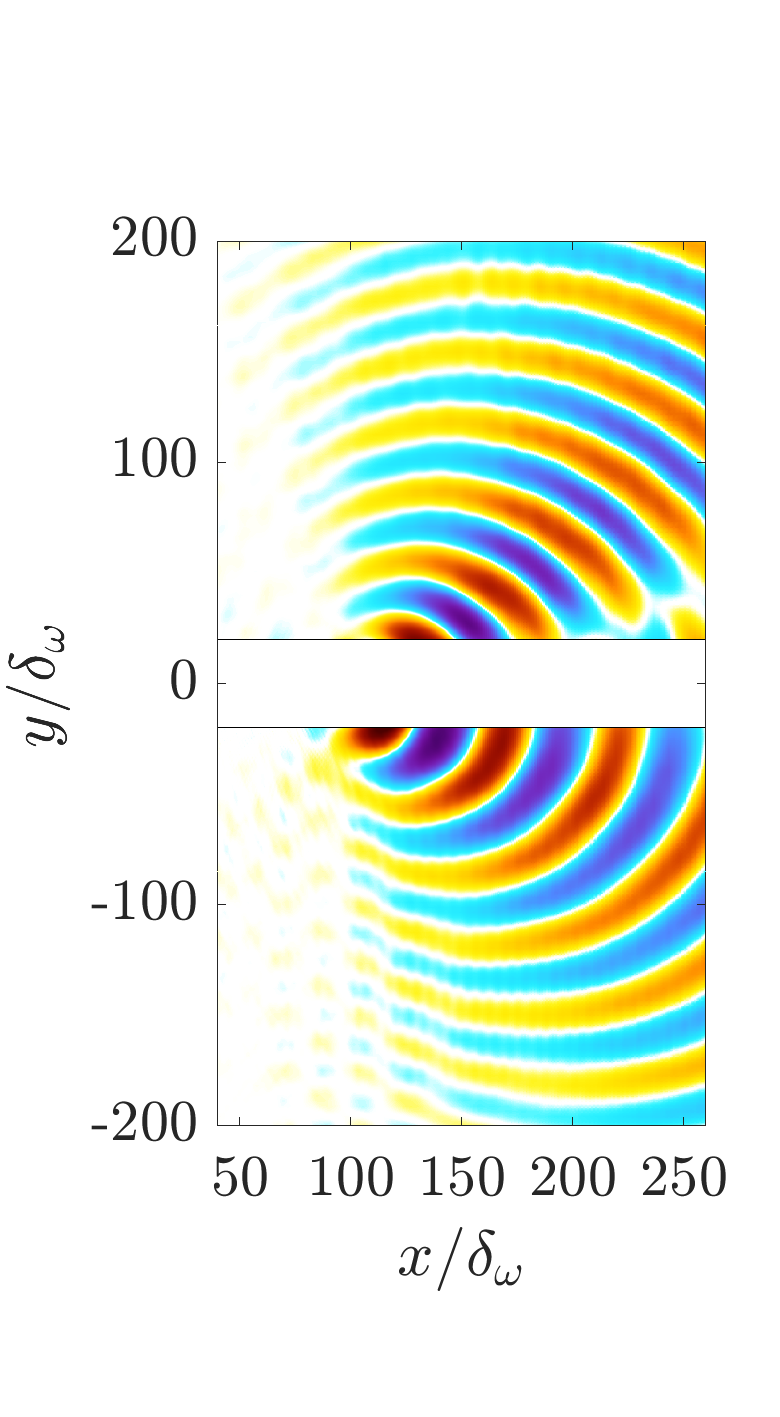}%
\put(10,99) {c)}%168_Sources_meanFlowAnalytic_wave_PML-Flow
%Better? 268_Sources_meanFlowAnalytic_wave_PML-Flow-Cutcell
\end{overpic}
\caption{Fluctuating pressure fields $p'/(\rho_0 c_0^2)$ a) DNS result (benchmark dataset), b) result of Lighthill's equation \cite{Lighthill1951}, and c) result of AWE-PO \cite{schoder2022aeroacoustic}.
The plots use color scale minimum and maximum values of $\pm 1.5 \cdot 10^{-4}$, from blue to red.}\label{fig:radiation}
\end{figure}

%\textbf{Attempts to resolve Q1:} 
Historically, hybrid models were established to quantify the energy transformation from the fluid motion to the acoustic field. These first hybrid models used the free-field Green's function to model the acoustic field and are called "Acoustic analogies." Lighthill \cite{Lighthill1951} recast the conservation equations into a wave equation and formulated Lighthill's analogy. For the first time, it was possible to describe the amount of energy converted from the fluid flow into the acoustic far-field. Figure \ref{fig:Lighthill}a shows the Lighthill sources for a mixing layer.
% \begin{figure}[ht!]
% \centering
% \hspace{0.1cm}
% \begin{overpic}[scale=0.6,trim=0 4 0 30,clip]{Figures/divdiv_Ttot_centered057.png}% 57
% \put(21,17) {a)}%
% \end{overpic}
% \begin{overpic}[scale=0.6,trim=55 4 0 30,clip]{Figures/vort057.png}%  92
% \put(6,19) {b)}%
% \end{overpic}
% \caption{Snapshot of a) the right-hand-side term of Lighthill's equation $S_\mathrm{LH}/(\rho_0 \Delta U^2/\delta_\omega^2)$ and b) the vorticity $\omega_z/(\Delta U/\delta_\omega)$. The color scales range between $\pm 0.2$ for the right-hand-side and between $\pm 1$
% for vorticity, from blue to red.\label{fig:Lighthill} }
% \end{figure}
%From now on, fluid dynamics and acoustics are two essential ingredients to tackle flow-induced sound problems. However, the direct physical relationship between the fluid flow and sound is essential to understand the complex energy conversion processes. 
Powell \cite{Powell1964} found that regions where vortical structures (see Fig. \ref{fig:Lighthill}b) are distorted contribute effectively to the sound generation mechanism. Despite the outstanding achievements, \textbf{the following fundamental barrier exists in physical understanding of the aeroacoustic energy conversion processes:}
\begin{learningobj}[Barrier 1]
Ffowcs Williams and Hawkings \cite{FfowcsWilliams1969} and Möhring \textit{et al.} \cite{Moehring1969} have shown that different equivalent distributions of sound sources consistently describe the same acoustic field. This built-in non-uniqueness of aeroacoustic wave equations has been pointed out by Tam \cite{Tam2001,Tam2002} as a fundamental obstacle to ever succeed in identifying the correct sound sources. Non-uniqueness of the underlying solution is known from inverse problems and is an issue of sparse data (typically encountered when inferring from measurement data). Having obtained highly-resolved simulation data of the flow field, the sound sources are determined reliably by the flow field.
\end{learningobj}
%\textbf{Attempts to resolve Q2:} 
Over the years, several attempts to model flow-induced sound in arbitrary flows have been made. A huge effort was invested in deriving a stringent definition of vortical (divergence-free) flow and acoustic (curl-free) waves inside arbitrary flow. This effort resulted in wave equations that post-process the flow simulation to obtain an acoustic field \cite{Ewert2003,Kaltenbacher2017,Munz2007,Seo2005} or an acoustic far-field \cite{Lighthill1951,Lilley1974,Powell1964}. %Additionally, this procedure quantifies the proportion of energy responsible for acoustic emissions and acoustic interactions with the underlying flow field. 
%Lighthill
%\cite{Lighthill1951,Lighthill1953} proposed his acoustic analogy in terms of the fluctuating density. 
For example, one achieves reasonable results when solving Lighthill's equation with appropriate boundary conditions~\cite{Curle1955,FfowcsWilliams1969}, especially for a low Mach number approximation of the source term \cite{Caro2009,Kaltenbacher2010,Kaltenbacher2008,Marretta2003,Oberai2002,Sandboge2006,Schaefer2010}. Figure~\ref{fig:radiation}b shows the fluctuating pressure obtained from Lighthill's equation for the mixing layer example~\cite{schoder2022aeroacoustic,Schoder2022b}. Several aspects were emphasized upon Lighthill's theory, like the importance of the vorticity as a source of sound~\cite{Howe1975,Moehring1978,Powell1964,Ribner1962} and the compressibility in the source term \cite{Crow1970}. %Goldstein \cite{Goldstein2003} proposed his generalized acoustic analogy, which shows that the Navier-Stokes equations can be rewritten as a set of linearized Euler equations (LEE).

A major drawback of these early models is that the fluctuating field (flow and acoustics) is computed, which converges to the acoustic field in steady flow regions only. Additionally, an ambiguity of these early models arises since the sources depend on the solution of the equations \cite{schoder2019} posing an implicit non-linear equation. Despite these achievements, \textbf{the following fundamental barrier exists in defining the acoustic field adequately:}
\begin{learningobj}[Barrier 2]
    An additional issue comes from the intricate distinction between the acoustic field and the non-acoustic part of the flow \cite{Fedorchenko2000,Morfey1973}. Noise generation mechanisms \cite{Chu1958} and propagation \cite{Doak1954,Golubev1998} are intrinsically coupled. %This fact is considered by the concept of acoustic-vortical-entropy waves in a moving fluid.
    At a low Mach number flows, the incompressible flow part (divergence-free or solenoidal) is a valid basis for constructing a decisive separation of the fields \cite{Kaltenbacher2017,Ribner1962}. The restriction to just a solenoidal flow may be oversimplifying \cite{Fedorchenko1997} but is a convincing direction chosen by the acoustic perturbation equation \cite{Ewert2003}. The clear \textbf{distinction between propagation effects and sound generation mechanism is tedious}. The state-of-the-art is that acoustic propagation is coupled and sound propagation cannot be addressed separately from its generation.  
\end{learningobj}
Therefore, the investigations of this project rely on an accurate DNS solution to cover all the effects in the high-resolved flow dataset. \textbf{Upon that highly resolved flow data, we expect to shed light on a precise definition of flow and acoustics for a general fluid flow and the energy conversion process.}

As first recognized by Phillips \cite{Phillips1960} and Lilley \cite{Lilley1974}, the terms responsible for mean flow-acoustics interactions should be part of the wave operator to bring out a more physical source definition.
Another approach for computing aeroacoustics is based on a systematic decomposition of the field properties. This circumvents that sources depend on the acoustic solution and provides a rigorous definition of acoustics. Ribner \cite{Ribner1962} formulated his dilatation equation such that the fluctuating pressure is decomposed in a pseudo pressure and an acoustic pressure part. Hardin and Pope \cite{Hardin1994} developed their viscous/acoustic splitting technique and derived the expansion about the incompressible flow (EIF). %, where they introduced a density correction
%$\rho_1$. 
The EIF formulation was modified over the years substantially \cite{Shen1999,Shen1999a,Slimon1999}. 
%For the sake of being more general and starting from the LEE, the field variables $(\rho, \bm
%u, p)$ are each decomposed in a temporal mean component and a fluctuating
%component. 
Bailly~\textit{et~al.} \cite{Bailly2000,Bogey2002} indicate significant aeroacoustic source terms on the momentum equation of the linearized Euler equations (LEE). Over the years, the LEE were modified to guarantee that only acoustic waves are propagated \cite{DeRoeck2009}. Significant contributions based on the LEE were derived in \cite{Munz2003,Munz2007,Seo2005,Seo2006}. 

Ewert and Schröder \cite{Ewert2003} proposed a different approach leading to the acoustic perturbation equations (APE). Instead of filtering the flow field, the source terms of the
wave equations are filtered according to the characteristic properties of the acoustic modes obtained from the LEE. Hüppe and Kaltenbacher \cite{Hueppe2014} derived a computationally efficient reformulation of the APE-2 system and named it perturbed convective wave equation (PCWE). Using PCWE, we developed relevant algorithms to build a solid theory for the hybrid aeroacoustic workflow \cite{schoder2019,schoder2019hybrid}. 
%A significant part comprised studies on conservative interpolation of aeroacoustics sources \cite{schoder2021application}. This conservative interpolation is essential to conserve energy from largely deviating discretization and guarantee accurate acoustic simulations. For instance, the flow grid has to resolve the boundary layers, which leads to highly stretched cells. In contrast, the acoustic propagation mesh is preferably uniform to minimize discretization errors \cite{schoder2021application}. 
We showed that using optimized discretization schemes significantly reduces the effort of the acoustic simulation without a relevant reduction in accuracy \cite{schoder2021application,schoder2020efficient}. Being fast and accurate enables us to use these algorithms for sound optimization and improved noise reduction measures. These fast and accurate algorithms are implemented in the open-source software \textit{openCFS} (\url{openCFS.org}, \cite{CFS}). Since then, several low Mach number flow applications have been addressed by computational aeroacoustics using the PCWE model successfully~\cite{falk20213d,kaltenbacher2021stable,lasota2021impact,schoder2019conservative,schoder2021aeroacoustic,schoder2020aeroacoustic2,schoder2020efficient,schoder2021application,tautz2019aeroacoustic,Valasek2019}. Recently, we proved that the fluid motion and acoustics can be separated inside the source region \cite{maurerlehner2022aeroacoustic} and showed that the definition of the acoustic variable is valid for incompressible flows \cite{maurerlehner2022aeroacoustic}. Our research resolved barrier 2 for incompressible flows (see Fig. \ref{fig:Zahlenstreifen}). 
% With a focus on a strict coupling and also relying on a strict definition of the variables,
% Ewert and Kreuzinger developed a computational workflow for low Mach numbers recently \cite{ewert2021hydrodynamic}. 

\noindent Developed by Spieser and me, the aeroacoustic wave equation based on Pierce's operator (AWE-PO) \cite{schoder2022aeroacoustic,Spieser2020,Schoder2022b} is our first attempt to establish a scalar wave equation for subsonic flows. \textbf{Highly relevant for the current topic}, I developed algorithms for Helmholtz's decomposition of DNS results to separate vortical and compressible effects \cite{schoder2019,Schoder2019N.2561,schoder2020helmholtz,schoder2020postprocessing}. 
%Using the theory of \cite{Spieser2020} and the implemented algorithms of HD in openCFS, Spieser and PI Schoder jointly developed the AWE-PO formulation and computational methods as a first attempt to derive a consistent wave equation for subsonic Mach number flows. 
The results of the fluctuating pressure fields compared to the one of Lighthill's equation and the one from the DNS are shown in Fig. \ref{fig:radiation}. The sound field's directivity, propagation, and convection effects are captured well. The acoustic intensities deviate less than 2~dB from the DNS result. %This error is comparable with one reported in previous studies \cite{margnat2014compressibility}. %This recent work was carried out during the PI's research stay at the LMFA in Lyon in 2021 \cite{schoder2022aeroacoustic,Schoder2022b}. 
% \begin{figure}[ht!]
% \centering
% \begin{overpic}[scale=1.0,trim=0 30 0 51,clip]{Figures/figure_4a.png}%
% \put(29,99) {a)}%
% \end{overpic}
% \begin{overpic}[scale=1.0,trim=50 30 0 51,clip]{Figures/figure_4b.png}%
% \put(10,99) {b)}%33_Lighthill_wave_p_T_x2_pml4_spatialSourceBlending
% %Better? 233_Lighthill_wave_p_T_x2_pml4_spatialSourceBlending-Cutcell
% \end{overpic}
% \begin{overpic}[scale=1.0,trim=50 30 0 51,clip]{Figures/figure_4c.png}%
% \put(10,99) {c)}%168_Sources_meanFlowAnalytic_wave_PML-Flow
% %Better? 268_Sources_meanFlowAnalytic_wave_PML-Flow-Cutcell
% \end{overpic}
% \caption{\label{fig:radiation}Fluctuating pressure fields $p'/(\rho_0 c_0^2)$ a) DNS result, b) result of Lighthill's equation \cite{Lighthill1951}, and c) result of AWE-PO \cite{schoder2022aeroacoustic}.
% The plots use a color scale minimum and maximum values of $\pm 1.5 \cdot 10^{-4}$, from blue to red.}
% \end{figure} % 0.00015*c^2/rhoamb
%Therein, special attention was put on the visualization and interpretation of the right-hand side of both wave equations. 
The analysis of the right-hand side reveals an about 90\% magnitude reduction for AWE-PO source compared to Lighthill's source \cite{schoder2022aeroacoustic,Schoder2022b}. For the first time, this reduction was explained and is mainly attributed to the filtering property of the material derivative during the right-hand side computation. This work was only possible by the unique team composition of having the DNS results at LMFA\footnote{Laboratoire de Mécanique des Fluides et d’Acoustique - UMR 5509} and my expertise with numerical methods based on the FEM. This led further to the preliminary derivation of the cPCWE model and the ERC Starting grant application in fall 2022.

\begin{table}[ht!]
    \centering
    \caption{My contributions beyond state-of-the-art by the enclosed publications. Further details on additional publications one finds in the appendix.}
    \begin{tabular}{p{6.2cm}p{6.2cm}ll}
    \toprule
        Status quo & Contribution & Enclosed \\
         &  & publication \\
        \toprule
         A common method was to illustrate compressible effects by showing the fluid dilatation and vortical effects by the vorticity. & We showed how to obtain the compressible and vortical velocity on confined flow regions by solving elliptic equations based on Helmholtz's decomposition. & \cite{schoder2020helmholtz}\\
          &  &  \\
         A common method to account for free domains in elliptic equations is to use infinite elements or asymptotic boundary conditions. & We implemented a method that combines the ideas of the PML for hyperbolic equations and infinite elements to account for the free field in elliptic equations. This is used for instance in publication \cite{schoder2020helmholtz,schoder2022aeroacoustic}. &  \cite{schoder2019revisiting} \\
          &  &  \\
         Before that study it was not clear if viscous acoustic splitting works in the near field and can be reproduced by measurements. & Within this contribution, we showed that viscous acoustic splitting works in the near field and we can extract the acoustics based on the definition for low Mach number (incompressible) flows. & \cite{maurerlehner2022aeroacoustic} \\
          &  &  \\
         Scalar acoustic formulations using the acoustic potential have only be verified for low Mach number flows. & In this contribution, we show an aeroacoustic wave equation based on Pierce's operator which we investigated for a subsonic mixing layer in 2D. & \cite{schoder2022aeroacoustic} \\
         \toprule
    \end{tabular}
    \label{tab:TECH}
\end{table}

\paragraph{Beyond the state-of-the-art}

To this day, acoustics has been appropriately defined outside arbitrary fluid motion and outside the generating flow-induced sound sources. Recently and convincingly, preliminary work proved for the PCWE that flow effects and acoustics can be separated for low Mach number flows \cite{maurerlehner2022aeroacoustic}. Since this evidence is missing for arbitrary subsonic flows, it is highly interesting to address the fundamental barriers 1 and 2 by the selected goals 1 and 2.
\begin{learningobj}[Goals 1 and 2] 
\begin{enumerate}
    \item[G1] \textbf{Define flow regions' acoustics consistent with the existing definition.} Currently, it is only possible to define acoustics in flow regions whenever the ratio of the fluid vorticity magnitude and the acoustic frequency tends towards zero \cite{legendre2015interactions}.
    \item[G2] \textbf{Based on the rigorous definition of acoustics, define a consistent energy transformation process and highlight the mechanisms.} The physics of sound generation will be modeled by a scalar wave equation to deliver a clear picture of the effects of flow-induced sound.
\end{enumerate}
\end{learningobj}
With the use of Helmholtz' decomposition, we will extend the PCWE equation substantially for subsonic flows. We believe that a clear definition of the acoustic quantity in conjunction with an appropriate wave model is the root of a clear picture of the energy transformation process.

\begin{figure}[ht!]
\centering
\includegraphics[width=0.85\textwidth, trim=6.5cm 0.5cm 0.5cm 2.5cm,clip]{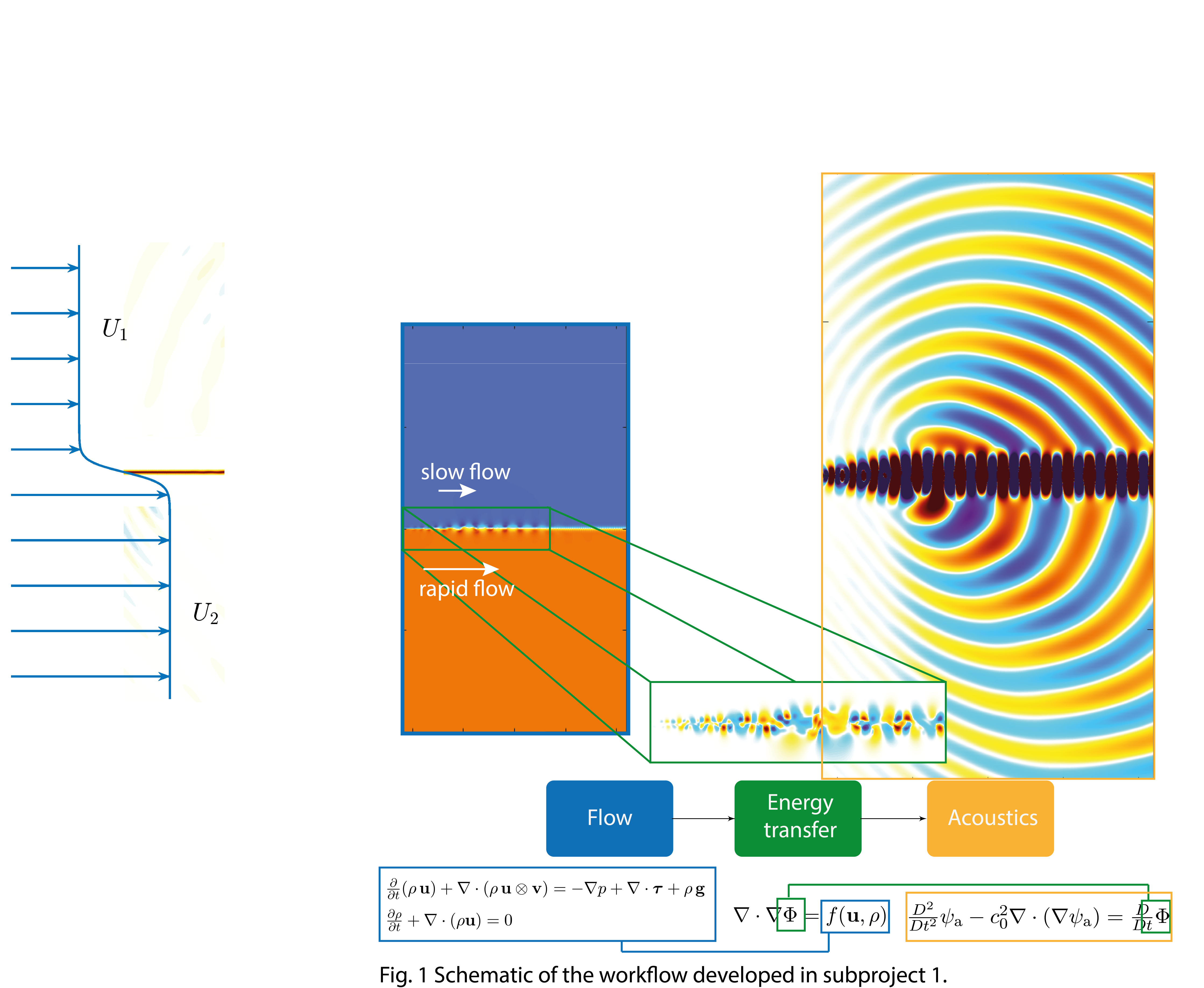}
\caption{Schematic of the hybrid workflow of the cPCWE used for the aeroacoustic simulations applied to the mixing layer example.}
    \label{fig:workflow}
\end{figure}

Regarding the state-of-the-art analysis, we will derive a wave equation based on the fundamental ideas of \cite{Ewert2003}. We collect these ideas, join them to a concise concept (methodological generalization) and further extend the concept to subsonic compressible flows (physical generalization). The derived aeroacoustic model should lead to a stable scalar wave equation that solves an (approximated) compressible potential for isothermal and thermal flows. The wave operator should only excite longitudinal wave modes (acoustic modes) and include convection. The physical goal of this new wave equation is that it holds for subsonic Mach number flows and recovers the PCWE in the incompressible limit. 
With respect to these priors, I preliminarily derived a convective wave equation based on compressible source data (so-called cPCWE~\cite{fb7a8e16805348f3af08e5941547bad2}). As used during the derivation of the APE system, we first apply linearization and then distinguish between vortical and acoustical perturbations.
The cPCWE can be derived efficiently from the APE-1 system \cite{Ewert2003} and the use of Helmholtz's decomposition \cite{schoder2020helmholtz}, such that
\begin{eqnarray}
p &=&p_0 + p'  = p_0 + p_{\mathrm{v}} + p_\ra \\
\rho &=& \rho_0 + \rho' \\%=  \rho_0 + \rho_1 + \rho_\ra\\
\bm u &=&\bm u_0 + \bm u' =  \bm u_0
+ \bm u_{\mathrm{v}} + \bm u_\ra = \bm u_0 + \nabla \times \bm A - \nabla \psi_\ra \,,
\end{eqnarray}
with the total pressure $p$, the mean pressure $p_0$, the perturbation pressure $p'$, the fluid dynamic perturbation pressure $ p_{\mathrm v}$, the acoustic perturbation pressure $p_\ra$, the total density $\rho$, the mean density $\rho_0$, the perturbation density $\rho'$, the total velocity $\bm u$, the mean velocity $\bm u_0$, the perturbation  velocity $\bm u'$, the fluid dynamic perturbation velocity $\bm u_{\mathrm v} = \nabla \times \bm A$, the acoustic perturbation velocity $\bm u_{\mathrm a} = - \nabla \psi_\ra$ and the vector potential $\bm A$. To eliminate the compressible part of the flow velocity, the Poisson equation of the compressible potential $\phi$
\begin{equation}
    \Delta \, \phi = \nabla \cdot \bm u'
\end{equation}
has to be solved and the vortical fluctuating velocity can be obtained by $ \bm u_\mathrm{v} = \bm u' - \nabla \phi \,.
$
We define the acoustic field as irrotational by the acoustic scalar potential $\psi_\ra$. %   as a gradient field of the acoustic particle velocity
%\begin{equation}
%$\bm u_\ra = -\nabla \psi_\ra $.
%\end{equation}
%We generalize the decomposition into a vortical and compressible acoustic part 
%\cite{schoder2019helmholtz}. 
Considering a general compressible flow, we arrive at the following perturbation equations
\begin{eqnarray}
\label{eq:APE1}
\frac{\partial p'}{\partial t} + \bm u_0 \cdot \nabla p' +
\rho_0 c_0^2 \nabla \cdot \bm u_\ra &=& 0 \\[2mm]
\rho_0 \frac{\partial \bm u_\ra}{\partial t} + \rho_0 \nabla (
\bm u_0 \cdot \bm u_\ra ) + \nabla p' &=& \rho_0 \nabla \Phi_\mathrm{p}\ \label{eq:momentum1}
\end{eqnarray}
with the isentropic speed of sound $c_0$ and the source potential $\Phi_\mathrm{p}$. In this preliminary derivation, we neglect thermal as well as viscous effects and discard the vorticity mode according to \cite{Ewert2003}. 
Rewriting equation (\ref{eq:momentum1}) yields the definition of the fluctuating pressure
\begin{equation}
\label{eq:pressAPE1}
p' = \rho_0 \frac{\partial \psi_\ra}{\partial t} +  \rho_0\, \bm u_0\cdot
\nabla \psi_\ra +  \rho_0 \Phi_\mathrm{p} = \underbrace{\rho_0 \frac{\mathrm{D}\psi_\ra}{\mathrm{D} t}}_{p_\ra} + \underbrace{\rho_0 \Phi_\mathrm{p}}_{p_{\mathrm v}} \, .
\end{equation}
%without further constraints on the mean velocity. 
The first part accounts for the acoustic pressure $p_\ra$ and the second part is a result of the following Poisson equation \cite{Ewert2003}
\begin{equation}
    \Delta \Phi_\mathrm{p} = - \nabla \cdot  \left[  ( (\bm u_{\mathrm v} \cdot \nabla) \bm u_{\mathrm v})' +    (\bm u_0 \cdot  \nabla ) \bm u_{\mathrm v} +   (\bm u_{\mathrm v} \cdot  \nabla)  \bm u_0 \right] \, .
    \label{eq:filter}
\end{equation}
The first term includes the self-noise of vortical structures. The second and third term account for the shear-noise interactions. Substituting (\ref{eq:pressAPE1}) into (\ref{eq:APE1}) yields the cPCWE
\begin{equation}
\frac{1}{c_0^2} \, \, \frac{\mathrm{D}^2\psi_\ra}{\mathrm{D} t^2} - \Delta \psi_\ra =
- \frac{1}{\rho_0 c_0^2}\, \frac{\mathrm{D} \Phi_\mathrm{p}}{\mathrm{D} t} \,.
\end{equation}
This convective wave equation describes acoustic sources
generated by flow structures and their wave propagation
through subsonically flowing media. In addition, instead of the original unknowns
$p_\ra$ and $\bm v_\ra$, we have just one scalar
$\psi_\ra$ unknown. %As shown in \cite{Spieser2020} and consistent with the pressure correction equation, the fluctuating vortical pressure in the overall domain can be recovered by
% $
% p_\mathrm{v} = \rho_0 \Phi_\mathrm{p}\,.
% $
%Finally, we have derived a scalar formulation that separates the source generation processes of vortical flows, including self-noise and shear-noise, and the linear acoustic propagation.
Figure \ref{fig:workflow} shows a schematic of the hybrid workflow applied to the mixing layer example in three steps:
\begin{enumerate}
    \item The flow field is simulated by a DNS.
    \item The DNS results are further processed to obtain the sources of the cPCWE and model the energy transfer to the acoustic equation.
    \item Finally, the acoustic wave propagation is computed.
\end{enumerate}  

\newpage
\section{Technical Applications and Benchmarking}
% TODO 5 Pages
The workflow involves three steps \cite{schoder2019hybrid}. First unsteady incompressible resolved LES typically based on finite volume scheme is computed. The flow quantities (global and local) statistics are used to detect convergence (limit cycle for most fluctuating properties). Mostly, a grid study based on grid convergence index is carried out.  Based on GCI\footnote{Grid convergence index, e.g., have a look at \cite{steffen1995analysis} or \cite{roache1998verification}.}, the quantities of interest are validated and compared systematically. For instance, to obtain a limit cycle for fan noise applications, about 5-8 revolutions (pre-computations) are necessary. This is a rule of thumb experienced by former research peers. During the last few years, we used \textit{StarCCM+} \cite{schoder2020computational,falk20213d}, \textit{OpenFOAM} \cite{lasota2023anisotropic}, \textit{Fluent} \cite{schoder2019hybrid}, \textit{Fire} \cite{maurerlehner2022aeroacoustic}, \textit{Flexi} \cite{schoder2019hybrid} and several internal research codes \cite{schoder2022aeroacoustic,maurerlehner2022aeroacoustic} for flow computations.

After having obtained the flow quantities, the sources are computed on the flow grid by user routines of the flow simulation software (UDFs) %. We showed that this is most accurate if the source terms are grid-dependent (spatial derivatives are necessary to compute the sources) 
\cite{6a9f9f55f945498f966b894a625c2418}. Then, we interpolate the CAA sources from the flow cells to the coarser FEM mesh. We showed that this interpolation must be energy conserving \cite{schoder2021application}. Otherwise, the simulation results will suffer from tremendous errors. If the source relocation (weighting to other grid points) is acoustically compact, we found that the influence on the sound signal is negligible (concerning measurement uncertainty).  The methods are implemented in the open-source software \textit{openCFS}. Compared to the flow data storage amount, the interpolated sources reduce the storage requirements by 95\% \cite{schoder2020efficient}.
At the end of this step, the sources are visualized and qualitatively checked. If compact, we found that odd-looking checkerboard sources do not influence the acoustic spectrum in the far-field \cite{schoder2021application}. The computed sources can be validated by microphone array source visualizations.

Finally, the acoustic propagation is computed. Instead of an analogy method (greens function), we use FEM to discretize the wave-PDE. Thus, we can predict the sound in any geometry and have the best approximation properties of FEM (being consistent and converging). Additionally, convergence analyses and GCI routines are applied to find the most efficient computational setup. Based on the acoustic radiation, we compare the solution pointwise to measured signals. This validates the whole CAA workflow \cite{lasota2023anisotropic,maurerlehner2022aeroacoustic,falk20213d,schoder2021aeroacoustic,schoder2020computational}. During the last years, the following statements were found by our research:
\begin{itemize}
\item The PCWE computes the acoustic sound pressure \cite{maurerlehner2022aeroacoustic}.
\item The PCWE has an easy-to-compute source term based on the incompressible pressure \cite{schoder2019hybrid}.
\item The source term requires relatively low storage amount compared to four equation acoustic models \cite{kaltenbacher2017aiaa}.
\item Vibrating surfaces can be included via the boundary condition of the acoustic potential \cite{1ccbd3dfdc534dc080870d7f2c452f51}.
\item The coupling to vibro-acoustics is straightforward over the boundary interface of the acoustic potential \cite{maurerlehner2022aeroacoustic}.
\item The scalar source is beneficial for machine learning algorithms \cite{tieghi2023machine}.
\item The scalar source term is easy to visualize \cite{schoder2021aeroacoustic,schoder2020computational}.
\item For flows with low mean velocities and medium pressure gradients, the time derivative source term is essential. Besides, Crighton's criterion is fulfilled by definition (source drift is avoided) \cite{schoder2021aeroacoustic}.
\item The space discretization of the source term in the acoustic model can be coarse (as long as the mesh spacing is acoustically compact) compared to the flow simulation \cite{schoder2021application}.
\item The PCWE formulation allows an ALE reference system for moving domains \cite{kaltenbacher2017aiaa}.
\item The convective source term only plays a role in the low-frequency range. Since the time derivative can be explained as a frequency weighting in the frequency domain \cite{schoder2021aeroacoustic}.
\item Using FEM instead of the acoustic analogy idea using Green's function, the PCWE is applicable for arbitrary domains and complicated ducted structures \cite{schoder2020efficient}.
\item The sources can be computed from any incompressible unsteady flow simulation, as it relies on the incompressible pressure \cite{lasota2023anisotropic,schoder2020computational,maurerlehner2022aeroacoustic}.
\end{itemize}

The presented method is exemplarily developed for the PCWE. This method can be applied to any wave equation model (Lighthill wave equation \cite{a7a5f9f90d4e407699bddb85b728e1c3}, Vortex sound \cite{schoder2019hybrid}, and other hybrid workflows). In most cases, these wave equations do not compute the acoustic sound signal. But other wave equation models have a different range of applicability (e.g. Mach number, entropy production) and are developed for specific applications \cite{schoder2019hybrid}. 

Furthermore, I addressed how to separate flow and acoustic quantities within a direct aeroacoustic computation \cite{schoder2020postprocessing,schoder2020helmholtz}. If a compressible flow is simulated, the flow and acoustic results (no matter if it is resolved) are hidden inside one physical quantity. Using Helmholtz decomposition, we showed that we can separate nonlinear acoustic waves from the flow quantities. This contribution led to new potential formulations based on Pierce wave operator \cite{schoder2022aeroacoustic}. %and using a Clebsh vector field decomposition. To conclude, this was a short introduction to my latest work in the field of aeroacoustic.

\begin{table}[ht!]
    \centering
    \caption{My contributions beyond state-of-the-art by the enclosed publications. Further details on additional publications can be found inside the appendix.}
    \begin{tabular}{p{6.2cm}p{6.2cm}ll}
    \toprule
        Status quo & Contribution & Enclosed \\
         &  & publication \\
        \toprule
        There was a benchmark case issued by our research group for low-pressure axial fans. & This contribution was the first systematic study on the hybrid aeroacoustic based on the PCWE and applied to rotating machinery. We showed in every detail how to perform acoustic predictions and validate them by the dataset in hand.  & \cite{schoder2020computational}  \\
        No systematic study existed on grid to grid interpolation from a CFD grid to an acoustic mesh. & This systematic study shows pitfalls and adresses them with the underlying methods, analysis their application limit and shows how to perform a grid convergence study for a whole aeroacoustics simulation. & \cite{schoder2021application}\\
        We knew that the scalar source term of the PCWE equation could be beneficial to investigate with machine learning techniques. & This study is the first study on analyzing the PCWE source term with machine learning techniques.& \cite{tieghi2023machine} \\
        From a very practical point of view, there are limited resources available during very early product assessment in startups. & We studied a low cost method and pointed out how to leverage the results of these to obtain acoustic measures for improving the devices. It is applied to an electric ducted fan unit.  &  \cite{fluids8040116} \\
         \toprule
    \end{tabular}
    \label{tab:Appl}
\end{table}

Additionally, I have to note that the promising investigations on the Aeroacoustic Fan benchmarking dataset led to the decision \cite{0db92a333fb042b1a3a8e7304eb7c84a}, that providing relevant datasets of experimental and computational results is a unique point how to distinguish my research group from any other research group in this field. To supply such cases, the generation of such datasets is ingrained into my research roadmap and planned to be part of any research activity in the future (voice benchmark dataset is currently planned). The current existing benchmarking platform of the EAA, will be transformed into new shape soon. With this initiative, I want to tackle future developments in understanding sound generation, model development towards machine learning in physics.

\newpage
\section{Fluid-Structure-Acoustic interaction}
% TODO 5 Pages

A number of low Mach number flow applications have been addressed by computational aeroacoustics using the PCWE model. This includes human phonation \cite{schoder2020efficient,schoder2021aeroacoustic,falk20213d,lasota2021impact,Maurerlehner,8a9a644042a1457089cb245ea872ffa5,valavsek2019application}, HVAC systems \cite{tautz2019aeroacoustic} and fan noise \cite{schoder2020computational,tieghi2023machine,schoder2019conservative,schoder2021application} using \textit{openCFS} \cite{CFS}. The PCWE is valid for incompressible flows and near-field acoustics \cite{maurerlehner2022aeroacoustic}. Recently, a convective wave equation was proposed for investigating jet-noise and installation effects \cite{schoder2022aeroacoustic}.

Regarding the state-of-the-art, we derive a wave equation that takes into account relevant second-order products omitted in the derivation of the APE-2 system to obtain a convective wave operator that depend on the instantaneous incompressible flow field. A similar derivation can be performed starting from the linearized perturbed compressible equations (LPCE) \cite{Seo2006}. Both equations are derived within this contribution and can be directly applied to fluid-structure-acoustic interaction problems of technical and biological flows.

\paragraph{Application to human phonation}

Voice disorders have high social and economic relevance. In 2000, a study estimated the US wealth losses of 186 billion USD annually \cite{ruben2000}. Speaking is often taken for granted and the most used communication tool for us humans. %If we catch a cold and get hoarse, we realize how dependent we are on a healthy voice. 
Voice issues restrict the quality of life, lead to isolation and depression \cite{smith1996,cohen2010, merrill2011}, and endanger the livelihood of people suffering from chronic voice disorders. According to Ruben, people with severe speech disabilities are more often found to be unemployed or in a lower economic class than people with a healthy voice \cite{ruben2000}. Furthermore, it is reported that this trend will increase in a knowledge-based society. Consequently, communication disorders will be one public health challenge for the twentyfirst century.

The human voice production (phonation) is a complex process consisting of the interaction between the tracheal airflow and the vocal folds. Phonation is only working under certain conditions properly:
\begin{enumerate}
\item Symmetric oscillation of the vocal folds. 
\item Periodic oscillation of the vocal folds.
\item Total closure of the glottal gap during every oscillation cycle. This requires full contact of the vocal folds. Otherwise, this kind of phoniatric malfunction is called the glottal gap or glottis closure insufficiency.
\end{enumerate}
Due to the expiration air stream self-excitation, the two elastic vocal folds are oscillating (typically between $100$ Hz and $300$ Hz) and interrupt the airflow periodically. The fluid dynamic forces and the turbulent structures form the primary acoustic voice signal inside the glottis. This acoustic signal is then modulated by the pharynx constrictions (movement of mouth and jaws), tongue motion, and oral-nasal cavity and emitted from the mouth as an audible voice.

The central objective is to develop a computational model \textit{simVoice} for clinical applicability. The incompressible CFD using an LES (Large Eddy Simulation) turbulence model will be based on prescribed vocal fold oscillations identified from high-speed imaging. A pressure-driven airflow with prescribed vocal fold motion will be used for the model. In this way, the fluid-solid interaction problem, whose accuracy critically depends on reliable geometrical and material parameters of the vocal folds, is circumvented. The acoustic model is based on the PCWE with the substantial derivative of the incompressible pressure as a source term. 

The \textit{simVoice} model reveals laryngeal interrelations between airflow, vocal fold dynamics, and resulting acoustics. Based on that knowledge, specific treatment strategies can be suggested. The innovative aspects are:
\begin{itemize}
\item the knowledge to which amount turbulent scales have to be
resolved for sustaining critical acoustic characteristics;
\item revealing the cause and effect chain of the phonation process;
\item first detailed numerical study on the dependencies of vocal
fold dynamics towards the acoustic quality;
\item developing a simulation model for teaching young medical
doctors and scientists;
\item providing a model for clinical treatment support in the long run.
\end{itemize}

The validated 3D virtual model \textit{simVoice} will be made available for scientific and clinical colleagues at no charge. The model will enable medical institutions to judge the effectiveness of therapy techniques, help to suggest changes in existing therapy, and help to suggest new therapies. The model will be a starting point for individualizing therapy approaches.% yielding to the point therapy techniques in the future. 

\begin{table}[ht!]
    \centering
    \caption{My contributions beyond state-of-the-art by the enclosed publications. Further details on additional publications one finds in the appendix.}
    \begin{tabular}{p{6.2cm}p{6.2cm}ll}
    \toprule
        Status quo & Contribution & Enclosed \\
         &  & publication \\
        \toprule
        There was a lack of understanding the aeroacoustic source terms and what or how to visualize them during the phonation process. & We investigated in detail the aeroacoustic source terms of the PCWE and showed benefits of either time-snapshot visualizations or time-harmonic visualizations. Furthermore, we showed at which location the source terms contribute and we observed a strong reduction of the full PCWE source term compared to the individual parts.& \cite{schoder2021aeroacoustic}\\
        The first study which investigated functional voice disorders based on computed acoustic results. & We showed the possibility of investigating functional voice disorders with means of computer simulation by a highly resolved 3D model called \textit{simVoice}.& \cite{falk20213d}\\
        The potential of the anisotropic minimum dissipation subgrid-scale model regarding the simulation of the human voice production process was unknown. & This study shows the applicability of our workflow to simulations using OpenFoam. Furthermore, the effects of the anisotropic minimum dissipation subgrid-scale model on the acoustic results using the PCWE are investigated with satisfying agreement.& \cite{lasota2023anisotropic} \\
         \toprule
    \end{tabular}
    \label{tab:FSAI}
    \end{table}

\paragraph{Beyond state-of-the-art}

Based on the preliminary work of the PCWE, we derive a convective wave equation based on the instantaneous incompressible velocity field. %As used during the derivation of the APE system, we first apply linearization and then distinguish between vortical and acoustical perturbations.
The PCWE can be derived efficiently from the APE-2 system \cite{Ewert2003} and the use of Helmholtz decomposition \cite{schoder2020helmholtz}
\begin{eqnarray}
p &=&p_0 + p'  = p_0 + p_{\mathrm{ic}} + p_\ra  \\ \rho &=& \rho_0 + \rho'\\
\bm u &=&\bm u_0 + \bm u' =  \bm u_0
+ \bm u_{\mathrm{ic}} + \bm u_{\mathrm a} = \bm u_0 + \nabla \times \bm A - \nabla \psi_\ra \,.
\end{eqnarray}
with the pressure $p$, the mean pressure $p_0$, the perturbation pressure $p'$, the incompressible perturbation pressure $ p_{\mathrm{ic}}$, the acoustic perturbation pressure $p_\ra$, the density $\rho$, the mean density $\rho_0$, the perturbation density $\rho'$, the velocity $\bm u$, the mean velocity $\bm u_0$, the perturbation  velocity $\bm u'$, the incompressible perturbation velocity $\bm u_{\mathrm{ic}} = \nabla \times \bm A$, the acoustic perturbation velocity $\bm u_{\mathrm a} = - \nabla \psi_\ra$ and the vector potential $\bm A$. We define the compressible acoustic field as irrotational field that can be modeled by the acoustic scalar potential $\psi_\ra$ as a gradient field of the acoustic particle velocity
%\begin{equation}
$\bm u_\ra = -\nabla \psi_\ra $.
We arrive from a general flow field at the following perturbation
equations
\begin{eqnarray}
\frac{\partial p_\ra}{\partial t} - c_0^2 \frac{\partial \rho'}{\partial t} &=& - \frac{\partial p_{\mathrm{ic}}}{\partial t} \, , \label{eq:p}\\[2mm]
\frac{\partial \rho'}{\partial t} + \nabla \cdot ( \rho_0 \bm u_{\mathrm a} + \bm u_0 \rho') &=& 0 \label{eq:rho} \, , \\[2mm]
\frac{\partial \bm u_\ra}{\partial t} + \nabla \big(
\bm u_0 \cdot \bm u_\ra \big) + \nabla \frac{p'}{\rho_0} &=& \frac{1}{\rho_0} \nabla p_{\mathrm{ic}}\ \label{eq:momentum} \, ,
\end{eqnarray}
with the isentropic speed of sound $c_0$. Combining equation (\ref{eq:p}) and (\ref{eq:rho}) results in the intermediate result
\begin{equation}
    \frac{\partial p_\ra}{\partial t} - c_0^2 \nabla \cdot ( \rho_0 \bm u_{\mathrm a} + \bm u_0 \rho') = - \frac{\partial p_{\mathrm{ic}}}{\partial t} \, .
\end{equation}
Before, proceeding with the derivation, the previously discarded (during the derivation of the APE) product of the $\bm u_{\mathrm{ic}} \rho'$ is added again
\begin{equation}
    \frac{\partial p_\ra}{\partial t} - c_0^2 \nabla \cdot ( \rho_0 \bm u_{\mathrm a} + \bm u_0 \rho' + \bm u_{\mathrm{ic}} \rho') = - \frac{\partial p_{\mathrm{ic}}}{\partial t} \, .
\end{equation}
Now the fluctuating density can be eliminated by using the relation of the pressure and density of $p' = c_0^2\rho'$ and apply the divergence to the term. The divergence of the incompressible flow is zero and therefore, the following result is obtained
\begin{equation}
    \frac{\partial p_\ra}{\partial t} - c_0^2 \nabla \cdot ( \rho_0 \bm u_{\mathrm a}) + c_0^2  (\bm u_0 + \bm u_{\mathrm{ic}})  \cdot \nabla p' = - \frac{\partial p_{\mathrm{ic}}}{\partial t}  \, .
\end{equation}
Splitting the pressure into its fluctuating parts yields the following equation
\begin{equation}
    \frac{\partial p_\ra}{\partial t}  +  (\bm u_0 + \bm u_{\mathrm{ic}})  \cdot \nabla p_{\mathrm a} - c_0^2 \nabla \cdot ( \rho_0 \bm u_{\mathrm a}) = - \frac{\partial p_{\mathrm{ic}}}{\partial t}  - (\bm u_0 + \bm u_{\mathrm{ic}})  \cdot \nabla p_{\mathrm{ic}} \, .
\end{equation}
At this stage, the material derivative for this convective wave equation is defined by $\mathrm{D}/\mathrm{D}t = \partial /\partial t  +  (\bm u_0 + \bm u_{\mathrm{ic}})  \cdot \nabla$ with $\bm u_0 + \bm u_{\mathrm{ic}}$ being the incompressible field of an incompressible flow simulation. This convective derivative is suitable for modeling acoustics on deformed grids and for fluid-structure interaction problems where one has no parametrization of the grid deformation before the simulation. In some cases, for instance fan simulations and with the use of an ALE reference frame, the original PCWE equation can be used \cite{kaltenbacher2017aiaa} but in the general FSAI case this is not possible. Using the definition of the material derivative, we arrive at
\begin{equation}
    \frac{\mathrm{D} p_\ra}{\mathrm{D} t}  - c_0^2 \nabla \cdot ( \rho_0 \bm u_{\mathrm a}) = - \frac{\mathrm{D} p_{\mathrm{ic}}}{\mathrm{D} t} \, . \label{eq:someResult}
\end{equation}
Rewriting equation (\ref{eq:momentum}) with the scalar potential and discarding the gradient yields the following intermediate result %Vertauschen von zeit und ort..theorem
\begin{equation}
    -\frac{\partial \psi_\ra}{\partial t} - 
\bm u_0 \cdot \nabla \psi_\ra  +  \frac{p'}{\rho_0} = \frac{1}{\rho_0} p_{\mathrm{ic}} \, .
\end{equation}
The second-order interaction $-\bm u_{\mathrm{ic}} \cdot \nabla \psi_\ra$ is again added to the equation 
\begin{equation}
    -\frac{\partial \psi_\ra}{\partial t} - 
(\bm u_0 + \bm u_{\mathrm{ic}}) \cdot \nabla \psi_\ra   = \frac{p_{\mathrm{ic}} - p'}{\rho_0} = - \frac{p_{\mathrm a}}{\rho_0}\, .
\end{equation}
Rewriting the left-hand side terms with the material derivative, we 
define the acoustic pressure
\begin{equation}
p_{\mathrm a} = \rho_0  \frac{\mathrm{D} \psi_\ra}{\mathrm{D} t} \, ,
\end{equation}
Finally, we insert the definition of the acoustic pressure and the definition of the acoustic potential into equation (\ref{eq:someResult}) and obtain the PCWE for FSAI applications
\begin{equation}
\frac{1}{c_0^2} \, \, \frac{\mathrm{D}^2\psi_\ra}{\mathrm{D} t^2} - \Delta \psi_\ra =
- \frac{1}{\rho_0 c_0^2}\, \frac{\mathrm{D}  p_{\mathrm{ic}}}{\mathrm{D} t} \,.
\end{equation}
This convective wave equation describes acoustic sources
generated by incompressible FSAI applications and their wave propagation
through flowing media. In addition, instead of the original unknowns
$p_\ra$ and $\bm v_\ra$, we have just one scalar
$\psi_\ra$ unknown. An additional benefit of this modified PCWE is that the boundary conditions can be described consistently with the scalar potential and are independent of the aeroacoustic source generation. Finally, it is emphasized that in comparison to the standard PCWE, the convective operator of the modified PCWE is based on the instantaneous incompressible velocity field.

 % \includepdf[pages=1]{publication/2019 online_nme.6298.pdf}

%%% BIBLIOGRAPHY %%%%%%%%%%%%%%%%%%%%%%%%%%%%%%%%%%%%%%%%%%%%%%%%%%%%%%%%%%%%%%%%
% \clearpage
% \printbibliography[heading=subbibintoc,title={References}] % print chapter bibliography

%%%%%%%%%%%%%%%%%%%%%%%%%%%%%%%%%%%%%%%%%%%%%%%%%%%%%%%%%%%%%%%%%%%%%%%%%%%%%%%%%
 \cleardoublepage %Comment in THE THEORY (69 PAGES)
 % !TeX spellcheck=en_US
% !TeX root=main.tex
%%%%%%%%%%%%%%%%%%%%%%%%%%%%%%%%%%%%%%%%%%%%%%%%%%%%%%%%%%%%%%%%%%%%%%%%%%%%%%%%%
\chapter{Theory and Computational Methods}
\label{chp:02-fundamentals}

\begin{learningobj}[Contributions and research roadmap]
The contributions of this chapter are four-fold:
\begin{itemize}
    \item We provided the computational theory of Helmholtz's decomposition for non-convex and restricted flow domains, including mathematical proofs of convergence for the FEM implementation. The theoretical foundation and the restrictions were subsequently empirically shown by a test case of a Mach 0.8 direct aeroacoustic simulation. These detailed evaluation were a first step towards a consistent implementation for aeroacoustics to post-process flows into compressible and vortical structures \cite{schoder2020helmholtz}.
    \item We solved the infinite domain modeling issue of elliptical problems. In some cases, Helmholtz decomposition is facing boundary conditions, set at far distant (at infinity). These algorithms based on characteristic decay functions were implemented in \textit{openCFS} and numerical convergence was proofed in this article \cite{schoder2019revisiting}.
    \item We applied hybrid algorithms implemented in the \textit{openCFS} software package to compute the aeroacoustic solution inside pipes using the PCWE. The computational results and the experimental work conducted showed the first time experimental evidence that the viscous acoustic splitting in aeroacoustics is not just a mathematical theory, but is indeed valid for very low Mach numbers. These results motivated me to design the ERC starting grant in 2022 \cite{maurerlehner2022aeroacoustic}.
    \item We made the first attempt to generalize the PCWE methodology to subsonic Mach numbers (AWE-PO) \cite{Schoder2022b,schoder2022aeroacoustic}. The first results are very promising and involve some detailed combination of previously studied theories like the Helmholtz decomposition and the hybrid workflow. This work motivated me to design a general algorithm in how to obtain convective wave equations based on a two stage source term filtering (Laplace equation and material derivative) and led to the idea of the cPCWE.
\end{itemize}
These contributions led to the promising ERC starting grant proposal based on the idea of the cPCWE. It should overcome some restrictions of the AWE-PO and further gain insight into sound generation mechanisms of turbulent flows. To this end, these future studies provide first learnings towards studying flow-induced sound in the transonic region.
\end{learningobj}

\subsection*{Enclosed original work}

\begin{center}
\begin{tabular}{ c p{13cm} }
 \cite{schoder2020helmholtz} & \textbf{Stefan Schoder}, Klaus Roppert, and Manfred Kaltenbacher. “Helmholtz’s decomposition for compressible flows and its application to computational aeroacoustics”. In: SN Partial Differential Equations and Applications 1.6 (2020), pp. 1–20. doi: 10.1007/s42985-020- 00044-w. 
 \\  
 \cite{schoder2019revisiting} & \textbf{Stefan Schoder}, Florian Toth, Clemens Freidhager, and Manfred Kaltenbacher. “Revisiting infinite mapping layer for open domain problems”. In: Journal of Computational Physics 392
(2019), pp. 354–367. doi: 10.1016/j.jcp.2019.04.067. \\  
  \cite{maurerlehner2022aeroacoustic} & Paul Maurerlehner, \textbf{Stefan Schoder}, Johannes Tieber, Clemens Freidhager, Helfried Steiner, Günter Brenn, Karl-Heinz Schäfer, Andreas Ennemoser, and Manfred Kaltenbacher. “Aeroacoustic formulations for confined flows based on incompressible flow data”. In: Acta Acustica 6 (2022), p. 45. doi: 10.1051/aacus/2022041. \\  
   \cite{Schoder2022b} &  \textbf{Stefan Schoder}, Étienne Spieser, Hugo Vincent, Christophe Bogey, and Christophe Bailly. “Acoustic modeling using the aeroacoustic wave equation based on Pierce's operator”. AIAA Journal (2023) accessed June 8, 2023. doi: 10.2514/1.J062558.\\  
\end{tabular}
\end{center}

\newpage
\section[Helmholtz’s decomposition]{Helmholtz’s decomposition for compressible flows and its application to computational aeroacoustics}

\vspace{0.6cm}
\textbf{Stefan Schoder}, Klaus Roppert, and Manfred Kaltenbacher. “Helmholtz’s decomposition for compressible flows and its application to computational aeroacoustics”. In: SN Partial Differential Equations and Applications 1.6 (2020), pp. 1–20. doi: 10.1007/s42985-020- 00044-w. 

\vspace{0.6cm}
\textbf{Abstract:} The Helmholtz decomposition, a fundamental theorem in vector analysis, separates a given vector field into an irrotational (longitudinal, compressible) and a solenoidal (transverse, vortical) part. The main challenge of this decomposition is the restricted and finite flow domain without vanishing flow velocity at the boundaries. To achieve a unique and $L_2$-orthogonal decomposition, we enforce the correct boundary conditions and provide its physical interpretation. Based on this formulation for bounded domains, the flow velocity is decomposed. Combining the results with Goldstein’s aeroacoustic theory, we model the non-radiating base flow by the transverse part. Thereby, this approach allows a precise physical definition of the acoustic source terms for computational aeroacoustics via the non-radiating base flow. In a final simulation example, Helmholtz’s decomposition of compressible flow data using the finite element method is applied to an overflowed rectangular cavity at Mach 0.8. The results show a reasonable agreement with the source data and illustrate the distinct parts of the Helmholtz decomposition.

\vspace{0.6cm}
\begin{learningobj}[My contribution]
This work is a substantial extension and description of Helmholtz decomposition started during my PhD. The theory and proofs are carried out and the method is evaluated on a testcase of a concave flow condition being an overflown cavity at flow speeds close to the transonic flows. I managed and took care of the manuscript writing and revised it based on the reviewers comments. With this publication, I sharpened the algorithms, theory and numerical analysis of the methods used based on initial investigation during my PhD Thesis. Finally, this study is the documentation of my implementations in \textit{openCFS}.
\end{learningobj}

%\includepdf[pages=-,pagecommand={},width=\textwidth]{publication/2020 Schoder2020_Article_HelmholtzSDecompositionForComp.pdf}

\newpage
\section{Revisiting infinite mapping layer for open domain problems}

\vspace{0.6cm}
\textbf{Stefan Schoder}, Florian Toth, Clemens Freidhager, and Manfred Kaltenbacher. “Revisiting infinite mapping layer for open domain problems”. In: Journal of Computational Physics 392
(2019), pp. 354–367. doi: 10.1016/j.jcp.2019.04.067.

\vspace{0.6cm}
\textbf{Abstract:} We revisit and investigate the application of an infinite mapping layer in the context of different physical fields. This layer remaps the infinite physical domain to a finite extension covering the boundary of the computational domain. Thus, infinite domain problems are solved efficiently by this advanced finite element method. Compared to infinite elements, this method has major practical advantages since it uses standard basis functions. However, the solution at the outer boundary of the infinite layer must be bounded. In principle, this method can be applied to arbitrarily shaped mapping layers.

The presented method is verified against three physical field problems and their analytic solution. Firstly, the computation of the electrostatic potential in an unbounded region is investigated. Secondly, a benchmark problem of mechanical engineering demonstrates the method for continuum mechanics. Finally, eigenmodes of deep water waves are solved accurately. For these three cases, we assess three different mapping functions of tangent, exponential, and rational type and comment on their suitability for different physical field problems. The method guarantees high efficiency and accuracy, as shown by the application examples.

\vspace{0.6cm}
\begin{learningobj}[My contribution]
This work was conducted to solve the free field radiation problem of Helmholtz decomposition. The idea of mapping the free space with characteristic decay functions in a similar way was planned, conducted and implemented by me in \textit{openCFS}. I organized further applications to other physical fields and conducted a literature study for elliptic equations and free field modeling. I managed the manuscript, performed the writing and revised the manuscript with the co-authors. With this publication, I sharpened the algorithms, theory, numerical analysis and applied the methods used to various physical fields based on some initial investigation during my PhD Thesis.
\end{learningobj}

%\includepdf[pages=-,pagecommand={},width=\textwidth]{publication/2019 jcp.pdf}

\section{Aeroacoustic formulations for confined flows based on incompressible flow data}

\vspace{0.6cm}
Paul Maurerlehner, \textbf{Stefan Schoder}, Johannes Tieber, Clemens Freidhager, Helfried Steiner, Günter Brenn, Karl-Heinz Schäfer, Andreas Ennemoser, and Manfred Kaltenbacher. “Aeroacoustic formulations for confined flows based on incompressible flow data”. In: Acta Acustica 6 (2022), p. 45. doi: 10.1051/aacus/2022041.

\vspace{0.6cm}
\textbf{Abstract:} The hybrid aeroacoustic approach is an efficient way to address the issue of the disparity of scales in Computational AeroAcoustics (CAA) at low Mach numbers. In the present paper, three wave equations governing propagation of flow-induced sound of low Mach number flows, namely the Perturbed Convective Wave Equation (PCWE), Ribner’s Dilatation (RIB) equation, and Lighthill’s wave equation, are applied using the Finite Element Method (FEM). An airflow through a circular pipe with a half-moon-shaped orifice at three operating flow speeds is considered, where validation data from measurements on a dedicated test rig is available. An extensive analysis of the flow field is provided based on the results of the incompressible flow simulation. The resulting acoustic source terms are investigated, and the relevant source term contributions are determined. The results of the acoustic propagation simulations revealed that the PCWE and RIB are best suited for the present task. The overall deviation of the predicted pressure spectra from the measured mean values amounted to 2.26 and 2.13 times the standard deviation of the measurement compared to 3.55 for Lighthill’s wave equation. Besides reliably predicting the flow-induced sound, the numerical procedure of source term computation is straightforward for PCWE and RIB, where the source term contributions, shown to be relevant, solely consist of time derivatives of the incompressible pressure. In contrast, the Lighthill source term involves spatial derivatives and, thus, is strongly dependent on the spatial resolution and the numerical method actually used for approximating these terms.

\vspace{0.6cm}
\begin{learningobj}[My contribution]
I am a senior author of this paper, i.e., the technical advisor of the PhD student P. Maurerlehner (formally supervised by M. Kaltenbacher). I contributed to the definition of the scientific questions behind this work. I co-authored this study and took care of editing major parts of the manuscript. The study is a first grain on my roadmap to prove the applicability of viscous acoustic splitting of flow and acoustics inside the flow source domain. The algorithms used for this paper were implemented by me.
\end{learningobj}

%\includepdf[pages=-,pagecommand={},width=\textwidth]{publication/aacus220017.pdf}

\section{Acoustic modeling using the aeroacoustic wave equation based on Pierce's operator}

\vspace{0.6cm}
\textbf{Stefan Schoder}, Étienne Spieser, Hugo Vincent, Christophe Bogey, and Christophe Bailly. “Acoustic modeling using the aeroacoustic wave equation based on Pierce's operator”. AIAA Journal (2023) accessed June 8, 2023. doi: 10.2514/1.J062558.

\vspace{0.6cm}
\textbf{Abstract:} The capabilities of an aeroacoustic wave equation based on Pierce's operator (AWE-PO) for modeling subsonic flow-induced sound and for sound prediction are investigated. The wave equation is applied to an isothermal two-dimensional mixing layer computed by direct numerical simulation.
In contrast to a direct numerical simulation, providing the acoustic fluctuations directly, the simulations based on Lighthill's wave equation and the AWE-PO rely on a hybrid workflow to predict the generated sound field. 
Special attention is put on the interpretation of the right-hand side of both wave equations. Comparing the terms on the right-hand side in Lighthill's theory and AWE-PO suggests a source amplitude for AWE-PO that is 90\% smaller.
This reduction is attributed to the filtering property of the material derivative. Finally, the results of the acoustic far-field pressure are compared. It is shown that the radiated sound field's directivity, propagation, and convection effects are well captured for both wave equations.
The computations using Lighthill's equation and AWE-PO are found to provide acoustic intensities within 1.8~dB from the reference direct numerical simulation.
This error is comparable with the errors reported for Lighthill's equation in previous studies.

\vspace{0.6cm}
\begin{learningobj}[My contribution]
The work is a first attempt to go with the PCWE equation beyond the incomprehensible flow regime. It involves algorithms I previously designed (hybrid aeroacoustics and Helmholtz decomposition in \textit{openCFS}). I took care of writing the manuscript and the process to be published. The work was carried out partly during my research stay at LMFA in Lyon.
\end{learningobj}

%\includepdf[pages=-,pagecommand={},width=\textwidth]{publication/1.j062558.pdf}

%%% BIBLIOGRAPHY %%%%%%%%%%%%%%%%%%%%%%%%%%%%%%%%%%%%%%%%%%%%%%%%%%%%%%%%%%%%%%%%
%\clearpage
%\printbibliography[heading=subbibintoc,title={References}] % print chapter bibliography

%%%%%%%%%%%%%%%%%%%%%%%%%%%%%%%%%%%%%%%%%%%%%%%%%%%%%%%%%%%%%%%%%%%%%%%%%%%%%%%%%
 \cleardoublepage %Comment in THE Benchmarks (69 PAGES)
 % !TeX spellcheck=en_US
% !TeX root=main.tex
%%%%%%%%%%%%%%%%%%%%%%%%%%%%%%%%%%%%%%%%%%%%%%%%%%%%%%%%%%%%%%%%%%%%%%%%%%%%%%%%%

\chapter{Technical Applications and Benchmarking}
\label{chap:chapter_3}

\begin{learningobj}[Contributions and research roadmap]
The contributions of this chapter are four-fold:
\begin{itemize}
    \item We issued several benchmark datasets of theoretical and practical relevance in a database. One of the datasets of high practical value in today's economy is the Dataset FAN-01 for validating aeroacoustic simulations of HVAC systems for, e.g., battery cooling or engine cooling systems. In this work, we precisely described every step of how to reach a reliable flow-induced sound prediction \cite{schoder2020computational}. This resulted in resonance within the relevant scientific community and industry, which further motivated us to progress this topic.
    \item We resolved one of the most common pitfalls when applying a hybrid aeroacoustic simulation workflow by the energy-conservative source interpolation and supported it with a theoretical basis. It allows performing large scale numerical simulation with relatively little computational cost of the acoustic simulation part \cite{schoder2021application}.
    \item We applied machine learning techniques to the theory of the PCWE to detect noise sources of a simulation based on the FAN-01 setup. It was the first in-depth study performing this task \cite{tieghi2023machine}.
    \item We studied the noise emissions of an electric ducted fan unit experimentally using a low-cost testing routine that can be applied during the very early stage of prototyping within very limited personal and monetary resources. We show how this simply measurement approach already could be leveraged to assess noise-signature improvements of a fan prototype \cite{fluids8040116,schoder2022quantification}. %Electirfied air mobility ...
\end{itemize}
These contributions lead to the idea to supply benchmarking datasets online combining measurements, high-fidelity simulations and performed studies within a data-frame website in the future. It will be one of the USPs of the research groups involved and gains international visibility. Already with the first benchmarking dataset FAN-01 \cite{0db92a333fb042b1a3a8e7304eb7c84a} international OEMs are requesting projects, see this as a lighthouse of my research activity. 
\end{learningobj}

\subsection*{Enclosed original work}

\begin{center}
\begin{tabular}{ c p{13cm} }
   \cite{schoder2020computational} & \textbf{Stefan Schoder}, Clemens Junger, and Manfred Kaltenbacher. “Computational aeroacoustics of the EAA benchmark case of an axial fan”. In: Acta Acustica 4.5 (2020), p. 22. doi: 10.1051/aacus/2020021. \textbf{Under the 5 most cited papers in the journal 2021.}\\ 
   \cite{schoder2021application} & \textbf{Stefan Schoder}, Andreas Wurzinger, Clemens Junger, Michael Weitz, Clemens Freidhager,
Klaus Roppert, and Manfred Kaltenbacher. “Application limits of conservative source
interpolation methods using a low Mach number hybrid aeroacoustic workflow”. In:
Journal of Theoretical and Computational Acoustics 29.1, 2050032 (2021). doi: 10.1142/
S2591728520500322. \\ 
   \cite{tieghi2023machine} & Lorenzo Tieghi, Stefan Becker, Alessandro Corsini, Giovanni Delibra, \textbf{Stefan Schoder}, and Felix Czwielong. “Machine-learning clustering methods applied to detection of noise sources in low-speed axial fan”. In: Journal of Engineering for Gas Turbines and Power 145.3 (2023), p. 031020. doi: 10.1115/1.4055417. \\ 
   \cite{fluids8040116} & \textbf{Stefan Schoder}, Jakob Schmidt, Andreas Fürlinger, Roppert Klaus, and Maurerlehner Paul. “An Affordable Acoustic Measurement Campaign for Early Prototyping Applied to Electric Ducted Fan Units”. Fluids 8, no. 4: 116 (2023), doi: 10.3390/fluids8040116. \\ 
\end{tabular}
\end{center}

\newpage
\section{Computational aeroacoustics of the EAA benchmark case of an axial fan}

\vspace{0.6cm}
\textbf{Stefan Schoder}, Clemens Junger, and Manfred Kaltenbacher. “Computational aeroacous- tics of the EAA benchmark case of an axial fan”. In: Acta Acustica 4.5 (2020), p. 22. doi: 10.1051/aacus/2020021.

\vspace{0.6cm}
\textbf{Abstract:} This contribution benchmarks the aeroacoustic workflow of the perturbed convective wave equation and the Ffowcs Williams and Hawkings analogy in Farassat’s 1A version for a low-pressure axial fan. Thereby, we focus on the turbulence modeling of the flow simulation and mesh convergence concerning the complete aeroacoustic workflow. During the validation, good agreement has been found with the efficiency, the wall pressure sensor signals, and the mean velocity profiles in the duct. The analysis of the source term structures shows a strong correlation to the sound pressure spectrum. Finally, both acoustic sound propagation models are compared to the measured sound field data.

\vspace{0.6cm}
\begin{learningobj}[My contribution]
The work aim is twofold. Firstly, I aimed to present the usefulness of the algorithms I designed and implemented in \textit{openCFS}. Secondly, to advertise the highly useful aeroacoustics benchmark dataset co-hosted with my collaboration in Erlangen and by my research group. I took care of the writing and shaping the article to the published version. The algorithms used for this paper were implemented by me. During the data generation I assisted the Clemens Junger to apply the algorithms.
\end{learningobj}

%\includepdf[pages=-,pagecommand={},width=\textwidth]{publication/2020 acta fan schoder.pdf}

\section[Application limits of conservative source
interpolation methods]{Application limits of conservative source
interpolation methods using a low Mach number hybrid aeroacoustic workflow}

\vspace{0.6cm}
\textbf{Stefan Schoder}, Andreas Wurzinger, Clemens Junger, Michael Weitz, Clemens Freidhager,
Klaus Roppert, and Manfred Kaltenbacher. “Application limits of conservative source
interpolation methods using a low Mach number hybrid aeroacoustic workflow”. In:
Journal of Theoretical and Computational Acoustics 29.1, 2050032 (2021). doi: 10.1142/
S2591728520500322.

\vspace{0.6cm}
\textbf{Abstract:} In low Mach number aeroacoustics, the well-known disparity of scales allows applying hybrid simulation models using different meshes for flow and acoustics, which leads to a fast computational procedure. The hybrid workflow of the perturbed convective wave equation involves three steps: (1) perform unsteady incompressible flow computations on a subdomain; (2) compute the acoustic sources and (3) simulate the acoustic field, using a mesh specifically suited. These aeroacoustic methods seek for a robust, conservative and computational efficient mesh-to-mesh transformation of the aeroacoustic sources. In this paper, the accuracy and the application limitations of a cell-centroid-based conservative interpolation scheme is compared to the computationally advanced cut-volume cell approach in 2D and 3D. Based on a previously validated axial fan model where spurious artifacts have been visualized, the results are evaluated systematically using a grid convergence study. To conclude, the monotonic convergence of both conservative interpolation schemes is demonstrated. Regarding arbitrary mesh deformation (for example, the motion of the vocal folds in human phonation), the study reveals that the computationally simpler cell-centroid-based conservative interpolation can be the method of choice.

\vspace{0.6cm}
\begin{learningobj}[My contribution]
In precursor of each acoustic fan simulation it is important to know how to establish an acoustic mesh based on physical principles. The article summarizes these principles I found and structures them into easy guidelines. I took care of presenting the data and writing the manuscript with my co-authors. The algorithms used for this paper were implemented by me.
\end{learningobj}

%\includepdf[pages=-,pagecommand={},width=\textwidth]{publication/2021 Schoder Conservative JTCA.pdf}

\section{Machine-learning clustering methods applied to detection of noise sources in low-speed axial fan}

\vspace{0.6cm}
Lorenzo Tieghi, Stefan Becker, Alessandro Corsini, Giovanni Delibra, \textbf{Stefan Schoder}, and Felix Czwielong. “Machine-learning clustering methods applied to detection of noise sources in low-speed axial fan”. In: Journal of Engineering for Gas Turbines and Power 145.3 (2023), p. 031020. doi: 10.1115/1.4055417.

\vspace{0.6cm}
\textbf{Abstract:} The integration of rotating machineries in human-populated environments requires to limit noise emissions, with multiple aspects impacting on control of amplitude and frequency of the acoustic signature. This is a key issue to address and when combined with compliance of minimum efficiency grades, further complicates the design of axial fans. The aim of this research is to assess the capability of unsupervised learning techniques in unveiling the mechanisms that concur to the sound generation process in axial fans starting from high-fidelity simulations. To this aim, a numerical dataset was generated by means of large eddy simulation (LES) simulation of a low-speed axial fan. The dataset is enriched with sound source computed solving a-posteriori the perturbed convective wave equation (PCWE). First, the instantaneous flow features are associated with the sound sources through correlation matrices and then projected on latent basis to highlight the features with the highest importance. This analysis in also carried out on a reduced dataset, derived by considering two surfaces at 50\% and 95\% of the blade span. The sampled features on the surfaces are then exploited to train three cluster algorithms based on partitional, density and Gaussian criteria. The cluster algorithms are optimized and their results are compared, with the Gaussian Mixture one demonstrating the highest similarity (>80\%). The derived clusters are analyzed, and the role of statistical distribution of velocity and pressure gradients is underlined. This suggests that design choices that affect these aspects may be beneficial to control the generation of noise sources.

\vspace{0.6cm}
\begin{learningobj}[My contribution]
I am a senior author of this paper, i.e., the advisor of the manuscript regarding the PCWE and the source term formulation connected to the PCWE model. I contributed to the definition of the scientific questions behind this work with respect to the acoustic results, and contributed to the writing and editing of the manuscript.
\end{learningobj}

%\includepdf[pages=-,pagecommand={},width=\textwidth]{publication/gtp_145_03_031020.pdf}

\section[Affordable Acoustic Measurement Campaign for Early Prototyping]{An Affordable Acoustic Measurement Campaign for Early Prototyping Applied to Electric Ducted Fan Units}

\vspace{0.6cm}
\textbf{Stefan Schoder}, Jakob Schmidt, Andreas Fürlinger, Roppert Klaus, and Maurerlehner Paul. “An Affordable Acoustic Measurement Campaign for Early Prototyping Applied to Electric Ducted Fan Units”. Fluids 8, no. 4: 116 (2023), doi: 10.3390/fluids8040116.

\vspace{0.6cm}
\textbf{Abstract:} New innovative green concepts in electrified vertical take-off and landing vehicles are currently emerging as a revolution in urban mobility going into the third dimension (vertically). The high population density of cities makes the market share highly attractive while posing an extraordinary challenge in terms of community acceptance due to the increasing and possibly noisier commuter traffic. In addition to passenger transport, package deliveries to customers by drones may enter the market. The new challenges associated with this increasing transportation need in urban, rural, and populated areas pose challenges for established companies and startups to deliver low-noise emission products. The article’s objective is to revisit the benefits and drawbacks of an affordable acoustic measurement campaign focused on early prototyping. In the very early phase of product development, available resources are often considerably limited. With this in mind, this article discusses the sound power results using the enveloping surface method in a typically available low-reflection room with a reflecting floor according to DIN EN ISO 3744:2011-02. The method is applied to a subsonic electric ducted fan (EDF) unit of a 1:2 scaled electrified vertical take-off and landing vehicle. The results show that considerable information at low costs can be gained for the early prototyping stage, despite this easy-to-use, easy-to-realize, and non-fine-tuned measurement setup. Furthermore, the limitations and improvements to a possible experimental setup are presented to discuss a potentially more ideal measurement environment. Measurements at discrete operating points and transient measurements across the total operating range were conducted to provide complete information on the EDF’s acoustic behavior. The rotor-self noise and the rotor–stator interaction were identified as primary tonal sound sources, along with the highest broadband noise sources located on the rotor. Based on engineering experience, a first acoustic improvement treatment was also quantified with a sound power level reduction of 4 dB(A). In conclusion, the presented method is a beneficial first measurement campaign to quantify the acoustic properties of an electric ducted fan unit under minimal resources in a reasonable time of several weeks when starting from scratch.

\vspace{0.6cm}
\begin{learningobj}[My contribution]
This manuscript is a summary of a low cost experimental acoustic evaluation method for electric ducted fan units. I helped to design and structure the tests, organized the testing facilities and helped the second author during the measurements and data acquisition. Especially the introduction was written by me. Together with the second author, I drafted, wrote and revised the manuscript.
\end{learningobj}

%\includepdf[pages=-,pagecommand={},width=\textwidth]{publication/fluids-08-00116.pdf}

%%% BIBLIOGRAPHY %%%%%%%%%%%%%%%%%%%%%%%%%%%%%%%%%%%%%%%%%%%%%%%%%%%%%%%%%%%%%%%%
% \clearpage
% \printbibliography[heading=subbibintoc,title={References}] % print chapter bibliography

%%%%%%%%%%%%%%%%%%%%%%%%%%%%%%%%%%%%%%%%%%%%%%%%%%%%%%%%%%%%%%%%%%%%%%%%%%%%%%%%%
 \cleardoublepage %Comment in THE FSAI (59 PAGES)
 % !TeX spellcheck=en_US
% !TeX root=main.tex
%%%%%%%%%%%%%%%%%%%%%%%%%%%%%%%%%%%%%%%%%%%%%%%%%%%%%%%%%%%%%%%%%%%%%%%%%%%%%%%%%

\chapter{Fluid-Structure-Acoustic interaction with application to human phonation}

\begin{learningobj}[Contributions and research roadmap]
The contributions of this chapter are three-fold:
\begin{itemize}
    \item We studied the PCWE right-hand-side terms and model simplifications in detail \cite{schoder2021aeroacoustic}. It was especially interesting that the right-hand-side term is cancelling out partly due to the convective derivative, which was later in detail investigated during the study of the AWE-PO and explained by the cancelation of frozen convected sources \cite{schoder2022aeroacoustic}. It was already acknowledged in the literature that this might happen, but for the first time a case was investigated and it was observed in reality.
    \item We showed that the simVoice model, derived in numerous studies before, is able to predict and investigate different functional voice disorder characteristics \cite{falk20213d}. The model is based on the PCWE and it is possible to observe the acoustic field also inside the human airways, which are not easy accessible with measuring routines at living subjects.
    \item We investigated the impact of different subgrid-scale models for hybrid simulations using \textit{openFoam}. The workflow developed showed excellent agreement of the vowels in the formant chart \cite{lasota2023anisotropic}.
\end{itemize}
These contributions lead to Weaves (DFG-FWF) proposal to advance the FSAI Modeling of aeroacoustics and consistently describe the energy transformation process from the fluid flow, mechanical and acoustic energy. The application to human phonation results in algorithms that solve numerical challenges that are also present in other medical applications, such as the application to cardiological problems of the heart valve (full contact).
\end{learningobj}

\subsection*{Enclosed original work}

\begin{center}
\begin{tabular}{ c p{13cm} }
   \cite{schoder2021aeroacoustic} & \textbf{Stefan Schoder}, Paul Maurerlehner, Andreas Wurzinger, Alexander Hauser, Sebastian Falk, Stefan Kniesburges, Michael Döllinger, and Manfred Kaltenbacher. “Aeroacoustic sound source characterization of the human voice production-perturbed convective wave equation”. In: Applied Sciences 11.6 (2021), p. 2614. doi: 10.3390/app11062614. \textbf{Under the 5 most cited papers in the journals field 2022.}\\ 
   \cite{falk20213d} &  Sebastian Falk, Stefan Kniesburges, \textbf{Stefan Schoder}, Bernhard Jakubaß, Paul Maurerlehner, Matthias Echternach, Manfred Kaltenbacher, and Michael Döllinger. “3D-FV-FE aeroacoustic larynx model for investigation of functional based voice disorders”. In: Frontiers in physiology 12 (2021), p. 226. doi: 10.3389/fphys.2021.616985.\\ 
   \cite{lasota2023anisotropic} &  Martin Lasota, Petr Šidlof, Paul Maurerlehner, Manfred Kaltenbacher, and \textbf{Stefan Schoder}. "Anisotropic minimum dissipation subgrid-scale model in hybrid aeroacoustic simulations of human phonation." The Journal of the Acoustical Society of America 153.2 (2023): 1052-1063. doi: 10.1121/10.0017202.\\ 
\end{tabular}
\end{center}

\newpage
\section[Aeroacoustic sound source characterization of the human voice]{Aeroacoustic sound source characterization of the human voice production-perturbed convective wave equation}

\vspace{0.6cm}
\textbf{Stefan Schoder}, Paul Maurerlehner, Andreas Wurzinger, Alexander Hauser, Sebastian Falk, Stefan Kniesburges, Michael Döllinger, and Manfred Kaltenbacher. “Aeroacoustic sound source characterization of the human voice production-perturbed convective wave equation”. In: Applied Sciences 11.6 (2021), p. 2614. doi: 10.3390/app11062614.

\vspace{0.6cm}
\textbf{Abstract:} The flow-induced sound sources of human voice production are investigated based on a validated voice model. This analysis is performed using a hybrid aeroacoustic workflow based on the perturbed convective wave equation. In the first step, the validated 3D incompressible turbulent flow simulation is computed by the finite volume method using STARCCM+. In a second step, the aeroacoustic sources are evaluated and studied in detail. The formulation of the sound sources is compared to the simplification (neglecting the convective sources) systematically using time-domain and Fourier-space analysis. Additionally, the wave equation is solved with the finite element solver openCFS to obtain the 3D sound field in the acoustic far-field. During the detailed effect analysis, the far-field sound spectra are compared quantitatively, and the flow-induced sound sources are visualized within the larynx. In this contribution, it is shown that the convective part of the sources dominates locally near the vocal folds (VFs) while the local time derivative of the incompressible pressure is distributed in the whole supra-glottal area. Although the maximum amplitude of the time derivative is lower, the integral contribution dominates the sound spectrum. As a by-product of the detailed perturbed convective wave equation source study, we show that the convective source term can be neglected since it only reduces the validation error by 0.6\%. Neglecting the convective part reduces the algorithmic complexity of the aeroacoustic source computation of the perturbed convective wave equation and the stored flow data. From the source visualization, we learned how the VF motion transforms into specific characteristics of the aeroacoustic sources. We found that if the VFs are fully closing, the aeroacoustic source terms yield the highest dynamical range. If the VFs are not fully closing, VFs motion does not provide as much source energy to the flow-induced sound sources as in the case of a healthy voice. As a consequence of not fully closing VFs, the cyclic pulsating velocity jet is not cut off entirely and therefore turbulent structures are permanently present inside the supraglottal region. These turbulent structures increase the broadband component of the voice signal, which supports research results of previous studies regarding glottis closure and insufficient voice production.

\vspace{0.6cm}
\begin{learningobj}[My contribution]
This work was my first study on the aeroacoustics sources of the PCWE and gave me the first hints on a strong reduction of the source strength by the material derivative. Based on these results, I further conducted research in this direction. I structured the idea based on the graphics and simulations created by the second author, and wrote the paper in close collaboration with the second author. I provided the algorithms for this study. Together with the co-authors, I revised the manuscript and successfully published it.
\end{learningobj}

%\includepdf[pages=-,pagecommand={},width=\textwidth]{publication/2021applsci-11-02614.pdf}

\section{3D-FV-FE aeroacoustic larynx model for investigation of functional based voice disorders}

\vspace{0.6cm}
Sebastian Falk, Stefan Kniesburges, \textbf{Stefan Schoder}, Bernhard Jakubaß, Paul Maurerlehner, Matthias Echternach, Manfred Kaltenbacher, and Michael Döllinger. “3D-FV-FE aeroacoustic larynx model for investigation of functional based voice disorders”. In: Frontiers in physiology 12 (2021), p. 226. doi: 10.3389/fphys.2021.616985.

\vspace{0.6cm}
\textbf{Abstract:} For the clinical analysis of underlying mechanisms of voice disorders, we developed a numerical aeroacoustic larynx model, called simVoice, that mimics commonly observed functional laryngeal disorders as glottal insufficiency and vibrational left-right asymmetries. The model is a combination of the Finite Volume (FV) CFD solver Star-CCM+ and the Finite Element (FE) aeroacoustic solver CFS++. simVoice models turbulence using Large Eddy Simulations (LES) and the acoustic wave propagation with the perturbed convective wave equation (PCWE). Its geometry corresponds to a simplified larynx and a vocal tract model representing the vowel /a/. The oscillations of the vocal folds are externally driven. In total, 10 configurations with different degrees of functional-based disorders were simulated and analyzed. The energy transfer between the glottal airflow and the vocal folds decreases with an increasing glottal insufficiency and potentially reflects the higher effort during speech for patients being concerned. This loss of energy transfer may also have an essential influence on the quality of the sound signal as expressed by decreasing sound pressure level (SPL), Cepstral Peak Prominence (CPP), and Vocal Efficiency (VE). Asymmetry in the vocal fold oscillations also reduces the quality of the sound signal. However, simVoice confirmed previous clinical and experimental observations that a high level of glottal insufficiency worsens the acoustic signal quality more than oscillatory left-right asymmetry. Both symptoms in combination will further reduce the quality of the sound signal. In summary, simVoice allows for detailed analysis of the origins of disordered voice production and hence fosters the further understanding of laryngeal physiology, including occurring dependencies. A current walltime of 10 h/cycle is, with a prospective increase in computing power, auspicious for a future clinical use of simVoice.

\vspace{0.6cm}
\begin{learningobj}[My contribution]
As the primary coauthor from the collaborating research group. I provided the algorithms for the aeroacoustics model. I helped in the editing of the aeroacoustics details in the manuscript and provided detailed information during the revision.
\end{learningobj}

%\includepdf[pages=-,pagecommand={},width=\textwidth]{publication/2021fphys-12-616985.pdf}

\section[Anisotropic minimum dissipation subgrid-scale model in hybrid aeroacoustic]{Anisotropic minimum dissipation subgrid-scale model in hybrid aeroacoustic simulations of human phonation}

\vspace{0.6cm}
Martin Lasota, Petr Šidlof, Paul Maurerlehner, Manfred Kaltenbacher, and \textbf{Stefan Schoder}. "Anisotropic minimum dissipation subgrid-scale model in hybrid aeroacoustic simulations of human phonation." The Journal of the Acoustical Society of America 153.2 (2023): 1052-1063. doi: 10.1121/10.0017202.

\vspace{0.6cm}
\textbf{Abstract:} This article deals with large-eddy simulations of three-dimensional incompressible laryngeal flow followed by acoustic simulations of human phonation of five cardinal English vowels, /u, i, a, o, \ae/. The flow and aeroacoustic simulations were performed in OpenFOAM and in-house code openCFS, respectively. Given the large variety of scales in the flow and acoustics, the simulation is separated into two steps: (1) computing the flow in the larynx using the finite volume method on a fine moving grid with 2.2 million elements, followed by (2) computing the sound sources separately and wave propagation to the radiation zone around the mouth using the finite element method on a coarse static grid with 33000 elements. The numerical results showed that the anisotropic minimum dissipation model, which is not well known since it is not available in common CFD software, predicted stronger sound pressure levels at higher harmonics, and especially at first two formants, than the wall-adapting local eddy-viscosity model. The model on turbulent flow in the larynx was employed and a positive impact on the quality of simulated vowels was found.

\vspace{0.6cm}
\begin{learningobj}[My contribution]
I am a senior author of this paper, i.e., the technical advisor of the PhD student M. Lasota visiting our research institute (formally hosted by M. Kaltenbacher). I contributed to the definition of the scientific questions behind this work, guided part of the acoustic model based on my human phonation experience, conceived and directed the execution of the simulation workflow, and contributed to the writing of the manuscript.
\end{learningobj}

%\includepdf[pages=-,pagecommand={},width=\textwidth]{publication/1052_1.pdf}

%%% BIBLIOGRAPHY %%%%%%%%%%%%%%%%%%%%%%%%%%%%%%%%%%%%%%%%%%%%%%%%%%%%%%%%%%%%%%%%
% \clearpage
% \printbibliography[heading=subbibintoc,title={References}] % print chapter bibliography

%%%%%%%%%%%%%%%%%%%%%%%%%%%%%%%%%%%%%%%%%%%%%%%%%%%%%%%%%%%%%%%%%%%%%%%%%%%%%%%%%
 %\cleardoublepage
 %\include{05_PINN}

%%%%%%%%%%%%%%%%%%%%%%%%%%%%%%%%%%%%%%%%%%%%%%%%%%%%%%%%%%%%%%%%%%%%%%%%%%%%%%%%%
\cleardoublepage
% !TeX spellcheck=en_US
% !TeX root=main.tex
%%%%%%%%%%%%%%%%%%%%%%%%%%%%%%%%%%%%%%%%%%%%%%%%%%%%%%%%%%%%%%%%%%%%%%%%%%%%%%%%%
% \chapter{Simulation Theory}

% \begin{learningobj}
% In this chapter, you will learn\dots
% \begin{itemize}
%     \item ....
% \end{itemize}
% \end{learningobj}

\chapter{Conclusions}
\label{chap:chapter_5}

% \section{Diversity}

% The theories can ...

\section{Provided value}
Creating value is given by a relevant, distinct, relatively simple and lasting piece of work. In doing so, we provide the computational methods developed as open-source software.

\paragraph{openCFS}
Beside about 20 research-code-developers, I am one of the driving forces of openCFS \cite{CFS}. 
\textit{"Open Coupled Field Simulation (openCFS) is a finite element-based multi-physics modelling and simulation tool. With about 20 years of research driven development, the core of openCFS is used in scientific research and industrial application. The modelling strategy focuses on physical fields and their respective couplings."}\\

\textit{openCFS} is a finite element-based simulation tool that provides special modeling and simulation capabilities in the following research fields:
\begin{itemize}
\item Acoustics
\item Electrostatics
\item Mechanics
\item Magnetics
\item Piezoelectric
\item Thermal and Heat transfer
\end{itemize}
In addition to this single physical field computations, \textit{openCFS} has become known for coupling multiple physical fields:
\begin{itemize}
\item Magneto-Mechanics
\item Vibroacoustics
\item Piezo-Acoustics
\item Magneto-Mechanic-Acoustics
\item Thermo-Mechanics
\end{itemize}
Static Analysis, Transient Analysis, Harmonic Analysis, Multi-Harmonic Analysis, and Eigenfrequency Analysis  are part of \textit{openCFS} simulation repartior. Furthermore, special FEM topics are addressed by the software including:
\begin{itemize}
\item Nonconforming Grid
\item Model Order Reduction
\item Structural Optimization
\item Data processing and API for hybrid aeroacoustics
\end{itemize}
Regarding these essentials, I am driving the software development of the Acoustic, Aeroacoustic, and Vibroacoustic part of the \textit{openCFS} project. During the last year, research in connection to the Helmholtz decomposition as a tool for aeroacoustic post-processing has been developed. The computation of Helmholtz decomposition involves solving elliptic PDEs and therefore the consideration of the infinite space. A specific finite element formulation based on a coordinate transformation was published in 2019. Additionally, we showed that computing the incompressible part of a compressible flow field is the more accurate version regarding non-convex domains. Therefore, we used the curl-curl like equation and edge-FEM to discretize the PDE problem. The method presented as non-conforming Nitsche Interfaces strongly improves the mesh generation for edge-FEM-based solution algorithms. Overall, the \textit{openCFS} implementations increase the capability of existing methods and add value to the developed software.

 % \includepdf[pages=1]{publication/2019 jcp.pdf}
 % \includepdf[pages=1]{publication/2020 ieee NC Interfaces schoder.pdf}

\paragraph{openCFS-Data}
\textit{openCFS-Data} \cite{schoder2023opencfsdata} is a data processing and API for the hybrid aeroacoustics workflow. The hybrid workflow was already presented during the introduction of this thesis. 

Since the beginning of computational aeroacoustics (CAA), hybrid methodologies have been established as the most practical methods for aeroacoustic computations. The workflow of these approaches is based on three steps (see Publication \cite{schoder2019hybrid}): (1) perform unsteady flow computations based on an appropriate turbulence model on a restricted sub-domain; (2) compute the acoustic sources with a conservative algorithm; (3) compute the acoustic field. Analogously to CAA, decoupled vibroacoustic simulations use a similar three-step approach, aiming to compute the acoustic field due to structural deformations.

Both computational schemes require a robust transformation of the discrete fluid field (or in case of vibro-acoustics the discrete mechanical acceleration) to the acoustic simulation grid. Based on the framework of radial basis functions (RBF), we implemented a general method for the computation of source terms. Compared to different interpolation algorithms implemented in \textit{openCFS-Data} (nearest neighbor interpolation -- NNI, partial differential equation solver interpolation -- PDEI, polynomial interpolation -- PI, global RBF -- gRBF), the investigated local RBF (lRBF) interpolation has promising capabilities: 
\begin{itemize}
    \item An accurate \textbf{(consistent)} and fast interpolation technique \textbf{(scalable)}. 
    \item An interpolation technique that handles special grids, e.g. grids to resolve boundary layers \textbf{(physical)}. 
    \item A method to compute accurate derivatives of the primary flow variables, such as pressure $p$, velocity $\bm{u}$, density $\rho$, temperature $T$, and entropy $s$ \textbf{(derivative)}.
    \item A \textbf{flexible} algorithm that can be integrated into a standard product development cycle.
    \item A \textbf{generic} algorithm that assembles different hybrid aeroacoustic source terms, e.g. the divergence of the Lamb vector $\nabla \cdot(((\nabla \times \bm{u}) \times \bm{u})')$ or the divergence of an entropy source $\nabla \cdot (T'\nabla \overline{s} - s'\nabla \overline{T})$. %A prime denotes that just the fluctuating part of the physical quantity is considered.
\end{itemize}
A comparison of the different techniques with respect to some relevant parameters (consistent, local, physical, derivative, flexible, generic) is illustrated in Tab~\ref{tab:compare}. 
\begin{table}[ht!]
\centering
\caption{\label{tab:compare}Comparison of the different interpolation methods.}
\begin{tabular}{lrrrrrr}
\toprule
Method & consistent & scalable & physical & derivative & flexible & generic\\
\midrule
NNI & - - & ++ & - - & - - & ++ & -\\
PI & - & - & 0 & + & ++ & +\\
gRBF & - & - - & 0 & + & ++ & ++\\
\textbf{lRBF} & \textbf{-} & \textbf{++} & \textbf{+} & \textbf{+} & \textbf{++} & \textbf{++}\\
PDEI & ++ & - - & ++ & ++ & 0 & +\\
\bottomrule  
\end{tabular}
\end{table}
We compared the local RBF interpolation (lRBF) to different interpolation techniques. Overall, the lRBF is a good compromise between accuracy and flexibility. Although lRBF is not consistent with the used interpolation functions of typical PDE based solvers, the interpolation and derivatives are accurate and converge with increasing resolution. The special construction of lRBF provides a scalable algorithm that is tuned to flow phenomena including boundary layers. 
But most importantly, the lRBF is a generic and flexible toolbox that can connect different PDE solvers and exchange data efficiently. A further improvement would be the online coupling of solvers (data exchange during the calculation of the individual physical fields).
Overall, the method serves a tool that can connect different simulation software with a minimum required data description consistency.

By distributing the software \textit{openCFS} and \textit{openCFS-Data} to our students, the community, and the society, we provide outstanding value to the receivers. My students (Master, PhD) learn how to program, model, simulate, visualize their theoretical ideas. They built digital twins and are finally well suited for research or industry to deliver the highest quality products and serve a competitive advantage to their employees. Providing this knowledge under the \textit{MIT License}, we serve a valuable platform for building new businesses upon the initial software. These new business cases will drive employment in the automotive cluster around Graz and in general Austrian companies have a competitive advantage at low license cost.

\section{Significance}

Besides the research activities (publications and open-source software), we provide significant economic impact for Austria and the southern German region. As a core competence, we assist startups, small, mid, and international companies to develop their aeroacoustic workflows and digital twins. We help them to increase their social stake, fortify research units located in this region, and deliver knowledge to make outstanding innovations. These competencies are mostly transferred through international research projects with key partners all over the world. Consequently, we provide tools for noise reduction purposes and reduce the environmental impact of machinery and transportation (applied successfully in the fan noise projects, compressor noise projects and duct acoustics projects). I am proud of assisting people with voice disorders by providing a tool helping to educate young voice surgeons. These surgeons will increase a patient's life comfort and communication ability after suffering years.

\section{A step into the future}

Noise has been widely recognized as pollution that has to be avoided, controlled, regulated, and reduced. Its negative impact causes releasing stress hormones and changes in body functions, like insomnia. The socio-economic impact in a modern world of accurate modeling of acoustic emissions for prospective mitigation is evident. A detailed understanding of sound in fluid flows and the energy transfer from the flow to acoustic emissions are essential to reach the noise mitigation targets of the proposed directives for technical and medical applications. My group's central objective is to rigorously define, verify, and validate the acoustic variables and correctly model the aeroacoustic energy transformation. For the most relevant flow configurations, high-fidelity benchmarking datasets, theoretical models and computational pipelines will be issued to make flow-induced sound predictions more accessible. To address the objective, an aeroacoustic theory will be derived and implemented into the in-house-developed open-source software openCFS using the finite element method. The theory relies on key concepts from established models for incompressible flows, which we successfully validated recently. It is extended to a physical regime where previously neglected nonlinearity could be significant. For the first time in aeroacoustics, a framework will be derived using physics-informed neural networks to test modeling limitations involved by the linear theory. It will allow the integration of simulation results, mathematical formulations and experimental datasets seamlessly. We will prove the reliability of the classical theory by comparison with the data-driven model for the aviation-relevant flow configurations.
The innovative scientific aspect of understanding “what part of the chaotic subsonic flows is acoustics” is completely unsolved today. We will unveil a complete picture of the causal effects of the sound generation process and close existing knowledge gaps for subsonic flows near the transonic regime. 

%%% BIBLIOGRAPHY %%%%%%%%%%%%%%%%%%%%%%%%%%%%%%%%%%%%%%%%%%%%%%%%%%%%%%%%%%%%%%%%
% \clearpage
% \printbibliography[heading=subbibintoc,title={References}] % print chapter bibliography

%%% APPENDIX %%%%%%%%%%%%%%%%%%%%%%%%%%%%%%%%%%%%%%%%%%%%%%%%%%%%%%%%%%%%%%%%
\cleardoublepage
\appendix
% !TeX spellcheck=en_US
% !TeX root=main.tex
%%%%%%%%%%%%%%%%%%%%%%%%%%%%%%%%%%%%%%%%%%%%%%%%%%%%%%%%%%%%%%%%%%%%%%%%%%%%%%%%%

% !TeX spellcheck=en_US
% !TeX root=main.tex
%%%%%%%%%%%%%%%%%%%%%%%%%%%%%%%%%%%%%%%%%%%%%%%%%%%%%%%%%%%%%%%%%%%%%%%%%%%%%%%%%
\chapter{Appendix}

With the submission of this Habilitation thesis, I apply for the teaching license (\textit{Venia Docendi}) in the subject area “Aeroacoustics”. In this appendix, I show how my research and teaching contributions in the last 8 years, especially those since my appointment as a tenure-track assistant professor at Graz University of Technology (11/2020), cover this broad subject area. I describe all these achievements in detail after defining the “Aeroacoustics” subject area and highlighted how my contributions provide a consistent coverage of its main aspects regarding the communities active in aeroacoustics. These contributions encompass several scientific publications, my active participation in the scientific community, as well as my independent teaching activities and advisorship of Bachelor, Master and PhD students. 

\begin{learningobj}[Motivation and Impact]
In 2022 during the preparation of the hearing to the status of the tenure track position, at the TU Graz, I faced myself with the question ‘How to add lasting value for students?’\footnote{As Peter Drucker was concerned about how to add value for customers \cite{drucker2012practice}.}.
During the research, while reading \cite{den2011innovation} it became clear that the term ‘value’ has more dimensions than a particular group of people. Different meanings in different contexts. Actually, the question is not ‘How to add value for students?’, but should rather be ‘How to add value for students, the institute, the partner institutions, and the society?’. Although the book \cite{den2011innovation} is attracting a different audience, I will adapt it to illustrate value generation in the scientific research and teaching of 'Aeroacoustic', holistically.
\end{learningobj}

This work is dedicated to people in non-profit teaching organizations (university institutes): students, teachers, lecturers, full professors, and of course organizational staff.
I hope it will provide them further insight into what meaningful development is,
what value can be created from teaching in their
own basic research field. 
\begin{figure}[ht!]
\centering
\includegraphics[width=0.98\textwidth]{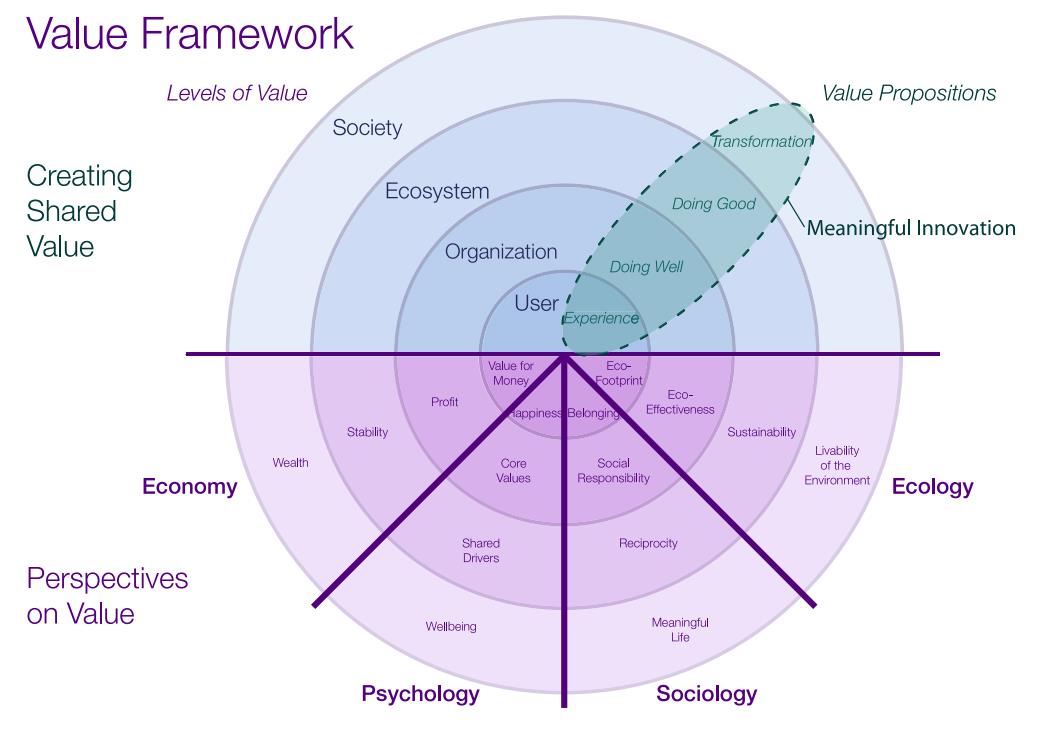}
\caption{Holistic value framework of creating shared values in platforms. \cite{den2011innovation}}
\label{fig:human_phonation}
\end{figure}
Based on the levels of creating shared value, I would like to introduce the following specific groups:
\begin{itemize}
\item The users are basically the \textit{students} itself and are subject to the teaching activity at the university.
\item The organization in the value framework of the university is represented by the \textit{institute or division} that is responsible for specific, contracted, and additional lectures fitting to the institutes' vision and responsibilities inside the university frame. 
\item Hence, part of the \textit{ecosystem} is the own and other universities, other institutes, scientific research partners, corporate partners, funding organizations, student unions, government, and exchange program organizations. All these organizations for the university ecosystem, including teaching and research activities.
\item And last but not least the \textit{society} itself.
\end{itemize}
%User = Students
%Orgnization = Institute
%Ecosystem = University, other institutes, scientific research partners, corporate partners, funding organizations, Student unions, Government, exchange programs
%Society = The society itself (direct - education, indirect - other values)s

The value of good teaching and research has several dimensions and will be evaluated in this preface to demonstrate the significance of teaching "Aeroacoustics". Starting with the core user, the students walk through the dimensions of shared value. From an economical perspective, students get value for investing their time and paying future taxes. The time investment will most generally lever their wages, as proven by several studies. Besides the economic impact, the satisfaction of achieving and the self-fulfillment will result in happiness in the long term. Students learn resilience and the belonging to the university and to their teaching group (not only by making the same experience) as learning reinforces teams. Currently, a very important issue is that students are taught to provide future knowledge to reduce the ecological footprint of society. 

Secondly, the organization itself may be non-profit but aims for additional financial capabilities to grow and advance research, teaching, and the stability of the contracting terms. Often not explicitly named work-life balance if institute members are sometimes explicitly influenced by (contracting issues). Therefore, a solid financial backbone (simply say profits) increases the value of the institute for employees and students. With this solid background, explicit social core values are learned and adopted by the students. Curiosity, creativity, adaptiveness, stand up for the silent, and many more core values are encouraged at the institutional level of shared value. With these psychological values, the social responsibility of an employer is to some point self-evident but creates a mindset for doing socially well. This combination of supporting money, core values, and social responsibility can foster the ecological development of the future at a much larger scale than the ecological footprint of a student.

Thirdly, defocusing further from the organizational level to the ecosystem level additional stakeholder are involved. The institutional role in this ecosystem is a tiny part but equally important as anyone else in the ecosystem. The ecosystem's economical scale is based on stability. Stability is important since it provides a framework for lasting, and predictable returns, profit, wealth, and connectivity. Besides financial stability, legal and relational stability are important for long-lasting businesses. Stability reduces opportunity costs and other explicit or implicit cost drivers. It is further enhanced with shared drivers, which are shared visions partly adopted by members in the ecosystem (e.g. trying to build low sound-emitting electrical driven fans) and at the peak, we have a combination of demand and solution that is reciprocal. This means for one enterprise inside the ecosystem they have a demand for a solution and on the other side, there is a demand to provide this solution (e.g a research cooperation - the institute has the demand for interesting applications and provides theoretical solutions, whereas the corporation has the demand for scientific background on its solution and provides the application details). This system can create sustainable development for ecology and also a sustainable business environment. Within this system "doing good" for society can be provided. 

Finally and in most global contexts, the society benefits from a functioning cluster and increases its wealth, well-being and provides a meaningful life experience while respecting the environment. Providing meaningful innovation is based on meaningful teaching and meaningful research (as outlined here).

Meaningful teaching is that the students have their experience to do well in their future organization and contributing to a good sustainable ecosystem to lead the transformation into prosperous times. This is a very shiny picture but describes what I expect of meaningful teaching. The same conclusion holds for meaningful research that it provides first-hand experience (risk is taken by risk-takers that can do that without significant market exposure). This enables organizations in doing well and contributing to a good sustainable ecosystem and again to lead the transformation into prosperous times.

\section{Council of European Aerospace Societies (CEAS)}
\label{chap:CaaResearchTech}

The research activities in general aeroacoustics have a wide variety and 70 years history. Since 2014, these activities have been mostly concentrated on technical applications \cite{camussi2020aeroacoustics,gely2019aeroacoustics,
enghardt2019aeroacoustics,wilson2018aeroacoustics,jivrivcek2016aeroacoustics}. 

\textit{"Thereby, the Council of European Aerospace Societies (CEAS) Aeroacoustics Specialists Committee (ASC) supports and promotes the interests of the scientific and industrial aeroacoustics community on a European scale and European aeronautics activities internationally. In this context, “aeroacoustics” encompasses all aerospace acoustics and related areas. Each year the committee highlights some of the research and development projects in Europe. \cite{camussi2020aeroacoustics}"}
\begin{figure}[ht!]
\centering
\includegraphics[width=0.98\textwidth]{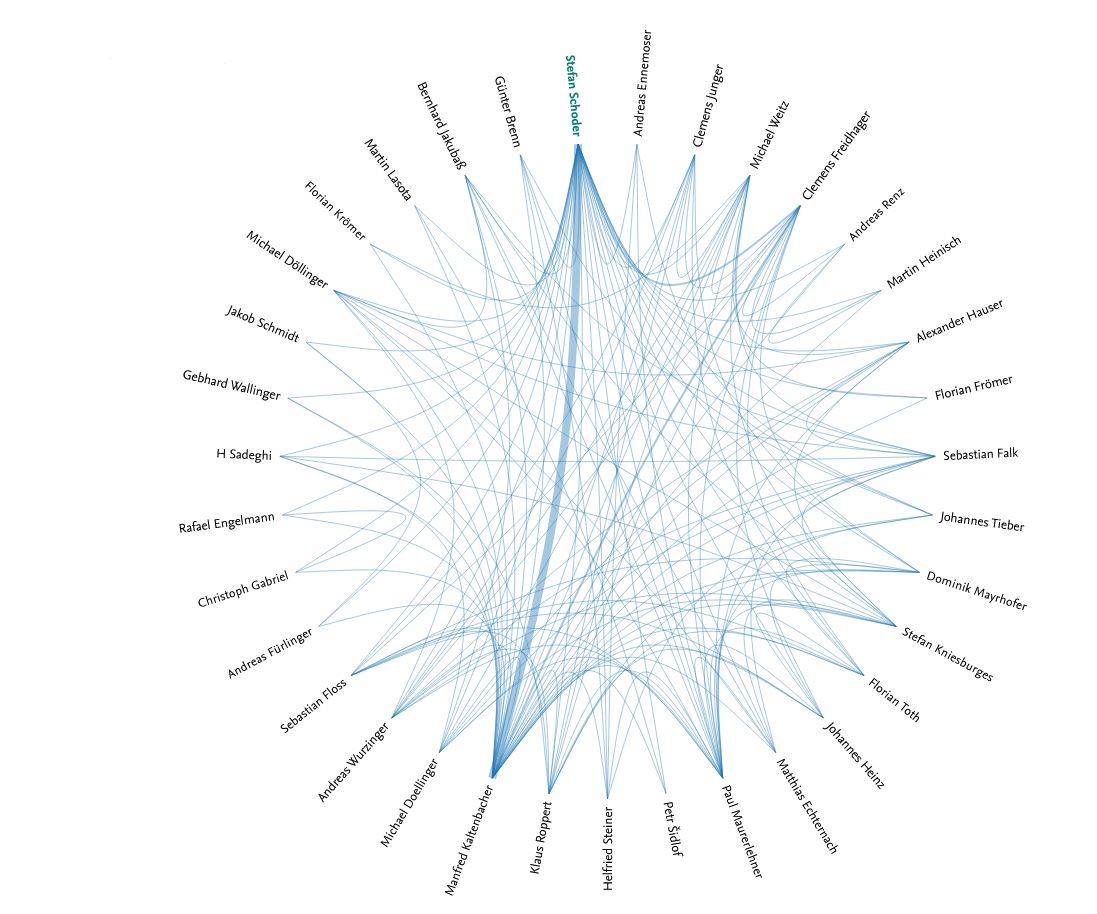}
\caption{Research network of my collaborations.}
\label{fig:res_net}
\end{figure}
According to these last six reports, the research landscape in aeroacoustics community in Europe is summarized as a preface. 

\paragraph{Year 2015 and 2016}
The year 2015 and 2016 were devoted to emerging topics in aircraft design, like "Broadband noise of rotors and airframes" \cite{wilson2018aeroacoustics,jivrivcek2016aeroacoustics}.
\begin{itemize}
\item Aeroacoustic simulation of a complete H145 helicopter in descent flight - University of Stuttgart
\item Empirical model of wall pressure one-point spectra beneath zero and adverse pressure gradient boundary layers - DLR (Deutsche Zentrum für Luft- und Raumfahrt)
\item Investigation of the generation of tones and the feedback loop in supersonic impinging jets using large-eddy simulation - ECL (École Centrale de Lyon)
\item Sound source localization on axial fans using a microphone array method - FAU (Friedrich-Alexander-Universität Erlangen-Nürnberg)
\item CAA prediction of broadband interaction noise in a turbofan stage using synthetic turbulence - ONERA (Office National d'Etudes et de Recherches Aérospatiales)
\item A numerical approach to trailing edge noise reduction by porous materials - DLR
\item Perturbed Convective Wave Equation for the precise computation of fan noise - TU Wien
\item Tailored Green's functions for flow noise predictions in ducted diaphragms - TU München
\item Combustion noise of premixed flames - RWTH Aachen
\item Acoustic multi-port characterization of modal filters containing micro-perforated panels - KU Leuven 
\item Aerofoil–turbulence interaction noise reductions by using wavy leading edges - University of Southampton
\item Direct aeroacoustic simulation of acoustic feedback phenomena on a side-view mirror - University of Stuttgart
\end{itemize}

\paragraph{Year 2017}
The year 2017 was devoted to a very important topic in aircraft design "Aircraft Noise Generated from Ducted or Un-Ducted Rotors" \cite{enghardt2019aeroacoustics}. 
\begin{itemize}
\item Cylinder noise reduction with plasma actuators - TsAGI Russia
\item Airframe and engine noise emission model for time-step simulation programs - DLR
\item Numerical analysis of the impact of the interior nozzle geometry on low Mach number jet acoustics - RWTH Aachen University
\item Identification of hydrodynamic and acoustic nature of the pressure POD modes in the near field of a subsonic jet - Université de Poitiers
\item Comprehensive experimental investigation of mode transmission through stator vane rows - DLR
\item Fan broadband noise predictions - EXA Powerflow
\item A thin active acoustic liner aiming at reducing the low-frequency tonal noise of UHBR architectures - EPFL
\item Broadband noise prediction of a low-noise serrated OGV using 3D CAA with synthetic turbulence approach - ONERA
\item A novel zonal approach for sound source prediction - TU Braunschweig
\item The modelling of the flow-induced vibrations of periodic flat and axial-symmetric structures using a wave finite element method - ECL
\item Nonlinear characterization of acoustic resonators based on neural networks - TU München
\item Non-intrusive measurement of the correlation between the turbulent density and the radiated acoustic pressure in jets - ECL
\item Strouhal number dependency of wall perforation impedance - TU Eindhoven
\item model for indirect combustion noise generation with entropy shear dispersion - Université Paris-Saclay
\item Acoustic modelling of holes and Helmholtz resonators - Imperial College
London
\end{itemize}

\paragraph{Year 2018}
The year 2018 was devoted to the topic "Future Aircraft Design and Noise
Impact", based on the recent developments of future aircraft designs \cite{gely2019aeroacoustics}.
\begin{itemize}
\item Numerical study of fan noise installation effects using the Immersed Boundary Method - ONERA
\item Full scale nose landing gear analysis - Trinity College Dublin
\item Numerical analysis of the impact of variable porosity on trailing-edge noise - RWTH Aachen University
\item RANS-based trailing-edge noise prediction using Amiet’s theory - VKI (von Karman Institute)
\item Multi-port eduction of installation effects applied to a small axial fan -  VKI
\item The conditions of quadrupole moment conservation in the evolution of small perturbations of stationary flows - TsAGI Russia
\item Uncertainty quantification for direct aeroacoustics of cavity noise - University of Stuttgart
\item Vehicle cabin noise - Chalmers
\item Nonlinear asymptotic impedance model for a Helmholtz resonator of finite depth - TU Eindhoven
\end{itemize}

\paragraph{Year 2019}
The year 2019 was mostly devoted to the topic "New materials for applications in aeroacoustics", based on urgent needs for sustainable aviation technologies \cite{camussi2020aeroacoustics}.
\begin{itemize}
\item Physics based approaches to design porous materials for noise reduction - TU Delft
\item Noise generation of a circulation control airfoil with Coanda flap and droop nose - DLR
\item Multi-approach study on nose landing gear noise - TU Delft, Trinity College Dublin
\item Flyover noise source localization during acoustic flight tests of advanced passenger aircraft - TsAGI Russia
\item Turbofan broadband noise predictions based on a ZDES calculation of a fan-OGV stage - ONERA
\item Experimental investigation of fan broadband noise generation and transmission using sparse sensor arrays - DLR
\item Study of noise generation by large-scale coherent structures in high-speed jets using conditional averages - ECL
\item Hybrid LES-RANS of serrated nozzle jets installed under wing - University of Cambridge
\item Wall-pressure fluctuations induced by a compressible jet flow over a flat plate at different Mach numbers - Univeristy of
Roma Tre
\item Effect of turbulent boundary layer induced coherence loss on beamforming measurements in industrial scale wind tunnel tests - University of Twente
\item Post-processing for direct aeroacoustic simulations using Helmholtz’s decomposition - TU Wien \cite{schoder2020postprocessing}
\item Numerical assessment of open-rotor noise shielding with a coupled approach - University of Stuttgart
\item A hybrid zonal prediction method for noise from a porous trailing-edge - DLR
\item Direct noise computation for automotive cabin noise predictions - TU Braunschweig
\item Unsteady wall pressure measurements using digital MEMS for airfoil trailing edge noise application - University Twente
\end{itemize}

\paragraph{Year 2020 and 2021}
In the year 2020 and 2021, the following topics were discussed \cite{nagy2022aeroacoustics}:
\begin{itemize}
\item Propeller noise (Roma Tre Univ., TU Delft, Univ. Toulouse) 
\item Techniques and methods in aeroacoustics (KTH, DLR, ONERA, NASA, Roma Tre Univ., TU Eindhoven, TU Graz \cite{1eecc4ba89524ed88a414f13c3e7620d,schoder2021application}, TU Braunschweig, EC Lyon, Univ. Poitiers)
\item Airframe noise (TU Delft, Univ. Bristol, Univ. Dublin, VKI, Univ. Poitiers, BTU, TU Braunschweig)
\item Fan and jet noise (ONERA, DLR, Univ. Southampton, Roma Tre Univ., VKI, Univ. Birmingham)
\item Aircraft interior noise (DLR, ECL)
\end{itemize}

\section{Activities in selected Research Groups}
Furthermore, research activities of selected research institutions are presented. To present a clear picture about the institutions, they are selected based on my research stays and collaboration partners of the last years.

\subsection{FAU - Prof. S. Becker}
The groups focus is on investigations in flow systems and machines. The fluid mechanical processes in their complex interaction with structural mechanics, heat transfer and acoustics are treated. High priority is given to a synthesis of basic research and industrial developments. In the treatment of the research topics, a complementary approach is chosen, which sensibly couples experimental, numerical and analytical work.

During my career, I was part of several collaboration projects including Prof. S. Becker. Therein, we studied fan noise, cavity noise, and new simulation methods for flow-induced sound. With the common focus on aeroacoustics, the software package \textit{CFS-Data} was started years ago to ensure a high quality hybrid workflow coupling.
\subsubsection{Research topics}
\begin{itemize}
\item Efficient low-noise turbomachinery
\item Automatic optimization of flow systems
\item Development of modern design methods for axial and radial impellers
\item Flow acoustic calculation methods and experimental investigation procedures
\end{itemize}

\subsection{Center Acoustique at LMFA - Prof. C.  Bailly}

During my research stay in Lyon (ECL), I had the chance to work with C. Bailly, C. Bogey and E. Spieser. They are interested in the deep physical understanding of the sound generation process of aeroacoustic. These interests peaked in the doctoral thesis of Etienne Spieser, investigating new and efficient techniques for computing flow-induced sound. Based on this topic, we studied extensions of his theory for a wide Mach number range. Apart form theoretical derivations, the LMFA is internationally known for its excellent researching competence, jet noise simulations, and experimental ground work for flow-induced sound benchmark cases. To conclude, I very much enjoyed the knowledge exchange at LMFA.

\subsubsection{Research topics}
\begin{itemize}
\item Aeroacoustics, direct noise computation (DNC)
\item Turbulence, numerical simulation, compressible flow
\item Subsonic and supersonic jet noise
\item Boundary layer noise
\item Propagation of sound in moving inhomogeneous media
\end{itemize}

\subsection{CTR at Stanford University - Prof. S. Lele}

Group profile \cite{CTR}: \textit{"At CTR, we study turbulence physics in complex multi-scale flows, including effects of combustion, acoustics, multi-phase interfaces and shock waves. Current research projects include: particle-laden flow in radiatively heated environment as part of the Stanford PSAAP-II Exascale Computing Engineering Center; combustion and turbulent reacting flows; aeroacoustics; two-phase flow, including microbubble generation by breaking waves and particle-laden flows; drag reduction using superhydrophobic surfaces; supersonic jet noise and airframe noise; shock-induced mixing in multi-material flows; thermoacoustic instabilities; hypersonics; active flow control; uncertainty quantification; large-eddy simulation; sub-grid-scale modeling for multi-physics flows; wall-modeling for large-eddy simulation of high-Reynolds number flows; computational linear algebra. Cross disciplinary research is emphasized at CTR, especially in the areas of numerical analysis and parallel computing of turbulent flows and validation with physical experiment."} One of my future aims is to participate at the CTR summer program.

\subsubsection{Research topics}
\begin{itemize}
\item Turbulence physics in complex multi-scale flows
\item Supersonic jet noise and airframe noise
\item Thermoacoustic instabilities
\item Uncertainty quantification
\end{itemize}

\section{My research portfolio}
Regarding the CEAS, the following research disciplines have been concerning flow-induced sound and are extended to biomedical applications. Within these disciplines, I listed the articles in Table \ref{tab:discipliens} two show the coverage of the scientific field.

\begin{table}[ht!]
    \centering
    \begin{tabular}{lllr}
        \toprule
         Technical Topics (CEAS)  & peer-reviewed & other contributions & \# \\
         \toprule
         Airframe noise  & \cite{schoder2019,58d4cf7021044654a5336023dcd859db} &  \cite{a770f65ab8b24dc7afd97a1b76090cd7,134164067f34451198bcfdd7b56237a2} & 4\\
         Fan and jet noise  & \cite{tieghi2023machine,schoder2022aeroacoustic,Schoder2022b,schoder2021application,schoder2020computational}  & \cite{0db92a333fb042b1a3a8e7304eb7c84a,1fe7f84c03de495489c030384a4f6fe4,7153b3e649af469dacb0adda2c7a5aa8,3e326e56062043229a102cf5e5492c46,02ab3dec3a04468eb8916c539a7fd9e3,1eecc4ba89524ed88a414f13c3e7620d,50d225209f8841afa18f851a89eeff7e} & 12\\
         Turbomachinery noise  & \cite{freidhager2021predicting,a7a5f9f90d4e407699bddb85b728e1c3,51c83e75474641818913eee7cd6e37a0} & \cite{c30ced592fc843ceb113af11e9b39b45} & 4 \\
         Combustion noise  & &  \\
         Helicopter noise  & &  \\
         Aircraft interior noise  & \cite{maurerlehner2022aeroacoustic,2a58295d76844cda89050563e83e0130,14263a89ed974d3ba725c5a0fac1905b,bf151632ea464ac5af366aa6387de915}& \cite{2a58295d76844cda89050563e83e0130}  & 4\\
         Propeller noise  & &  \\
         Vehicle (flow) noise  & \cite{maurerlehner2022aeroacoustic,47be353769e44b299b1e18af1ac56695,a7b148f2fca54a53a36676eb70bdc984,51c83e75474641818913eee7cd6e37a0} & \cite{9e7e9d6d05f94f74828c542019adec7d,5eb8154ec41d4d899bc73cdab9792105,67eeb2e4ab724d4cbaba0cd3071f8a23,df62da484c094a6daf9d7ec46269aae3,45fed07bd70d442e94f5d55dda73d856,93efca328af24b83b072f92fb67cdb7f,548db82fa73a4cf3906a188f23a6a6e2,c30ced592fc843ceb113af11e9b39b45,63da93f652c34f06ba215f0fcabcf617,134164067f34451198bcfdd7b56237a2} & 14\\
         HVAC Systems & \cite{tieghi2023machine,schoder2022aeroacoustic,Schoder2022b,schoder2021application,schoder2020computational}  & \cite{0db92a333fb042b1a3a8e7304eb7c84a,1fe7f84c03de495489c030384a4f6fe4,7153b3e649af469dacb0adda2c7a5aa8,bf151632ea464ac5af366aa6387de915}  & 9\\
         Techniques and methods  & \cite{schoder2019hybrid,schoder2020postprocessing,schoder2020aeroacoustic2,schoder2020radial} & \cite{dc655348e9c94901bbfd35150102f0d9,bb42f1314c734915add2b003d76893fc,4ace218804ef499d957c976c81c8e6c8,6a9f9f55f945498f966b894a625c2418,12d19d2781a1459ba820e62a68c1ffe9,666b14dcc447451f96e73bdce6ea5a95,9da4ca468a7c43f4a03f9e31e85ee981}  \\
         		 & \cite{schoder2019revisiting,schoder2022aeroacoustic,schoder2019conservative,schoder2021application,b0c32a02d34e4bcd8a9ce6ca2d10d1a5,ae778a626b8245d98920cfbab1c7f681} & \cite{kaltenbacher2021stable,toth2017infinite,1ccbd3dfdc534dc080870d7f2c452f51,1e8d2e93f98e49338108d1f15fb792aa,fb7a8e16805348f3af08e5941547bad2,494d0533a5fe4df880c8c300aae1cf04,3f42fca0fc5b4249991356b51090fb04,5a0294bbe2094cdc928b37f9a9590c2d} &  \\
          	 &  \cite{Schoder2022b,schoder2019,CFS,schoder2020helmholtz} & \cite{95d29d0088f14de3a8b2f34be37cef01,3e744d95c31c419b9d6a5baf0927054b,beaaa34fd5c34014af02d8b9f614d013,e0dfbced7b4b45698990afca6ba78f10,ce37c85896e34fd59aac9e19c70f66d6} & 34 \\
         Electric drives & \cite{76984270d1584f6c8fab3b2cc351efe2,a3363d67611f41eabc3f1d89a68d9086}  & \cite{bf151632ea464ac5af366aa6387de915} & 3 \\
         Acoustics &  & \cite{5a0294bbe2094cdc928b37f9a9590c2d,2a58295d76844cda89050563e83e0130} & 2\\
         \toprule
         Biomedical Topics & peer-reviewed & other & \# \\
         \toprule
         Phonation & \cite{lasota2023anisotropic,lasota2021impact,schoder2021aeroacoustic,schoder2020efficient,falk20213d} & \cite{8d2f5c22d71848ef9d69deb8a3ce63dd,f37b44cdf1ef4f5a872f31e9a75dabf2,dae115d0f8ee4abe854457575e0e6a62,86283181a7ae42b8b1e4cb326d6c8204}  &  \\
         Phonation & \cite{8a9a644042a1457089cb245ea872ffa5,71a8c9b2695d4d87828c5361aab32b14,78c5825cc00343b5964189a8e7e14b9d} &   & 12 \\
         Wind exposure &  &   &  \\
         Medical devices &  & \cite{6de0d7b24869404085ff2f59dde6ce0c,adc39ab9b3aa4afab442d3e5bbd18bf9,9e62472eccba4036a662df6eb7b76dc6,4cfd9ce6ce4f44f3904a59d93de43ef9} & 4 \\
         \toprule
    \end{tabular}
    \caption{Summary of my contributions in relation to the field aeroacoustics.}
    \label{tab:discipliens}
\end{table}

\subsection{Contributions as Co-Applicant of Submitted Research Grants}
\begin{itemize}
\item BMW AG – Entwicklungsvertrag (2019): Development of the SNGR\footnote{Stochastic noise generation and radiation.} methodology for the exhaust system of internal combustion engines. Using SNGR to efficiently calculate the transient turbulent effects based on statistical methods in conjunction with a RANS simulation. Compared to conventional transient flow acoustic simulation, this methodology reduces the computational effort by up to 95\%. The methodology is directly applicable to fuel cell and battery cooling systems.
\item Otto Bock Healthcare Products GmbH (2019) - Modeling and simulation of flow acoustics in a knee joint. This research project was a kickstarter for the project application to the FFG-Bridge funding. Our results show a significant reduction possibility of fluid dynamic sound for the GENIUM Prosthesis of Otto Bock.
\item Volare GmbH (2020) - Acoustics Optimisation of an Electric Ducted Fan Unit through Absorber Design and Placement. The idea of the kickstarter project was to measure the impact of absorber placements at an EDF Unit of Volares eVTOL Apeleon. This successful project was an initiation of a continuing collaborations at TU Graz.
\item ÖAW Grant, DR. ANTON OELZELT-NEWIN'SCHE STIFTUNG - Understanding Voice Disorders (2021). The goal of the project is to investigate the physical process, the cause and effect of human voice disorders. Our previously developed simulation model simVoice models the speaker. Based on simulation data and patient data, disease patterns are examined systematically. The results will provide a deep understanding of the consequences of vocal fold anomalies on the sound source generation in the larynx and how these occur in the vocal spectrum. 
\item FFG-Bridge with Otto Bock Healthcare Products GmbH - Modeling and simulation of flow acoustic in a knee joint (2021-2024). Design methods to significantly reduce the possibility of fluid dynamic sound for future Prosthesis of Otto Bock.
\item Aeroacoustics project (contractor confidential) of evaluating the wind-noise generation around the human head (2023).
\end{itemize}

\subsection{Contributions as Senior Researcher}
\begin{itemize}
%\item Human Voice 1 (details will be presented in the next chapter)
\item BMW Steyr GmbH – FFG-Bridge: Modellierung und Simulation der Strömungsakustik von Verdichtern in Turboladern von Fahrzeugmotoren, Nr.867971 (2018-2021). 
The aim of this project was to investigate special flow acoustic effects in the compressor and compressor of the exhaust gas turbocharger. Characteristic stabilization measures (connection with acoustic active structures such as cavities), hiss and rotational sound are analyzed in detail. In this project, Helmholtz decomposition was applied both in the context of acoustic continuation and as a post-processing tool.
\item AVL List GmbH – FFG-Bridge: Aeroakustik von strömungsführenden Baugruppen in Elektrofahrzeugen Nr. 874784 (2019-2022). 
As the last ongoing project initiated by my preliminary work, I would like to mention the project with AVL List. Here, the theoretical approach is taken further and the effects of turbulence modeling in pipe systems on the subsequent vibro-aero-acoustic simulation are investigated and how these can be represented computationally effectively using modern methods.
\item FWF Grant: Numerical computation of the human voice source Funding period 1 (2018-2021) - The voice is the carrier signal of speech. The process of voice production, also called phonation, can be described by the interaction between the tracheal airflow and the two elastic vocal folds in the larynx which are excited to periodical oscillations. Thus, the two oscillating vocal folds (normally between 100 Hz and 300 Hz) periodically interrupt the expiration air stream forming the primary acoustic voice signal. Although we use our voice continuously and take it for granted, the exact causalities between airflow, vocal fold dynamics, and resulting acoustic voice signal, especially for disturbed or dysphonic voice, are still not fully understood.

Our central objective was to develop an aeroacoustic computational model “simVoice” for clinical applicability in future. The “simVoice” model is a hybrid 3D-FVM (computational fluid with driven structural dynamics) and 3D-FEM (aeroacoustics) model, being optimized in computing time due to reduced complexity but still able to resolve the phonatory components to the needed degree.

Innovative scientific aspects of this project included the knowledge to which amount turbulent scales have to be resolved for sustaining critical acoustic characteristics, revealing the cause and effect chain of dynamics-airflow-acoustics for the phonation process and the first detailed numerical study on the dependencies of vocal fold dynamics towards the acoustic quality. %The expected clinical valuable outcomes of “simVoice” are to (1) help understanding pathological and physiological voice production processes, (2) identify new treatment approaches and to (3) simulate conservative and surgical treatment outcome.
\end{itemize}

\section{Scientific excellence and visibility in the furture}

\begin{figure}[ht!]
\centering
\includegraphics[width=0.79\textwidth,trim={2cm 6cm 5cm 4cm},clip]{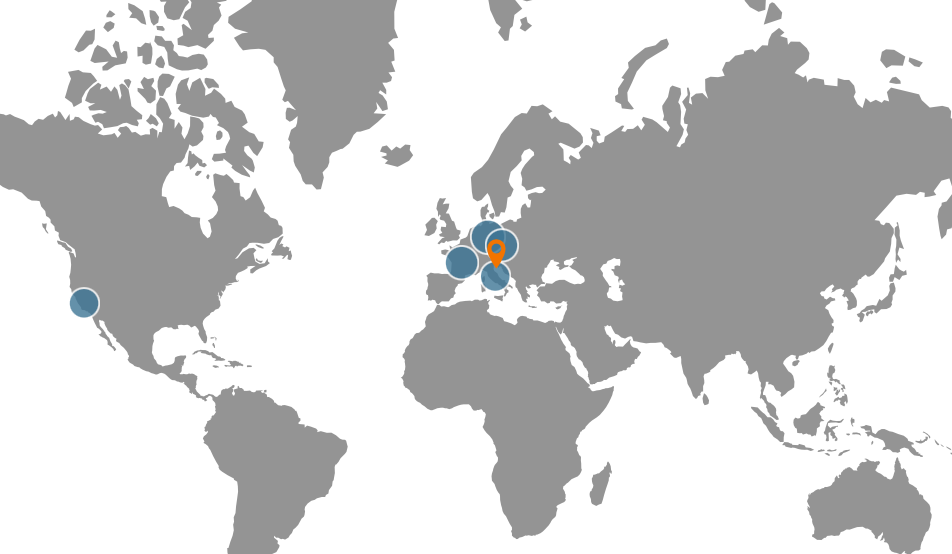}
\caption{Research map of my collaborations worldwide.}
\label{fig:res_map}
\end{figure}

The future of my research group is devoted to enable the Graz tech-clusters innovation ability and connect them to international research institutions and know-how clusters (see Fig.~\ref{fig:res_map} for my current network). Help the Austrian companies to provide better products to their premium segment customers all over the world.
Looking at the scientific content of my research, I would like to develop sound/noise prediction tools and increase understanding for specific physical applications and needs.
\begin{itemize}
\item Waves in fluids and definition of sound in fluid dynamics. (ERC Starting grants application)
\item Fluid-Structure-Acoustics interaction and the application to the human voice. (first steps of the Weave fund application)
\item Connecting FEM to fast far-field prediction methods. (PhD of F. Kraxberger)
\item Using virtualization of CFD sources like recursions and deep learning of CFD to provide CAA sources. (part of my initiative of new simulation pipeline in julia for HPC)
\item Providing a solid background for aero-acoustic model, like PCWE, cPCWE and the acoustics propagation. (ERC Starting grants application)
\item Providing benchmarking datasets for highly relevant industrial, technical and medical applications. (a starting point is the ERC Starting grants application)
\item Developing inverse models (reciprocity) for judging sound fields based on an observer.
\end{itemize}

Apart from that, we (my team and I) would like to achieve more realistic human phonation model based on the following improvement:
\begin{itemize}
\item Using auralization to mimic the real voice.
\item Using moving mesh techniques to produce vowel transition and moving VFs in computational acoustics.
\item Using MRT to extract patient data and customize the simVoice model to a myVoice model.
\item Using recursion techniques and deep learning to improve simulation speed.
\item Try to tackle the challenging unsolved fluid structure interaction (FSI) contact problem of the VFs during the oscillation cycle. (yet unsolved)
\item Using signal processing, learning techniques and the simVoice model to detect and provide a treatment assistance of voice disorders.
\end{itemize}

\chapter{List of Publications}
%TODO - LAST one was the AIAA paper 2023
The following list includes all my peer-reviewed (27), non-peer-reviewed post-doctoral publications and working papers (software implementations, software usage guides, and theoretical papers) in progress.

\printbibliography[heading=none,keyword={inbib}]

\chapter{Teaching flow-induced sound generation}
This provides a selection of teaching in flow-induced sound generation (sometimes called aeroacoustics which suggests a limitation of the theory to the fluid air) all over the world. To be honest, most research institutions and universities do provide their scientific outcome and excellence but hardly mention their study program or the course content. Thus, I provide a rather European centric view on the aeroacoustic lectures and summarized it in table \ref{tab:lectures}. More details on the courses are given in what follows.

\begin{small}
\begin{table}[ht!]
    \centering
    \begin{tabular}{l l l}
    \toprule
         Institution & Course name \\
         \toprule
         TU Braunschweig & Fundamentals of Aeroacoustics \\
         TU Braunschweig & Numerical Methods in Comp. Aeroacoustics \\
         TU Braunschweig & Methods in Aeroacoustics \\
         TU Braunschweig & Aeroengine Noise \\
         Centre Acoustipue LMFA/ERCOFTAC & Computational Aeroacoustics \\
         University Southampton & Aeroacoustics \\
         TU Eindhofen & An Introduction to Acoustics \\
         FAU Erlangen & Aeroacoustics \\
         TU Delft & Aeroacoustics: Noise Reduction Strategies  \\
          & for Mechanical Systems \\
         TU Delft & Aeroacoustics: Measurement Techniques \\
         Cambridge University & Aeroacoustics \\
         KTH Stockholm & Aeroacoustics \\
         DTU Denmark & Aeroacoustics \\
         Stanford University & Fundamentals of Acoustics \\
         Stanford University & Topics in Aeroacoustics \\
        University of Illinois Urbana-Champaign & Aeroacoustics \\
        von Karman Institute for Fluid Dynamics & Basics of Aeroacoustics and Thermoacoustics \\
         \toprule
    \end{tabular}
    \caption{An overview of selected courses related to the topic aeroacoustics in the year 2022.}
    \label{tab:lectures}
\end{table}
\end{small}

\newpage
\section{Aeroacoustics at TU Braunschweig - Prof. J. Delfs}
A solid physical background is presented and computational methods to obtain the radiated sound field are presented.
\subsection{Fundamentals of Aeroacustics}
Course description 2020: \textit{"Students acquire fundamental knowledge about sound generated aerodynamically and about sound propagation in moving media. Students know the basic terms and analytical computation methods of classical acoustics. Students know about the combination of the basic terms of acoustics and aerodynamics to aeroacoustics as an interdisciplinary topic in engineering science. Students know the basic mechanisms of aerodynamic sound generation and can explain the various phenomena related to sound propagation. Students are able to reduce applied problems in the field of aeroacoustics to the relevant equations and can identify source mechanisms. Students are able to orient themselves independently in literature on aeroacoustics."} 
\subsection{Numerical Methods in Computational Aeroacoustics (CAA)}
Course description 2020: \textit{"The reduction of noise generated by turbulent flow has become an important topic in various engineering areas such as automotive, civil, or aeronautical engineering. The field of aeroacoustics concerns the noise generation and propagation in turbulent flow and has matured dramatically in the past three decades, especially through the increasingly stringent limits imposed on the noise emissions of civil aircraft. For a continuous progress in the reduction of aircraft noise emissions numerical tools will become essential in the future to achieve an optimized low-noise design of critical aircraft components. In the last years numerical techniques evolved in the framework of Computational Aeroacoustics (CAA) that are optimized for the simulation of wave-propagation and generation in non-uniform flows. CAA can be deemed to be a subdiscipline of Computational Fluid Dynamics (CFD). However, the main objective of CAA to understand the physics of noise generation and propagation differs considerably from that of CFD such that own numerical issues and methods became neccessary. This lecture aims at introducing into the new numerical concepts of CAA. Furthermore the lecture mediates the skills that are neccessary to enable further studies of the topic with the help of current scientifical literature. A prior attendance at the lecture ‘Grundlagen der Aeroakustik’ that introduces into the physical and mathematical concepts of aeroacoustics is useful but not mandatory for an understanding of the lecture."} 
\subsection{Methods of Aeroacoustics}
Course description 2020: \textit{"Students know the essential analytical, numerical and experimental methods for the solution of aeroacoustic problems in the engineering practice. Students are aware of the strengths and weaknesses of the various methods of analysis in aeroacoustics; they can select in a targeted way the appropriate method and can assess obtained results in a critical way.
Students have insight into the parametric dependencies of different aerodynamically caused tonal and broadband sources of sound. The students are informed about methods insofar as they may apply or develop respective procedures for prediction or measurements.
The excursion conveys the practical use of experimental measurement methods for sound generated aerodynamically to the students. The contents put the students in the position to further elaborate on the experimental methods presented in the lecture and to recognize the meaning of the aeroacoustic experiment as the basis for the validation of computational methods."} 
\subsection{Aeroengine Noise}
Course description 2020: \textit{"Students possess concepts and fundamentals of aeroengine noise. Students are able to use methods for the solution of problems within the engineering field aeroengine noise; they know the basics behind equations, the modeling, and assumptions solving them. Students have insight into the parametric dependencies of various aeroacoustic (tonal and broadband) noise sources."} 

\section{Aeroacoustics at Centre Acoustique LMFA - Prof. Christophe Bailly}
The ECL and the LMFA provide a Master of Science in Acoustics, including the topics Advanced acoustics, Aeroacoustics, Active control of noise and vibrations, Transportation noise, Fluid - structure interactions, Physics of turbulent flows and Numerical simulation of flows.
\subsection{Ercoftac course in Computational Aeroacoustics}
Course description: \textit{"This course is intended for researchers in industry and in academia including Ph.D. Students with a good knowledge in fluid mechanics, who would like to build up or widen their knowledge in the field of aeroacoustics (modeling, computational tools and industrial applications). It will first provide a comprehensive overview of recent insights of aeroacoustics theories (Lighthill’s analogy and vortex sound theory, extensive hybrid approaches and wave extrapolation methods, duct acoustics). A number of practical problems involving the coupling between CFD’s results and CAA will be also thoroughly discussed (e.g. how design a mesh size for aeroacoustics applications using large eddy simulation, inclusion of mean flow effects via hybrid formulations such as the acoustic perturbation equations, presence of surfaces, aeroacoustic couplings, …) and realistic applications performed by the instructors (aeronautics, car industry, propulsion, energy,…) will be discussed. Advanced computational aeroacoustics methods will be also presented as well as what we can learn from the direct computation of aerodynamic noise. Finally, specific topics reflecting participant interests will be discussed in a final round table session. "}
% \subsection{Master of Science in Acoustics}
% The following topics are part of the aeroacoustics module: \textit{"Advanced acoustics • Aeroacoustics • Active control of noise and vibrations • Transportation noise • Fluid - structure interactions • Physics of turbulent flows • Numerical simulation of flows "}

\section{Aeroacoustics at University of Southampton - Prof. P. Joseph}
\subsection{Aeroacoustics}
Joseph describes his course content by the following bullet points: \textit{
\begin{itemize}
\item Brief review of fluid mechanics: conservation laws, thermodynamics, vortex dynamics.
\item Propagation of linear waves in moving media: linearized Euler equations, acoustics, vortical and entropy waves, the convected wave equation, basic properties of sound waves in moving media, sound refraction by non-uniform flows.
\item Acoustic impedance with flow: definition and properties of acoustic impedance, Helmholtz resonator, Ingard and Myers conditions for impedance with flow.
\item Methods for solving the wave equations: Green’s functions, Green’s formula, far field approximations, compact sources, and interferences.
\item Noise radiation by simple sources: types of sources, effect of source motion, convective amplification, the Doppler effect.
\item Sound radiation by free shear flows: Lighthill’s analogy, application to noise from turbulence.
\item Noise radiation from solid surfaces: general theory of Ffowcs Williams Hawkings and application to wave extrapolation.
\item Rotor noise: description of source mechanisms from aerofoils,
\item Duct acoustics: sound field in ducts and wave guides, properties of duct modes.
\item Turbo-machinery noise: fan rotor-alone tones, interaction tones, buzz-saw noise.
\item Aeolian tones: cavity noise, flow-acoustic feedback loops.
\end{itemize}
}

\section{Aeroacoustics at TU Eindhofen - Prof. S. W. Rienstra \& Prof. A. Hirschberg}
\subsection{An Introduction to Acoustics}
Course description: \textit{"Acoustics was originally the study of small pressure waves in air which can be detected by the human
ear: sound. The scope of acoustics has been extended to higher and lower frequencies: ultrasound and
infrasound. Structural vibrations are now often included in acoustics. Also the perception of sound
is an area of acoustical research. In our present introduction we will limit ourselves to the original
definition and to the propagation in fluids like air and water. In such a case acoustics is a part of fluid
dynamics.
A major problem of fluid dynamics is that the equations of motion are non-linear. This implies that an
exact general solution of these equations is not available. Acoustics is a first order approximation in
which non-linear effects are neglected. In classical acoustics the generation of sound is considered to
be a boundary condition problem. The sound generated by a loudspeaker or any unsteady movement
of a solid boundary are examples of the sound generation mechanism in classical acoustics. In the
present course we will also include some aero-acoustic processes of sound generation: heat transfer
and turbulence. Turbulence is a chaotic motion dominated by non-linear convective forces. An accurate
deterministic description of turbulent flows is not available. The key of the famous Lighthill theory
of sound generation by turbulence is the use of an integral equation which is much more suitable to
introducing approximations than a differential equation. We therefore discuss in some detail the use
of Green’s functions to derive integral equations.
Next to Lighthill’s approach which leads to order of magnitude estimate of sound production by
complex flows we also describe briefly the theory of vortex sound which can be used when a simple
deterministic description is available for a flow at low Mach numbers (for velocities small compared
to the speed of sound).
In contrast to most textbooks we have put more emphasis on duct acoustics, both in relation to its
generation by pipe flows, and with respect to more advanced theory on modal expansions and approximation
methods. This particular choice is motivated by industrial applications like aircraft engines
and gas transport systems."}

\section{Aeroacoustics at FAU - Prof. S. Becker}
\subsection{Aeroacoustics}
The course at the Friedrich-Alexander University will cover:
\begin{itemize}
\item Fluid mechanics basics (basic equations, turbulence, vorticity, compressible flows)
\item Acoustics (wave equation, Green's function, monopole, dipole and quadrupole)
\item Acoustic analogy (Lighthill analogy, Curle theory, Vortex sound, Ape equations)
\item Aeroacoustic measurement methods and measurement equipment
\item Hybrid numerical CAA methods
\item Applications: Tonal sound: cylinder
\item Broadband noise: Stage free-jet noise, airfoil noise, turbomachinery
\item Wind turbines, acoustics of the human voice
\item Assessment of sound perception 
\end{itemize}
Course description: \textit{Students will understand the basics of flow-induced sound radiation, its influence on humans, gain competence in the combination of fluid mechanics, structural mechanics and acoustics in the treatment of complex physical systems, can apply experimental and numerical methods to solve aeroacoustic problems, analyze current problems and applications, ranging from medical technology to vehicle acoustics, process plants to the acoustics of wind turbines.}

\section{Aeroacoustics at TU Delft - Assoc.Prof. D. Ragni}

\subsection{Aeroacoustics: Noise Reduction Strategies for Mechanical Systems}
Course description: \textit{"Learn how to reliably evaluate noise in a variety of mechanical systems and develop basic noise reduction strategies to improve their aerodynamic design and performance.
Finding the right balance between achieving maximum aerodynamic performance and lowering the noise output to comply with regulations and ensure people's comfort and health is a complicated task.
Many companies are nowadays asking for combined aeroacoustic and aerodynamic knowledge and for many engineers it can be difficult to translate complex aeroacoustic theories into practical design applications.
In this course you will acquire the knowledge and tools required in this challenging field, including aerodynamic design guidelines to maintain or improve machine performance.
You will review fundamental aeroacoustic principles and propagation theories in order to estimate the sound generated by a variety of mechanical systems. You will also explore dedicated noise-reduction strategies to understand how to improve the aeroacoustics of potential design solutions."}

\subsection{Aeroacoustics: Measurement Techniques}
Course description: \textit{"Learn how to measure noise and experimentally analyze noise‑source and reduction mechanisms, tackling the relevant fluid dynamics and acoustic parameters.
Companies are increasingly in need of professionals who can design, prototype and test aerodynamic components with integrated noise-reduction technologies.
In this course you will acquire the knowledge and tools to analyze and measure noise sources from aerodynamic systems, including microphone and beamforming applications.
More advanced aeroacoustic analogies based upon velocimetry will also be discussed, including some basics on the uncertainty of the measurements. You will review fundamental aeroacoustic principles and propagation theories and their applications to open and closed wind‑tunnels as well as for in‑field measurements."}

\section{Aeroacoustics at Cambridge University - Prof. A. Dowling}
Course description: \textit{"The students are expected to analyse and solve a range of practical engineering problems associated with acoustics. Examples include modelling of noise sources from jets, fans, wind turbines, vacuum cleaners, etc. and exploring ways to reduce noise either at the source or through acoustic damping. Upon completion of this module, the students would be well placed to pursue research in the area of acoustics and related fields. Students would also be more employable (the topics covered in the course is of interest to GE, Rolls-Royce, Dyson, Mitsubishi Heavy Industries, automobile companies and acoustic consultancies)
\begin{itemize}
\item Classical Acoustics: The wave equation and simple solutions, Impedance, Energy, Generalised functions and Green’s function, Sound from simple sources (monopoles, dipole, compact sources).
\item Jet noise: Compact quadrupole, Sound from a single eddy, Sound from a random distribution of eddies, Lighthill’s eighth-power law, Convection and refraction effects.
\item Sound propagation: Ray theory, Snell’s law, Refraction by temperature gradients.
\item Trailing edge noise: Scattering and shielding, Scattering from a source near a sharp edge, Example: Wind turbine noise and the aeroacoustics of the owl.
\item Duct acoustics: Normal modes, Concept of cut-off modes, Damping/liner, Helmholtz resonator, Example: Thermoacoustic instability.
\item Rotor/Fan Noise: Rotor alone noise, Rotor/Stator interaction noise, Examples: Aircraft noise, fan and turbine noise
\end{itemize}
}
% Georg Maierhofer Skript? g.maierhofer@maths.cam.ac.uk

\section{Aeroacoustics at KTH Stockholm - Asoc.Prof. C. O'Reilly}
Course description: \textit{"The course will contain learning activities on introductory computational aeroacoustics, fundamental acoustics, discretisation of partial differential equations, higher-order accurate methods, aeroacoustic analogies, direct sound computation, linearised acoustic propagation, hybrid aeroacoustic methods and atmospheric propagation. Lectures will be accompanied by assignments where students apply and evaluate the numerical implementation of aeroacoustic theory. These assignments will be presented to all participants and discussed in a seminar at the end of the course."}

\section{Aeroacoustics at DTU Denmark - Prof. W. Z. Shen}
Course description: \textit{"To bring the student in a position to better interpret the phenomena related to rotor aeroacoustics and noise, which include noise generation, propagation and perception. Also, it is to bring the student in a position to apply the state-of-the-art noise modelling tools in real life cases. Further the competence of working in a team will be further developed."}

\textit{"Aero-acustics concepts, Linear wave equation and acoustic analogy, Noise mechamisms of wind turbines, noise prediction models, low noise wind turbine design, noise measurements, noise propagation, parabolic wave equations, noise propagation models, computational aero-acoustics, noise perception and regulations, noise reduction techniques "}

\section{Aeroacoustics at Stanford University - Prof. S. Lele}
\subsection{Fundamentals of Acoustics}
Course description: \textit{"Acoustic equations for a stationary homogeneous fluid; wave equation; plane, spherical, and cylindrical waves; harmonic (monochromatic) waves; simple sound radiators; reflection and transmission of sound at interfaces between different media; multipole analysis of sound radiation; Kirchoff integral representation; scattering and diffraction of sound; propagation through ducts (dispersion, attenuation, group velocity); sound in enclosed regions (reverberation, absorption, and dispersion); radiation from moving sources; propagation in the atmosphere and underwater."}

\subsection{Topics in Aeroacoustics}
Course description: \textit{"Acoustic equations for moving medium, simple sources, Kirchhoff formula, and multipole representation; radiation from moving sources; acoustic analogy approach to sound generation in compact flows; theories of Lighthill, Powell, and Mohring; acoustic radiation from moving surfaces; theories of Curl, Ffowcs Williams, and Hawkings; application of acoustic theories to the noise from propulsive jets, and airframe and rotor noise; computational methods for acoustics."}

\section{Aeroacoustics at University of Illinois Urbana-Champaign - Prof. D. J. Bodony}
Course description: \textit{"Physical mechanisms and mathematical modeling of sound generation
and flow-sound interaction; An overview of aeroacoustics theories and computational approaches;
Advanced turbulence simulation techniques (DNS, LES, unsteady RANS) for evaluation nonlinear
sound sources; Accurate numerical methods and boundary conditions for direct computation of sound
generation and propagation. Both engineering and biological systems (e.g., the human voice) will be
discussed."}

\section{Aeroacoustics at VKI}

\subsection{Basics of aeroacoustics and thermoacoustics - Prof. C. Schram}
Course description: \textit{"The accurate modelling of the stability properties of energy conversion processes, such as gas turbines, industrial heaters or domestic heating systems, is facing aero-acoustical and thermo-acoustical issues hindering their economic sustainability.}

Course description: \textit{"This Lecture Series has been organized within the framework of the European Marie Curie Research Training Network (RTN) project “AETHER” (AEro-acoustical and THermo-acoustical coupling in enERgy processes), focused on the development of innovative prediction, diagnostic and control techniques for thermo-acoustical and aero-acoustical problems. The research topics address system modelling and stability analysis, the modelling of reacting and non-reacting unsteady flows and their coupling with acoustics, vibro-acoustic coupling and fatigue analysis, passive \& active control techniques. The targeted fields of application encompass gas turbines, industrial heating systems as well as domestic/district heating systems and boilers.
The aim of these Lecture Series proceedings is to present the state-of-the-art in this multi-disciplinary engineering field. The course notes start with introductory lectures on the combustion modelling and on the aero-acoustics of internal flows. The second part is devoted to aero-acoustics with courses on unsteady flow modelling and modelling of noise source and propagation. System modelling and stability analysis of the energy conversion processes are developed in the third part. The emphasis of the next note is put on thermo-acoustics with courses on experimental methods in combustion and thermo-acoustics and on passive and active control of combustion instabilities. Additional lectures notes on biomass combustion, vibro-acoustical coupling and fatigue analysis conclude this volume. For each course notes, selected test cases illustrate the capabilities of the different approaches, allowing an evaluation of their performances and a quick application in various fields of research. The main objective of these proceedings is to allow an information transfer between well-known scientists, leaders in the aero-acoustics and combustion fields, and demanding industries and laboratories. For these reasons, the course notes should appeal not only to experts already working in the domain, but also to newcomers to the field."} %For more details of the AETHER project, see http://www.cerfacs.fr/aether.

% \textit{"
% The aim of these Lecture Series proceedings is to present the state-of-the-art in this multi-disciplinary engineering field. The course notes start with introductory lectures on the combustion modelling and on the aero-acoustics of internal flows. The second part is devoted to aero-acoustics with courses on unsteady flow modelling and modelling of noise source and propagation. System modelling and stability analysis of the energy conversion processes are developed in the third part. The emphasis of the next note is put on thermo-acoustics with courses on experimental methods in combustion and thermo-acoustics and on passive and active control of combustion instabilities. Additional lectures notes on biomass combustion, vibro-acoustical coupling and fatigue analysis conclude this volume. For each course notes, selected test cases illustrate the capabilities of the different approaches, allowing an evaluation of their performances and a quick application in various fields of research. The main objective of these proceedings is to allow an information transfer between well-known scientists, leaders in the aero-acoustics and combustion fields, and demanding industries and laboratories. For these reasons, the course notes should appeal not only to experts already working in the domain, but also to newcomers to the field."}

\newpage
\section{My teaching activities}

Mostly, the lectures and topics presented in aeroacoustics at different locations have in common that they integrate the complexity of starting from the fundamentals of fluid dynamics and acoustics as a linearization of the equations of fluid dynamics. Thus, the basic acoustic equations are a first order approximation of the non-linear equations of fluid dynamics.\footnote{As Durst shows, it is also possible to derive for 1D fluid dynamics a non-linear wave equation. This equation can be used to describe the wave-speed as a function of the wave amplitude.} As a consequence, the ordinary acoustic equation is a boundary value problem that can describe vibroacoustic effects. But to be honest, in most cases the vibration is not the driving energy source and is mostly a side effect of electrodynamics or fluid dynamic excitation. As a consequence, the idea of including the pure fluid dynamic excitation into the acoustic volume sources was logic and firstly promoted Lighthill's aeroacoustic theory. This type of theory and the idea of acoustic analogies is taught in the basic aeroacoustic courses as we do at TU Graz. Upon this theory, the applications and additional simulation models discussed in the lectures are mostly driven by research and industry need\footnote{As it is logic for us in the Graz region and to serve the students with values by providing distinctive and lasting education for automotive specific, we present applications connected to their future employment status.}.

\subsection{Newly conceived independent teaching}

\subsubsection{Computational Acoustics Lecture at TU Graz}
The course will concentrate on the following topics:
\begin{itemize}
\item Basic equations of acoustics and analytic solution
\item Finite difference modeling
\item Finite Element modeling
\item Boundary Element modeling
\item Physics based machine learning for acoustics
\end{itemize}

\subsubsection{Aeroacoustics Lecture at TU Graz}
There are many application of flow induced sound in daily life: aircraft noise (jet engine), traffic noise (flow around cars, exhaust noise), musical instruments (wind instrument, organ pipe), etc. The course will concentrate on the following topics:
\begin{itemize}
\item Basic equations of flow dynamics and acoustics
\item Aeroacoustic analogies: Lighthill-, Curle- sowie Ffowcs Williams \& Hawkings Analogie, acoustic perturbation equations
\item Sound propagation computation applying Green's functions
\item Applications: edge tone, car side mirror, airfoil, rotor blades, human phonation
\end{itemize}

\subsubsection{Aeroacoustics Exercises at TU Graz}
The students will independently carry out analytical estimations and numerical simulations of problems relevant to practice in the field of flow acoustics. The students will be interactively supervised in the execution of the tasks.

\subsubsection{FEM Programming in Python Practical course at TU Wien}
Building on the content covered in the module "Fundamentals of the Finite Element Method", general topics such as version control, planning and testing of an implementation are first introduced. After the presentation of possible topics for implementation projects, the students program FE routines or extend existing programs. The implementations are validated using test cases to be developed by the students themselves. Finally, the extended program is applied to concrete problems.

\subsubsection{Online tutorial openCFS}
According to the work in the openCFS community, I structured the educational activities on flow-induced sound using the open source software openCFS. The course content can be retrieved online via www.opencfs.gitlab.io/userdocu.

\subsection{Independent teaching}

\subsubsection{Multiphysics I Lecture and Excercise at TU Wien and Graz}
In general, physical quantities describing the effects in coupled field problems are a function of space and time, and the description leads to a system of partial differential equations (PDEs). In general, these PDEs cannot be solved analytically and require the application of numerical methods such as the finite element (FE) method. The main advantage of this approach is that the coupled field problem can be computed very precisely by the physical effects as well as by the geometry and the computer simulations at a high resolution both in space and time.
In this course the basic approach to solve coupled field problems using the finite element method is taught. In detail the mathematical / physical relationships as well as their finite element formulation of the following coupled field problems are discussed: (1) electric flow field-heat conduction- mechanics (forward coupling); (2) piezoelectricity (direct volume coupling between electrostatic and mechanical field); (3) mechanics-acoustic (interface coupling). The lecture is designed to be interactive.

The students will independently perform simulations of practically relevant coupled field problems with the finite element software openCFS:
\begin{itemize}
\item electric field - heat conduction - mechanics: power semiconductors
\item piezoelectricity: piezoelectric bimorph actuator
\item mechanics-acoustics: sound radiation of a loudspeaker.
\end{itemize}
The students are interactively supervised in the execution of the tasks.

\subsubsection{Multiphysics II Lecture at TU Wien and Graz}
The accurate modelling of mechatronic systems leads to so-called multi-field problems, which are described by a system of linear partial differential equations. These systems cannot be solved analytically and thus numerical calculation schemes have to be applied. Thereby, the finite element (FE) method has been established as the standard method for numerically solving the coupled system of partial differential equations describing the physical fields including their couplings.
In detail, the course will teach the physical / mathematical modelling and its FE formulations of the following coupled fields Electromagnetics-Mechanics, Electromagnetics-Heat, and Aeroacoustics.

The students will independently perform simulations of practically relevant coupled field problems with the finite element software openCFS.

% \subsubsection{Multiphysics I Exercises at TU Wien and Graz}

% \subsubsection{Multiphysics II Exercises at TU Wien and Graz}
% The accurate modeling of mechatronic systems leads to so-called multi-field problems, which are described by a system of non-linear partial differential equations. These systems cannot be solved analytically and thus numerical calculation schemes have to be applied. Thereby, the finite element (FE) method has been established as the standard method for numerically solving the coupled system of partial differential equations describing the physical fields including their couplings.
% Students will discuss and simulate various physical applications. In detail, the course will teach the physical / mathematical modelling and its FE simulation of the following coupled fields Electromagnetics-mechanics, Electromagnetics-Heat, and Aeroacoustics.

\chapter{List of Teaching}

\paragraph{Teaching}
\begin{itemize}
    \item WS2022: UE Electrodynamics (EDYN) 437.256, Theory of Electrical Engineering (TET) UE 437.252, Strömungsakustik 437.313, 437.314, ET Seminarprojekt 437.239, Masterseminar and TI-Project 442.191 
    \item SS2022: MP2-VO 437.235 (Part Strömungsakustik), MP2-UE 437.236  (Part Strömungsakustik), Masterseminar TI 442.191, EDYN GF LU 437.141, Computational Acoustics 437.315 %(LVA aufgebaut und gehalten 3 ECTS)
\item SS2021: 437.235, 437.236, MP2 
\item WS2021: 437.236 Masterseminar TI, 442.191 Übung Strömungsakustik
\item WS2020: 442.191 Übung Strömungsakustik, 437.235, 437.236, Multiphysics 2 (CAA example - OpenFoam) WS 2020, 437.311, 437.312, 437.313, 437.314, Strömungsakustik WS2020 (Exercise Beamforming)
    \item SS2020 at TU Wien: Multiphysics 2 VO  \href{https://tiss.tuwien.ac.at/course/courseDetails.xhtml?dswid=3068&dsrid=619&courseNr=325099}{LINK}, Multiphysics 2 UE SS2019 und SS2020 \href{https://tiss.tuwien.ac.at/course/courseDetails.xhtml?courseNr=325101&semester=2020S&dswid=3068&dsrid=243}{LINK}
\end{itemize}

\paragraph{Co-Advisor, PhD Thesis}
\begin{itemize}
\item Paul Maurerlehner. (Aeroacoustics of confined flows in electrical vehicles)
\item Andreas Wurzinger. (Development of a validated simulation model for the prevention of flow-induced noise in prosthesis)
\item Florian Kraxberger. (PINNs for Helmholtz equation)
    \item Clemens Freidhager. (Aeroacoustics of Automotive Turbochargers – Rigorosum 31.März 2022)
\end{itemize}

\paragraph{Co-Advisor, Master Thesis}
\begin{itemize}
\item Klaus Schiller, (started 1. Dezember 2022)
\item Kaspar Glattfelder, Simulation of absorbers using openCFS, 2022
\item Florian Kraxberger, Understanding Human Voice Disorders, 2021
    \item Alexander Hauser, Numerical and Experimental Investigations on Resonators in Automotive Air Ducts (FFG-Bridge Compressor Sound), 2021
    \item Andreas Wurzinger, Aeroacoustic simulation of flow parts in medical and automotive applications (Projekt Numerische Berechnung der menschl. Stimme), 2020
    \item Florian Eberhart, Aeroacoustic investigations of an artifical human knee joint (Projekt Otto Bock), 2020
    \item Jakob Schmidt, Acoustic optimisation of an electric ducted fan unit through absorber design and placement (Projekt Apeleon), 2020
    \item Paul Maurerlehner, Efficient FEM Model of Human Phonation, 2020
    \item Michael Weitz, An approach to compute cavity noise using stochastic noise generation and radiation, 2019
\end{itemize}

\paragraph{Co-Advisor, Bachelor Thesis}
\begin{itemize}
\item Markus Mayr, Experimentelle Modalanalyse einer Knieprothese, 2022
    \item Olivia Gehrer, Establishing an aeroacoustic toolbox, 2022
\item Philipp Mayr, Investigation of ow stabilization in inlets of turbocharger compressors, 2021
    \item Thomas Brenn, Investigating non-conform rotating Nietsche interfaces, 2021
    \item Rupert Hiebl, Investigating the influence of structure couplings onto acoustic simulations, 2021
    \item Markus Mauerlehner, Charakterisierung von akustischen Absorbern im verbauten Zustand, 2021
    \item Tobias Florian, Benchmark of hybrid aeroacoustic workflow, 2020
    \item Alexander Hauser, Efficient FEM Model for Human Phonation, 2019
\end{itemize}

\chapter{List of Public Talks}

\paragraph{2023}

\begin{itemize}
    \item Numerical methods of Computational Aeroacoustics with a  focus on the Hybrid
Approach (\textbf{Invited}), DAGA 2023 - 49. Jahrestagung für Akustik, Hamburg, Germany
    \item Validation of aeroacoustic sound predictions in confined flows (\textbf{Keynote}), CFC 2023, 26/04/2023 Cannes, France
    \item Post-processing subsonic flows using physics-informed neural networks, AIAA Aviation
San Diego, USA
    \item Computational Aeroacoustics of Confined Flows (\textbf{Plenary}), ICTCA 2023, Monterey, USA 
    \item Post-processing flows using physics-informed neural networks, Forum Acusticum 2023, Turino, Italy
    \item Schallausbreitung und Schallanalyse (\textbf{Invited}), DEGA - Deutsche Gesellschaft für Akustik e.V.
22/09/23 Erlangen, Germany
    \item Verdichter - Theorie und Praxis (\textbf{Invited}), DEGA - Deutsche Gesellschaft für Akustik e.V.
22/09/23 Erlangen, Germany
\end{itemize}

\paragraph{2022}

\begin{itemize}
    \item Aeroacoustics of phonation (\textbf{Invited}), FLOW Autumn School on “Fluid Dynamics and Physiological Flows”
21/11/22, 
Stockholm, Sweden
    \item Aeroacoustic workflow based on Pierce’s operator to predict mixing layer sound, 28th AIAA/CEAS Aeroacoustics Conference
14/06/22 - 17/06/22
Southampton, United Kingdom
    \item Quantification of the Acoustic Emissions of an Electric Ducted Fan Unit,
2022 Delft International Conference on Urban Air-Mobility: DICUAM 2022
22/03/22 - 24/03/22
Hybrider Event, Netherlands
    \item Revisiting the EAA Benchmark for a low-pressure axial fan, DAGA 2022 - 48. Jahrestagung für Akustik
21/03/22 - 24/03/22
Hybrider Event, Baden-Württemberg, Germany
    \item CAA of rotating machinery (\textbf{Invited}), DEGA - Deutsche Gesellschaft für Akustik e.V.
22/02/20 Erlangen, Germany
    \item Live demo CAA (\textbf{Invited}), DEGA - Deutsche Gesellschaft für Akustik e.V.
22/02/20 Erlangen, Germany
    \item Schallausbreitung und Schallanalyse (\textbf{Invited}), DEGA - Deutsche Gesellschaft für Akustik e.V.
22/02/20 Erlangen, Germany
    \item Verdichter - Praxis (\textbf{Invited}), DEGA - Deutsche Gesellschaft für Akustik e.V.
22/02/20 Erlangen, Germany
    \item Verdichter - Theorie (\textbf{Invited}), DEGA - Deutsche Gesellschaft für Akustik e.V.
22/02/20 Erlangen, Germany
\end{itemize}

\paragraph{2021}

\begin{itemize}
    \item Aeroacoustics, GCCE PostDoc Meeting
25/02/21
Graz
    \item Definition of the acoustic potential by Helmholtz decomposition (\textbf{Invited}), DAGA 2021 - 47. Jahrestagung für Akustik
15/08/21 - 18/08/21
Hybrider Event, Austria
    \item SIMVOICE – Visualization of aeroacoustic sources in The human voice production Process, The 14th International Conference on Advances in Quantitative Laryngology, Voice and Speech Research
7/06/21 - 10/06/21
Bogota, Colombia
    \item Research at the TU Graz CAA Group and some Insights on Helmholtz Decomposition, Seminar Talk at ECL
24/06/21, 
Lyon, France
    \item Projects in Aeroacoustics using openCFS, openCFS-Meeting,
Austria
\end{itemize}

\paragraph{2020}

\begin{itemize}
    \item Radial Basis Function Interpolation for Computational Aeroacoustics, 2020 AIAA AVIATION FORUM
15/06/20 - 19/06/20
Virtuell, United States
    \item simVoice – Efficient acoustic propagation model of the human voice source using finite element method, 12th International Conference on Voice Physiology and Biomechanics: ICVPB 2020
2/12/20 - 4/12/20
Virtuell, France
\end{itemize}

\paragraph{2019}

\begin{itemize}
    \item Aeroacoustic formulation based on compressible flow data applying Helmholtz´s decomposition, Waves 2019: 14th International Conference on Mathematical and Numerical Aspects of Wave Propagation
26/08/19 - 30/08/19
Wien, Austria
    \item Aeroakustik von Turboladern im Automobilbereich (\textbf{Invited}), Mechatroniktag 2019
15/11/19
Wien
    \item Computational Aeroacoustics of an axial fan with leading edge serrations, 23RD INTERNATIONAL CONGRESS ON ACOUSTICS: integrating 4th EAA Euroregio 2019
9/09/19 - 13/09/19
Germany
    \item Acoustics: Source Localization and acoustic absorbers (\textbf{Invited}), Doosan Bobcat Internal Conference \& Joint Workshop with the Group on Fluids Mechanics led by the Mathematical Institute of the Czech Academy of Sciences
19/11/19 - 21/11/19
Dobřís, Czech Republic
    \item Aeroacosutic lessons learned from axial fan modeling and meausurements (\textbf{Invited}), Doosan Bobcat Internal Conference \& Joint Workshop with the Group on Fluids Mechanics led by the Mathematical Institute of the Czech Academy of Sciences
19/11/19 - 21/11/19
Dobřís, Czech Republic
    \item Strömungsakustische Analogien basierend auf kompressiblen Strömungsdaten, 
Vienna University of Technology
Getreidemarkt 9, Objekt 1, 1060, Wien, Austria
\end{itemize}

%%% BIBLIOGRAPHY %%%%%%%%%%%%%%%%%%%%%%%%%%%%%%%%%%%%%%%%%%%%%%%%%%%%%%%%%%%%%%%%
\clearpage
\printbibliography[heading=subbibintoc,notkeyword={inbib},title={References}] % print chapter bibliography

% \chapter*{Danksagung}
% %\addcontentsline{toc}{chapter}{Abstract}
% An dieser Stelle möchte ich all jenen danken, die durch ihre fachliche und persönliche Unterstützung zum Gelingen dieser Habilitationsarbeit beigetragen haben.
% Ein großes Dankeschön gilt Barbara, meiner Tochter Katharina Juliana und meinem Sohn Paul Josef, meinen Eltern Michaela und Johann und meinen Geschwistern Liesa Maria, Maria-Sophie, Johannes Michael und Gregor Benedikt SCHODER, die mich während des Studiums und dem späteren Leben hervorragend unterstützt haben. Ich danke den Partnern meiner Geschwister Mathias, Markus, Fiona und Gregors Freundin, meiner Nichte Valerie Sophie, den Eltern und Geschwistern meiner Frau, Maria und Alois, Harald, Bettina, Karin, Andrea und deren Partnern Verena, Hannes, Rula, Thomas und meinem Neffen Leonhard Alois und meiner Nichte Anna Rosa.\vspace*{0.4cm}

% \section*{Acknowledgement}
% I acknowledge my family, friends, PostDocs, Ph.D. students, students, collaborators, and people who contributed to the papers and my work. Thanks to Andreas W., Florian K. and Paul M. for proofreading. Thanks to Hugo V. and Etienne S. for being involved in the idea and generation of the title picture.

%%%%%%%%%%%%%%%%%%%%%%%%%%%%%%%%%%%%%%%%%%%%%%%%%%%%%%%%%%%%%%%%%%%%%%%%%%%%%%%%%
% \printbibliography[heading=bibintoc,title={References}] % print one bibliography for the whole document

\end{document}